\documentclass[preprint, 5p, times]{elsarticle}
\usepackage[T1]{fontenc}
\usepackage[latin9]{inputenc}

\usepackage{amsmath}
\usepackage{amssymb}

\usepackage{graphicx}
\usepackage{esint}
\usepackage{rotating} 
\usepackage{array, makecell}

\usepackage[caption=false, font=footnotesize]{subfig}
\usepackage{booktabs}
\usepackage{xcolor}

\usepackage{bbding}

\usepackage{adjustbox}
\usepackage{scrlayer-scrpage}
\usepackage{caption}

\usepackage{tabularx}
\journal{NIMA}
\usepackage{siunitx}
\usepackage{upgreek}
\usepackage[bookmarks,bookmarksopen,bookmarksdepth=6]{hyperref}
\hypersetup{
    colorlinks=true,
    linkcolor=blue,
    filecolor=magenta,      
    urlcolor=cyan
}

\usepackage{comment}

\usepackage{cases}
\usepackage{url}
\usepackage{lineno}

\newcommand{\x}{{\em x}}
\newcommand{\y}{{\em y}}

\newcommand{\f}{Fig.}

\newcommand{\venticinque}{100 $\times$ \SI{25}{\square\micro\meter}}  
\newcommand{\yceci}{\SI{150}{\micro\meter}}

\newcommand{\yv}{\SI{25}{\micro\meter}} 
\newcommand{\phip}{$\phi_{\mathrm{eq}}$ = 2.1 $\times$ 10$^{15}$ cm$^{-2}$} 
\newcommand{\phin}{$\phi_{\mathrm{eq}}$ = 3.6 $\times$ 10$^{15}$ cm$^{-2}$} 

\newcommand{\thr}{6.6\%} 
\newcommand{\thrp}{12.7\%}
\newcommand{\thrn}{9.5\%}

\definecolor{lightblue}{rgb}{0.6,0.6,1.}
\definecolor{mediumblue}{rgb}{0,0.,1.}
\definecolor{darkblue}{rgb}{0,0.,.2}

\definecolor{rootred}{rgb}{1,0.,0.}

\makeatother

\begin{document}

\title{Position resolution with \yv \ pitch pixel sensors before and after irradiation}
\author[uni]{I.~Zoi\corref{cor1}}
\author[uni]{A.~Ebrahimi~\fnref{fn1}}
\author[uni]{F.~Feindt}
\author[uni]{E.~Garutti}
\author[uni]{P.~Gunnellini}
\author[uni]{A.~Hinzmann}
\author[uni]{C.~Niemeyer}
\author[desy]{D.~Pitzl}
\author[uni]{J.~Schwandt}
\author[uni]{G.~Steinbr\"{u}ck}

\ead{irene.zoi@cern.ch}

\cortext[cor1]{Corresponding author}
\address[uni]{Institut f\"{u}r Experimentalphysik, Universit\"{a}t Hamburg, Luruper Chaussee 149, 22761 Hamburg, Germany.}

\address[desy]{Deutsches Elektronen-Synchrotron DESY, Notkestr. 85, 22607, Hamburg, Germany.}

\fntext[fn1]{Present address: {\em Paul Scherrer Institut (PSI) ,  Forschungsstrasse 111, 5232 Villigen, Switzerland}}

\begin{abstract}
Pixelated silicon detectors are state-of-the-art technology to achieve precise tracking and vertexing at collider experiments, designed to accurately measure the hit position of incoming particles in high rate and radiation environments. 
The detector requirements become extremely demanding for operation at the High-Luminosity LHC, where up to 200 interactions will overlap in the same bunch crossing on top of the process of interest. Additionally, fluences up to \SI{2.3e16}{\per\square\centi\meter} \SI{1}{\mega\electronvolt} neutron equivalent at \SI{3.0}{\centi\meter} distance from the beam are expected for an integrated luminosity of \SI{3000}{\per\femto\barn}.
In the last decades, the pixel pitch has constantly been reduced to cope with the experiments' needs of achieving higher position resolution and maintaining low pixel occupancy per channel.
The spatial resolution improves with a decreased pixel size but it degrades with radiation damage. Therefore, prototype sensor modules for the upgrade of the experiments at the HL-LHC need to be tested after being irradiated.
This paper describes position resolution measurements on planar prototype sensors with \venticinque \ pixels for the CMS Phase-2 Upgrade. It reviews the dependence of the position resolution on the relative inclination angle between the incoming particle trajectory and the sensor, the charge threshold applied by the readout chip and the bias voltage. A precision setup with three parallel planes of sensors has been used to investigate the performance of sensors irradiated to fluences up to \phin. The measurements were performed with a \SI{5}{\giga\electronvolt} electron beam. A spatial resolution of $\ensuremath{3.2 \pm \SI{0.1}{\micro\meter}}$ is found for non-irradiated sensors, at the optimal angle for charge sharing. The resolution is $\ensuremath{5.0 \pm \SI{0.2}{\micro\meter}}$ for a proton-irradiated sensor at \phip \ and a neutron-irradiated sensor at \phin. The extrapolated resolution to infinite beam momentum, where the contribution of multiple scattering can be neglected, has also been evaluated. 
\end{abstract}

\begin{keyword}
Pixel \sep Silicon \sep Sensors \sep CMS \sep HL-LHC \sep Radiation hardness \sep spatial resolution.
\end{keyword}

\maketitle

\section{Introduction}
 \label{sec:intro}

At the Large Hadron Collider (LHC) protons are brought to collision with an instantaneous luminosity of \SI{2e34}{\per\square\centi\meter\per\second}. 
As a consequence, multiple proton-proton (pp) collisions overlap in the same bunch crossing (pileup). 
Pixel detectors are situated close to the interaction point, where the track multiplicity is highest. Thanks to the fine 2D segmentation, such detectors allow the reconstruction of particle tracks and the measurement of primary and secondary vertices, even inside jets.
The instantaneous luminosity will be further increased in the future High Luminosity HL-LHC upgrade~\cite{man:HLLHC,man:HLLHCpys} to  \SI{5e34}{\per\square\centi\meter\per\second}, up to \SI{7.5e34}{\per\square\centi\meter\per\second} in the ultimate scenario, to increase the potential discovery reach of the experiments at the LHC, to study rare processes and to improve the accuracy of many Standard Model measurements. The number of pp interactions in the same bunch crossing could therefore reach an average of 200, causing a higher track multiplicity than during operation at the LHC. The HL-LHC is foreseen to operate for 10 years, delivering an integrated luminosity of \SI{3000}{\per\femto\barn} (or even \SI{4000}{\per\femto\barn} in the ultimate scenario) to the experiments, corresponding to a \SI{1}{\mega\electronvolt} neutron equivalent fluence $\phi_{\mathrm{eq}}$ of \SI{2.3e16}{\per\square\centi\meter} at  \SI{3000}{\per\femto\barn} and an ionizing dose of \SI{12}{\mega\gray} in the innermost layer of the CMS~\cite{man:CMS} pixel detector at \SI{3.0}{\centi\meter} distance from the beam~\cite{man:TDR}. 
To operate in such conditions, the experiments will undergo major upgrades for which new generations of radiation tolerant, fine pitch pixel sensors are being developed. Such sensors should achieve an efficiency higher than 98\% for a $\phi_{\mathrm{eq}}$ of more than \SI{1e16}{\per\square\centi\meter}, whilst keeping the bias voltage and the power dissipation at a manageable level. The sensor thickness and the pixel design have to be optimised to satisfy such requirement~\cite{paper:georg}. To ensure efficient tracking in a high pileup environment, the pixel occupancy per channel should not exceed the per mille level. The two-track separation will be improved, compared to the current detector performance, to allow effective track finding in highly energetic jets in the HL-LHC environment.  

\begin{figure}[t]
\centering

\includegraphics[width=0.5\textwidth]{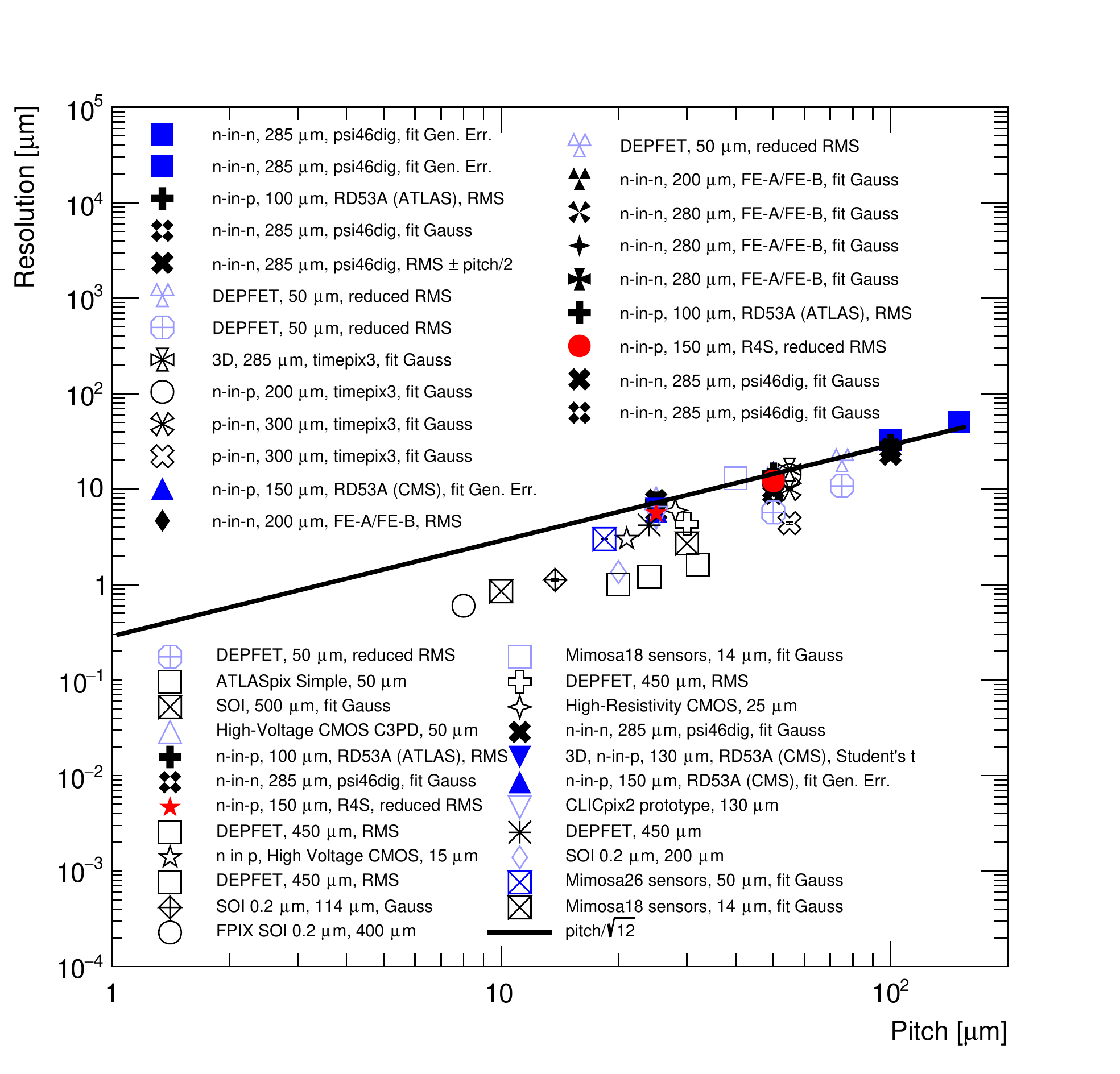}
\caption{Selection of spatial resolution measurements from the available literature performed using non-irradiated pixel detectors of different designs and pitch sizes. The measurements are ordered by decreasing pixel size. The legend reports the bulk type, the thickness, the readout chip (when applicable) and the residual width definition. The particle beams were perpendicular to the sensors. When in a paper more measurements are available, e.g.  different cluster sizes or algorithms used, the most simple and general one is reported here. Full (empty) markers refer to hybrid (monolithic) detectors. Black markers refer to hadronic high momentum beams, blue to low momentum leptonic beams, and light blue to unknown beam conditions. The red star (circle) refers to the measurement presented in this paper (samples of the same submission but different pitch) with a low momentum electron beam. The solid black line is the binary resolution of pitch~$/\sqrt{12}$. Details and references can be found in Table~\ref{table:resolution} of~\ref{appendix:reslit}. Figure~\ref{fig:literatureratio} allows to better compare the measurements.}  
\label{fig:literature}
\end{figure}

The segmentation of the detector used determines the precision of the particle's position measurement. Figure~\ref{fig:literature} shows a selection of spatial resolution measurements from the available literature performed using pixel detectors of different designs and pitch lengths.
The measurements are ordered by decreasing pitch sizes. The beam was perpendicular to the sensors for all measurements. Details and references can be found in Table~\ref{table:resolution} of~\ref{appendix:reslit}.

It is shown that reducing the pixel size greatly improves the resolution, which in many cases is better than the binary resolution of pitch~$/\sqrt{12}$. Full markers refer to hybrid detectors, where the pixel sensor is bump-bonded to a readout chip. This constitutes a limitation on the dimension of the pixel size: each channel has to be individually connected through bump-bonding to the corresponding channel of the readout chip. The reduction of the pixel cell area is therefore limited by the achievable logic density of the electronic needed to amplify, discriminate, and process the hit information~\cite{paper:reslimit}. The \SI{65}{\nano\meter} CMOS technology achieved a pixel size of $\ensuremath{50 \times \SI{50}{\square\micro\meter}}$ in the readout chip, allowing the realization of $\ensuremath{50 \times \SI{50}{\square\micro\meter}}$ and $\ensuremath{100 \times \SI{25}{\square\micro\meter}}$ sensor pixel cells, six times smaller than the one used in the current CMS pixel detector~\cite{paper:pixelp1}. The ultimate limit to the size reduction due to bump-bonding is considered to be at 5-\SI{10}{\micro\meter}~\cite{paper:bumpbonding}. 

These challenges can be avoided by employing monolithic detectors (empty markers in \f~\ref{fig:literature}), which are able to allocate high density CMOS circuitry in smaller pixel cells and therefore achieve an outstanding spatial resolution. However, this technology is not a suitable candidate for the CMS Phase-2 upgrade because of the relatively slow signal generation and limited radiation hardness.

\begin{figure}[ht]

\includegraphics[width=0.5\textwidth]{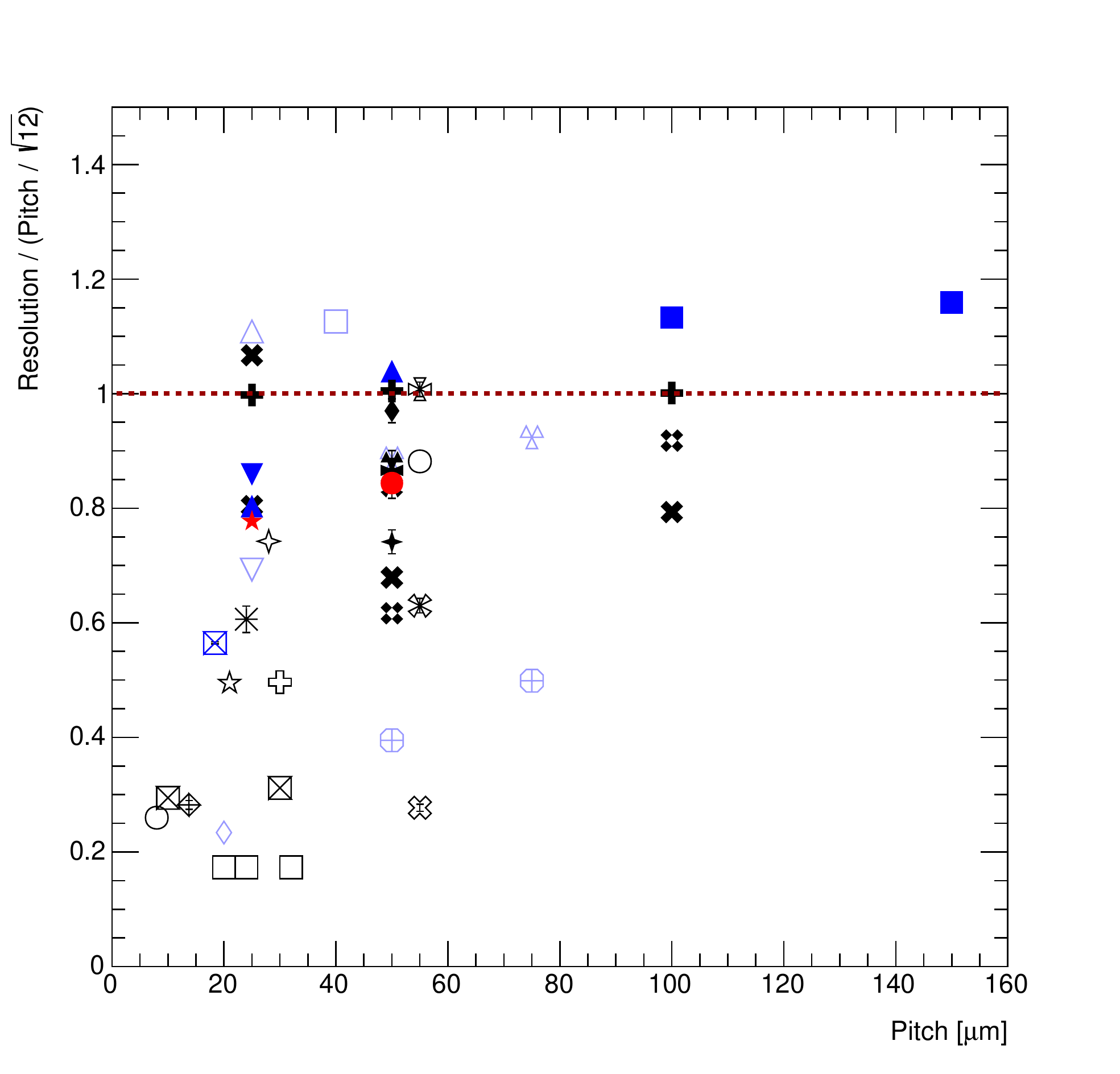}
\caption{The same measurements as in \f~\ref{fig:literature} are presented. The resolution has been divided by the binary resolution (pitch~$/\sqrt{12}$) to enhance the differences between measurements.}
\label{fig:literatureratio}
\end{figure}

Besides the size of the readout electronics, additional effects limit the reduction of the pixel pitch and the consequent improvement of the spatial resolution. Pixel pitches much smaller than the distance over which charges spread due to diffusion (4--\SI{8}{\micro\meter} for typical 150--\SI{200}{\micro\meter} thickness) would result in excessive charge sharing~\cite{paper:reslimit}. Furthermore, simulation studies~\cite{paper:ressim} on $\ensuremath{2 \times 2 \times \SI{10}{\cubic\micro\meter}}$ pixel cells show that the resolution becomes worse than pitch / $\sqrt{12}$ as the probability of a minimum ionizing particle traversing the pixel without any charge deposition is non-negligible with such small and thin pixels. The hit position resolution is additionally limited to approximately \SI{1}{\micro\meter} by the presence of delta electrons~\cite{paper:reslimit2}. Submicron precision is expected to be achieved, despite delta electrons, with stacked sensors~\cite{paper:stack1,paper:stack2} in applications where multiple scattering in the detector material does not affect the measurements.

 This paper aims to characterize the position resolution of $\ensuremath{100 \times \SI{25}{\square\micro\meter}}$ pitch pixel sensor candidates for the Inner Tracker (IT) Phase-2 upgrade of the CMS detector. To withstand the elevated radiation dose and to cope with charge carrier trapping due to radiation damage, the new sensors will have an active thickness of \yceci. Planar pixel sensors with different designs have been produced and tested for their suitability for the upgraded IT. The spatial resolution benefits from a reduced pixel size but it deteriorates with radiation damage. 
At the distance of 3.0--15.6 cm from the beam, where the barrel components will be placed, the majority of radiation comes from charged hadrons.
The pixel modules will be affected by both surface damage, mostly caused by ionizing energy loss in the readout chip, and bulk damage, due to non-ionizing energy loss in the sensor substrate. 
The creation of bulk defects changes the sensor macroscopic properties and reduces the charge collection efficiency. To choose suitable candidates for the Phase-2 upgrade, it is mandatory to evaluate the performance of irradiated sensors. Since radiation damage varies for different particle types, both proton and neutron irradiations have been performed. This paper presents the spatial resolution of a sensor irradiated with neutrons to \phin \  and of two sensors irradiated with protons to \phip, corresponding to more than 70\% of the full lifetime fluence of the second barrel layer  and to the full lifetime fluence of the third layer, respectively. 
 The purpose of this study is to measure the spatial resolution along the short \yv \ pitch direction of the $\ensuremath{100 \times \SI{25}{\square\micro\meter}}$ sensors. 
 The challenge of these measurements is to provide high resolution tracking and, at the same time, the possibility to operate the irradiated sensors at low temperature ($\approx$ \SI{-30}{\celsius}) as in the experiment. A dedicated setup consisting of three parallel sensors, which can be rotated simultaneously with respect to the beam axis, was used for the measurements. 
The spatial resolution has been measured for different fluences as a function of the beam incidence angle and momentum for various bias voltages and charge thresholds. 


\section{Pixel sensors and modules}
\label{sec:pixels}

\begin{figure}[t]
\centering
\subfloat[]{\label{fig:pstopdefault}
\includegraphics[width=0.45\textwidth]{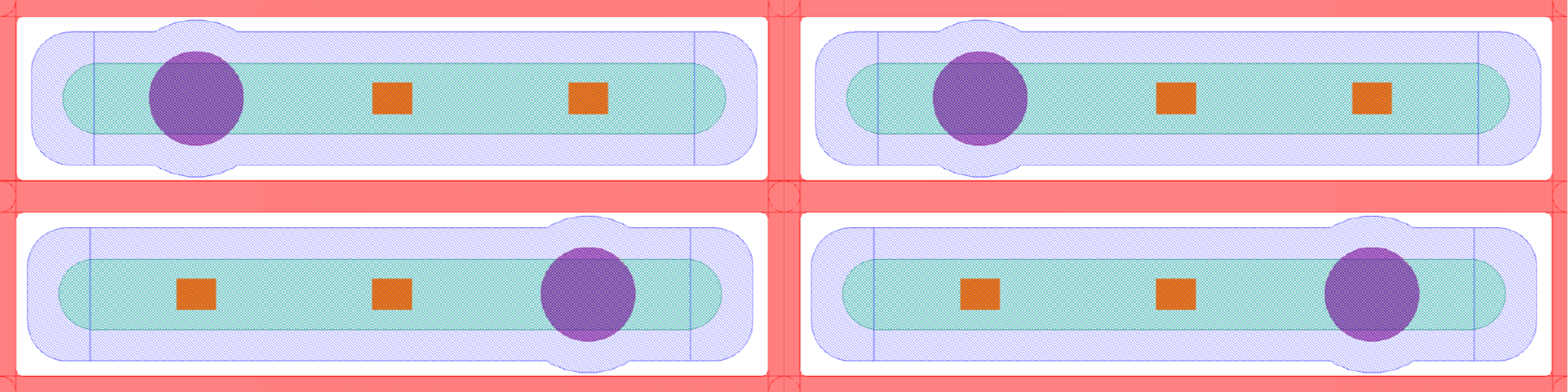}}\\
\subfloat[]{\label{fig:pstoprd53Apads}
\includegraphics[width=0.45\textwidth]{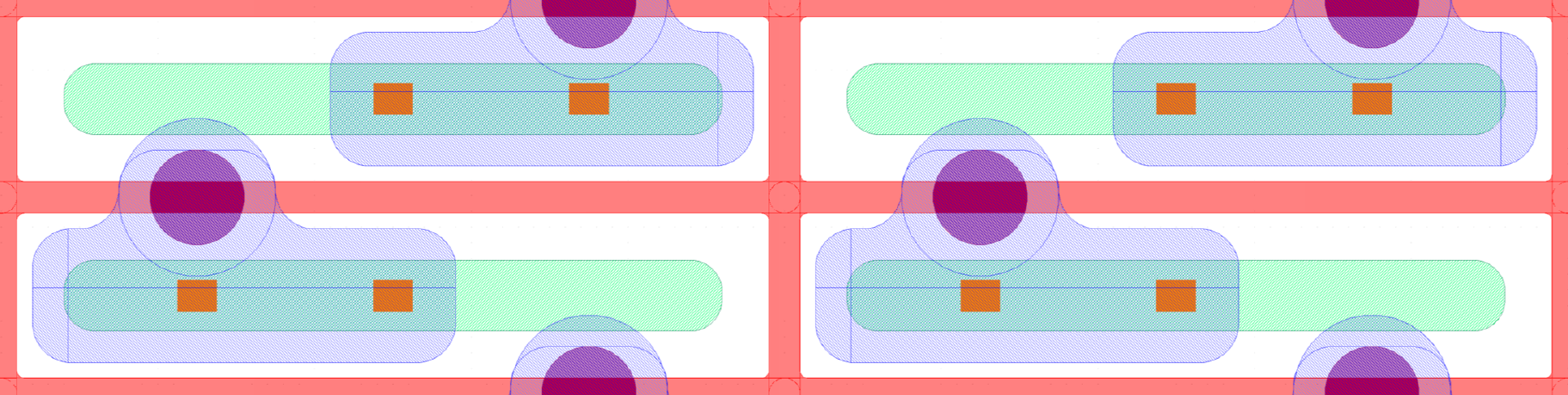}}\\

\subfloat[]{\label{fig:pspraydefault}
\includegraphics[width=0.45\textwidth]{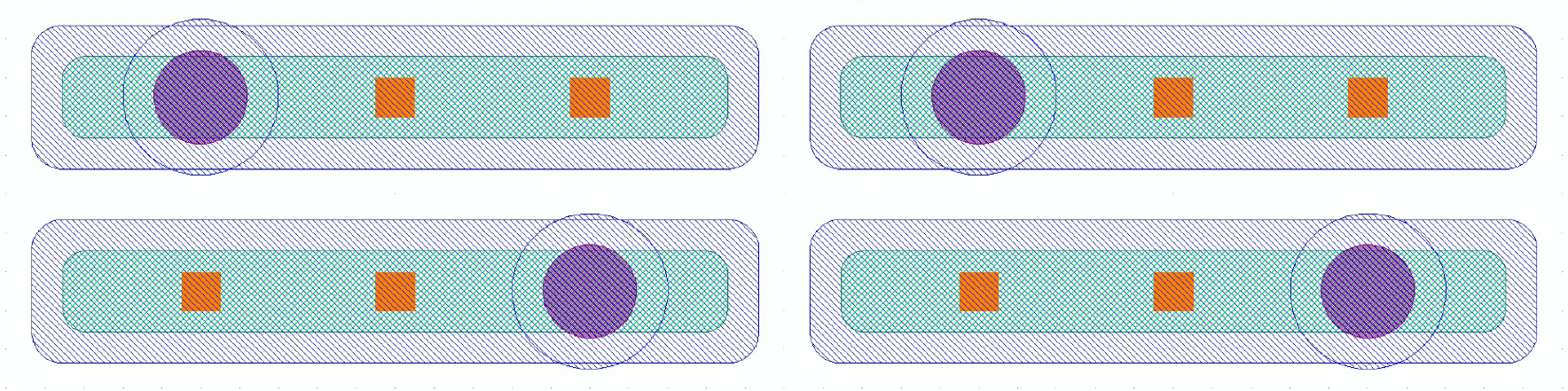}}\\
\subfloat[]{\label{fig:psprayrd53Apads}
\includegraphics[width=0.45\textwidth]{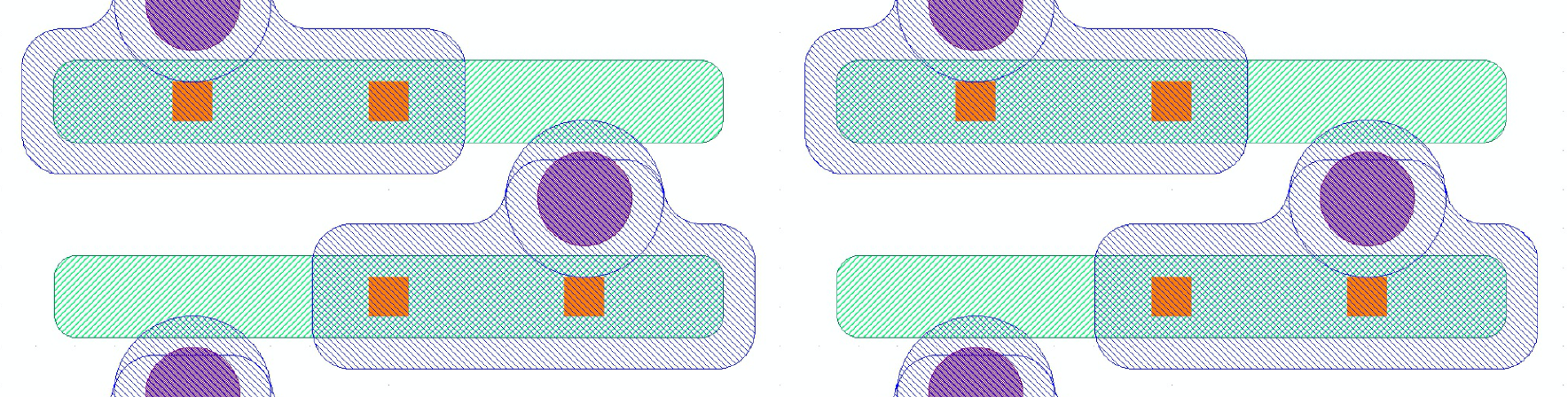}}

\caption{Top views of $2\times2$ pixel cells of the designs used as DUT for the resolution measurements. The bump-bond pads (purple circles) are arranged in a staggered pattern matching the ROC4Sens readout chip pattern. Red areas represent the p-stop, green dashed areas are n$^+$ implants and the purple dashed areas are made of metal. The orange squares identify the connection between the metal and the implant.
The layouts in \f~\ref{fig:pstopdefault} and \f~\ref{fig:pspraydefault} represent a basic design, with p-stop and p-spray isolations, respectively. The designs in \f~\ref{fig:pstoprd53Apads} and \f~\ref{fig:psprayrd53Apads} have structures intended to mimic the routing of the RD53A chip, the prototype for the final CMS readout chip, with p-stop and p-spray isolations, respectively.}
\label{fig:sensors}
\end{figure}

The sensors probed for this study~\cite{paper:designJoern} are planar n$^+$p sensors with a p$^+$ backside implant produced by Hamamatsu Photonics K.K. (HPK)~\cite{man:hpk}. They have an active thickness $t$ of \yceci \ and a pixel size of \venticinque; both p-stop and p-spray technology were tested for the pixel isolation. Several design variants have been produced to identify the ones satisfying the Phase-2 requirements. 
Results in Refs.~\cite{paper:georg,paper:designJoern,paper:finn} thoroughly describe the hit efficiency and other key properties for sensors irradiated up to a fluence of $\phi_{\mathrm{eq}}$ of \SI{14.4e15}{\per\square\centi\meter}.
All sensors used in this study have been bump-bonded to PSI ROC4Sens readout chips~\cite{thesis:R4S}. 
The ROC4Sens chip, developed at the Paul Scherrer Institute in Switzerland based on the PSI46 chip~\cite{man:psi46} (IBM 250 nm), allows for detailed charge collection studies due to its non-sparsified readout. The ROC4Sens chip has a thickness of \SI{700}{\micro\meter}. The readout chip's staggered 50$\times$\SI{50}{\square\micro\meter} bump-bond pattern matches the designs in \f~\ref{fig:sensors}, where four top views of $2\times2$ pixel cells of the designs used as Detector Under Test (DUT) in the resolution measurements presented in this study are shown. In Figs.~\ref{fig:pstopdefault} and \ref{fig:pstoprd53Apads} (\ref{fig:pspraydefault} and \ref{fig:psprayrd53Apads}) the variants have p-stop (p-spray) isolation. The designs in Figs.~\ref{fig:pstoprd53Apads} and \ref{fig:psprayrd53Apads} have structures intended to mimic the routing of the RD53A~\cite{man:RD53A} chip, the prototype for the final CMS readout chip (not covered in this study, more studies with the RD53A demonstrator chip can be found in Ref.~\cite{paper:georg}).

 Modules consisting of a single sensor bump-bonded to the readout chip are glued to a \SI{1.56}{\milli\meter} thick printed circuit board (PCB). To reduce the material budget, an open window is present in the PCB behind non-irradiated sensors. For irradiated modules, instead, \SI{50}{\micro\meter} of copper are deposited on the back of the PCB to ensure the thermal contact with the cooling elements.  

The spatial resolution of all the four variants of \f~\ref{fig:sensors} has been measured to be the same within uncertainties for non-irradiated sensors. The design in \f~\ref{fig:pstopdefault} has been measured also for two proton-irradiated sensors and a neutron-irradiated sensor. 

The proton irradiation to \phip \ was performed at the PS-IRRAD Proton Facility at CERN~\cite{man:cernps} with a beam momentum of \SI{24}{\giga\electronvolt}. The neutron irradiation to \phin \ was performed in the TRIGA Mark II reactor in Ljubljana~\cite{man:triga}. 
In both cases the readout chip was irradiated together with the sensor to which it was bump-bonded. 

To prevent annealing, the sensors are stored cold, except for irradiation, transport and handling.
To limit the leakage current during data taking, irradiated sensors were cooled to $\approx$ \SI{-24}{\celsius}, the lowest temperature achievable with the used setup. Non-irradiated sensors were tested at room temperature. 


\section{Spatial resolution measurement method}
\label{sec:method}

The measurements presented in this paper have been performed in the DESY II test beam facility~\cite{man:DESY} using an electron beam with momenta between 1 and \SI{6}{\giga\electronvolt}, a constant absolute momentum spread of $\ensuremath{\sigma_p = (158 \pm 6)}$~\SI{}{\mega\electronvolt} over the full momentum range and divergence $\ensuremath{d_{\alpha} = 1\text{\,mrad}}$. 
The EUDET-type beam telescope~\cite{man:eudet} available at the facility has a remarkable track resolution that allows measurements of the spatial resolution along the \SI{25}{\micro\meter} pitch of non-irradiated sensors~\cite{paper:georg,paper:marco}. When measuring the resolution of irradiated sensors, the material in the beam increases due to the additional cooling equipment. As a consequence, the effective track resolution is degraded, especially when rotating the DUT to perform measurements at different beam incidence angles. In fact, in the rotated DUT configuration, the material increases and a wider separation of the upstream and downstream telescope stations is required. The approach presented in this paper does not rely on an external reference tracking detector and allows for the measurement of the spatial resolution along the \SI{25}{\micro\meter} pitch of non-irradiated and irradiated sensors.

\begin{figure}[t]
\centering

\includegraphics[width=0.5\textwidth]{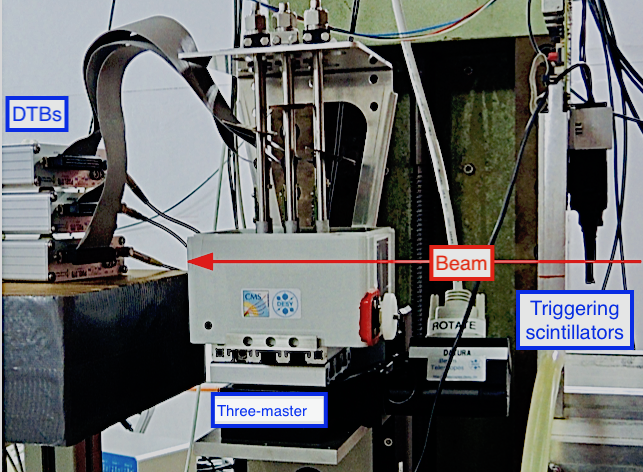} 
\caption{Image of the three-master in the DESY test beam area. The incoming beam is indicated by the red arrow. Elements of the data taking setup described in Sec.~\ref{sec:datataking} are also indicated.}
\label{fig:tb}
\end{figure}

\begin{figure}[t]
\centering

\includegraphics[width=0.5\textwidth]{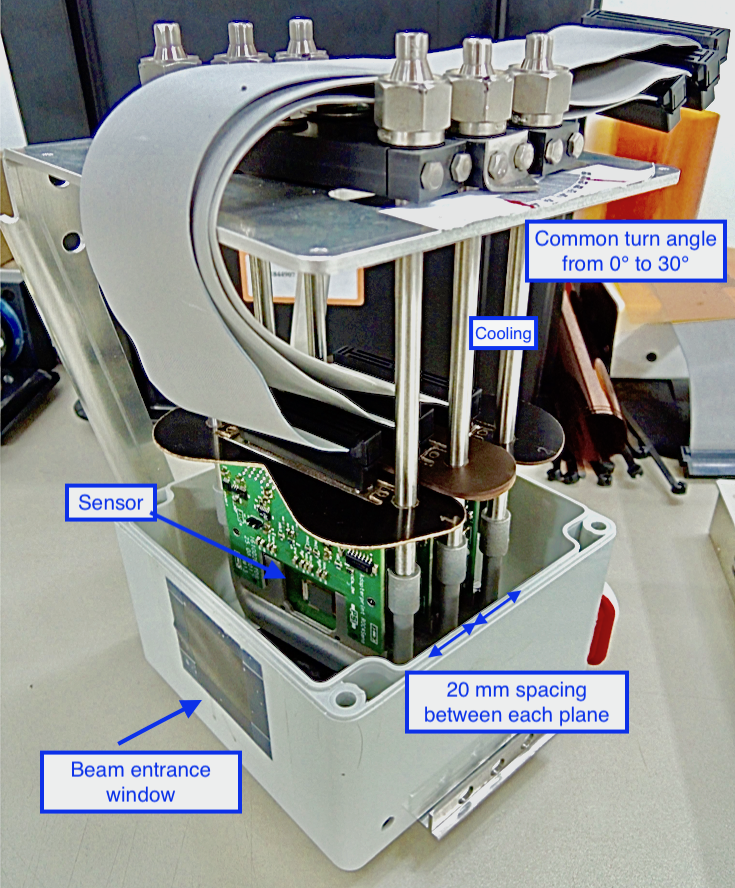} 
\caption{Image of the three-master. The three parallel equidistant planes of sensors are visible. The beam entrance window and the adjustable common turn angle are also indicated.}
\label{fig:box}
\end{figure}

Three parallel equidistant planes of sensors (referred to as three-master\footnote{The name comes from three-master, a ship with three masts.}) have been mounted along the beam axis, as shown in \f~\ref{fig:tb}. Figure~\ref{fig:box} is a close up on the three-master, where the \SI{20}{\milli\meter} spacing between planes, the adjustable common turn angle and the \SI{25}{\micro\meter} thick Kapton entrance window are visible. The metal tubes allow the circulation of a coolant liquid from an ethanol-based chiller to control the temperature of the irradiated DUTs. For thermal isolation and to prevent condensation, the plastic box can be closed, wrapped with ArmaFlex insulation and flushed with dry air. 

\begin{figure}
\centering
\includegraphics[width=0.5\textwidth]{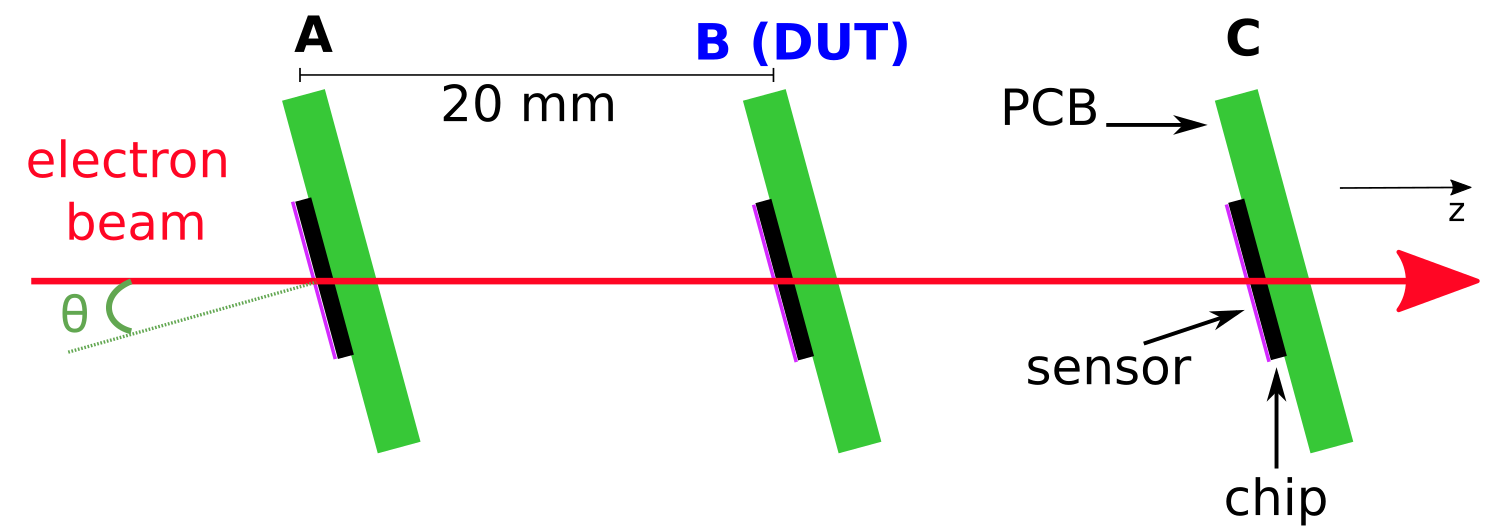} 
\caption{Sketch of the setup for the resolution measurements. Three parallel equidistant planes of modules (three-master) are mounted along the beam axis with a \SI{20}{\milli\meter} spacing between planes and an adjustable common turn angle. The two external planes (A and C) are used as reference to reconstruct the tracks of the incoming particles, while the central plane (B) is the DUT. The thin purple area represents the sensor, first element to be encountered by the beam (red arrow) in each module, followed by the readout chip (black area) and the PCB support (green area).}
\label{fig:setup}
\end{figure}

The material in front of each sensor is minimized to reduce the effect of multiple Coulomb scattering (MS) of the low energy electron beam as depicted in \f~\ref{fig:setup}, showing a sketch of the setup with the naming convention for the three planes. Thus, this method allows to perform spatial resolution measurements with a pixel size of \SI{25}{\micro\meter} at different beam incidence angles and while cooling irradiated devices. 

The two external planes, A and C, serve as reference to reconstruct the tracks of the incoming particles, while the central plane B is the DUT. \ref{sec:3mcombo} provides details on the three-master composition for the measurements described in this paper.

In the following, the method adopted to perform the resolution measurement is described. First, the data taking procedure is explained (Sec.~\ref{sec:datataking}), then the event reconstruction (Sec.~\ref{sec:recalign}) and offline analysis (Sec.~\ref{sec:selections}) are described, and finally the spatial resolution is defined (Sec.~\ref{sec:spres}).

\subsection{Data taking}
\label{sec:datataking}
The PSI ROC4Sens readout chip allows for data taking without zero suppression, making it a powerful tool for sensor studies. After receiving an external trigger, the analog signal of each pixel cell can be stored. An event rate of around \SI{150}{\hertz} is reached, limited by the USB2.0 connection between the Digital Test Board (DTB) and the data aquisition computer. The DTB is a dedicated FPGA board and features a digitizer with \SI{12}{bit} resolution. To save disk space, the digitized signals are stored for a 7$\times$7 pixel cell region-of-interest centred around a pixel above threshold.
Such seed pixel threshold is defined for each pixel as 4 times the noise, given by the Root Mean Square (RMS) of the pixel response in the absence of particles. 
The signal is corrected for baseline oscillations (pedestal) by subtracting the average pulse-height of the first and last pixel in the corresponding column of the region-of-interest.
The external trigger is provided for all the three sensors by scintillators upstream of the three-master. 
Each of the chips to which the sensors are bump-bonded is read out through a DTB that sends out a BUSY signal while the sensors' analog signals are read out. To ensure data integrity, a new acquisition can start only when none of the three DTBs is in BUSY state.

For each configuration 90k events were collected. Higher statistics was achieved in specific cases.

\subsection{Reconstruction and alignment}
\label{sec:recalign}
The track reconstruction is performed in several steps. First, contiguous pixels with a signal above the offline threshold (see \\Sec.~\ref{sec:selections}) are clustered together. The cluster charge is the sum of the charge collected by each pixel in the cluster. The cluster position is determined with the Center-of-Gravity (CoG) method from the row and column coordinates of the pixels in the cluster.
The $\eta$-algorithm~\cite{paper:stripeta1,paper:stripeta2} would yield better performance especially for small incidence angles, while the head-tail algorithm~\cite{paper:headtail} improves the resolution at large incidence angles. 
For this paper the CoG method was chosen and used consistently for all the analyses. The absolute results may therefore still be improved with optimized algorithms. Here emphasis is given to the relative variation of the resolution after irradiation and as a function of the pixel threshold around the optimal angle for resolution.

\begin{figure}[!t]
\centering
\includegraphics[width=0.5\textwidth]{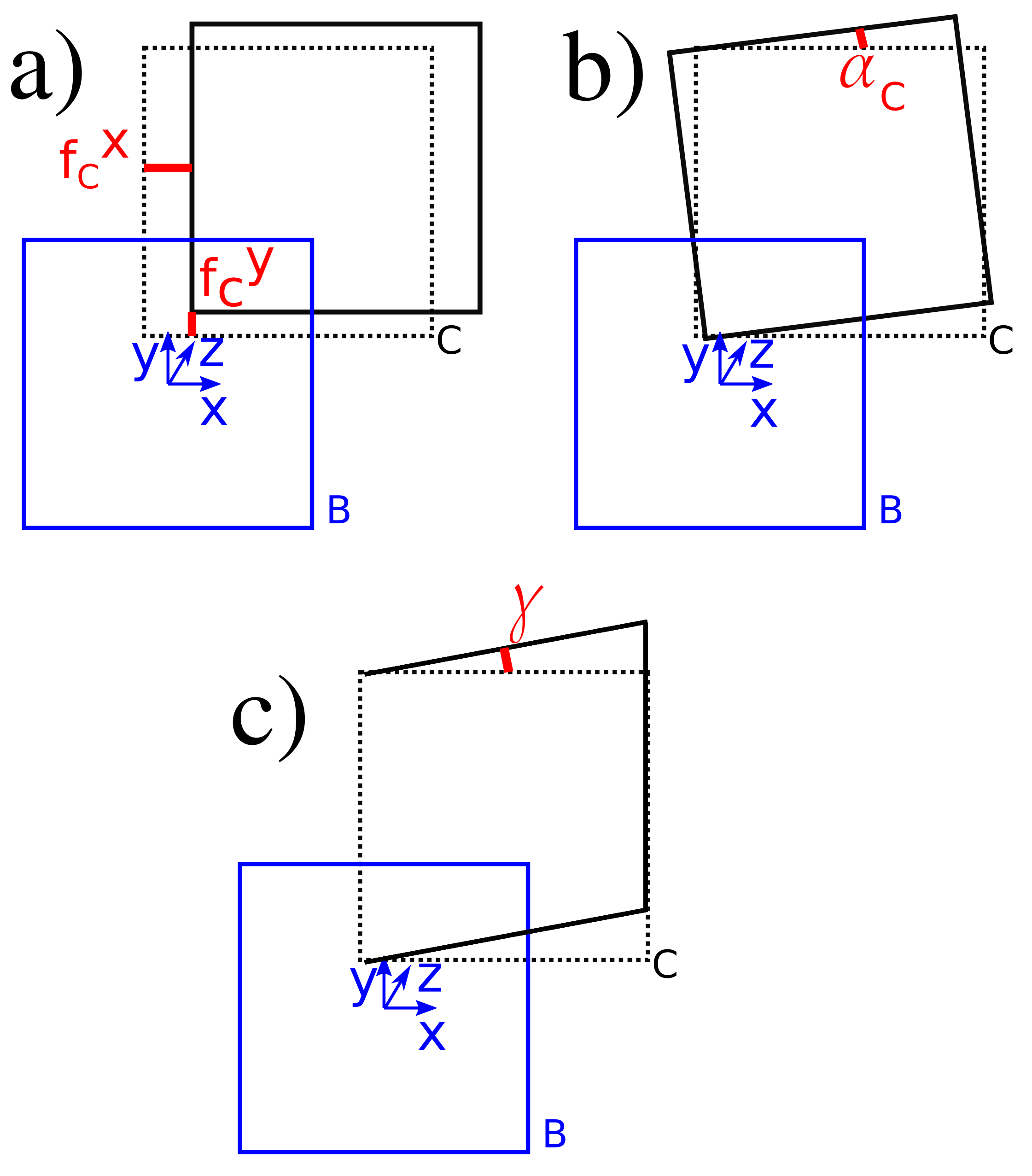} 
\caption{Schematic of the alignment parameters (in red) between planes B (in blue) and C (in black). The solid black line shows the assumed position of C while the dashed line shows the true position of C once aligned with B. The parameters with respect to B and A are similarly defined. In a), the alignment parameters $f_{\textrm{C}}^x$ and  $f_{\textrm{C}}^y$ taking into account the relative position in \x \ and \y \ of B and C are shown. In b), $\alpha_\textrm{C}$, parametrizing rotations around the $z$ axis, is indicated. In c), the correction $\gamma$, accounting for rotation around the $y$-axis, is shown.}
\label{fig:parameters}
\end{figure}

To reconstruct the tracks of the incoming particles, each plane's cluster local (row, column) coordinates are translated in the global (\x, \y) coordinate system. 
 The origin of the global coordinate system coincides with the center of sensor B, with the \x-axis defined along the \SI{25}{\micro\meter} pitch:
\begin{align*} 
 x & = \textrm{row} * \textrm{pitch}_x- l_x/2 \\ 
 y & = \textrm{column} * \textrm{pitch}_y- l_y/2, 
  \end{align*}
with the pitch$_x = $ \SI{25}{\micro\meter} and pitch$_y = $ \SI{100}{\micro\meter}, and $l_x=$ \SI{8}{\milli\meter} and $l_y=$ \SI{7.8}{\milli\meter} being the sensor's dimensions.
 The global coordinate system takes into account also the differences between the assumed and the actual location of the detectors, corrected through an iterative alignment procedure. 
 In the alignment procedure the three three-master planes are assumed to be parallel, 
the position of the central plane B is fixed and the two external planes A and C are aligned with respect to B.
The alignment parameters described in the following are drawn in red in \f~\ref{fig:parameters}, where for clarity only  planes B and C are shown. 
Four parameters ($f_{i}^x$, $f_{i}^y$ with $i$=A, C) take into account the relative position in \x \ and \y \ of the central and external planes (\f~\ref{fig:parameters}, a)) and two parameters ($\alpha_\textrm{A}$ and $\alpha_\textrm{C}$) allow for rotations around the beam axis $z$ (\f~\ref{fig:parameters}, b)).
One additional parameter $\gamma$ (\f~\ref{fig:parameters}, c)) is introduced once a preliminary alignment is in place to correct the \x \ coordinate for rotations around the \y-axis, initially excluded by the assumption of parallel planes: $x_i \rightarrow (1+\gamma)x_i$, with $i$=A, C. 
Once all transformations and corrections are applied to the coordinates, the resulting coordinate system is skewed and non-Cartesian.
The beam incidence angle $\theta$ in the $xz$-plane is measured with respect to the normal to the sensors.
The alignment procedure is repeated when sensors in the three-master are exchanged or when the beam incidence angle $\theta$ is varied, as this is achieved through mechanical movements of the setup. Each time the procedure is iterated until the variation of the alignment parameters with respect to the preceding iteration is stable and consistent with zero. A sub-micron (\SI{0.1}{\micro\meter} in $x$ and \SI{0.5}{\micro\meter} in $y$) precision is achieved for $f_{i}^x$ and $f_{i}^y$, while a sub-mrad precision is reached for the rotation parameters (0.09-0.2 mrad). The precision has been evaluated for a proton-irradiated sensor at the optimal angle and measurements performed with three different thresholds are consistent with the reported values.

\begin{figure}[!t]
\centering

\includegraphics[width=0.5\textwidth]{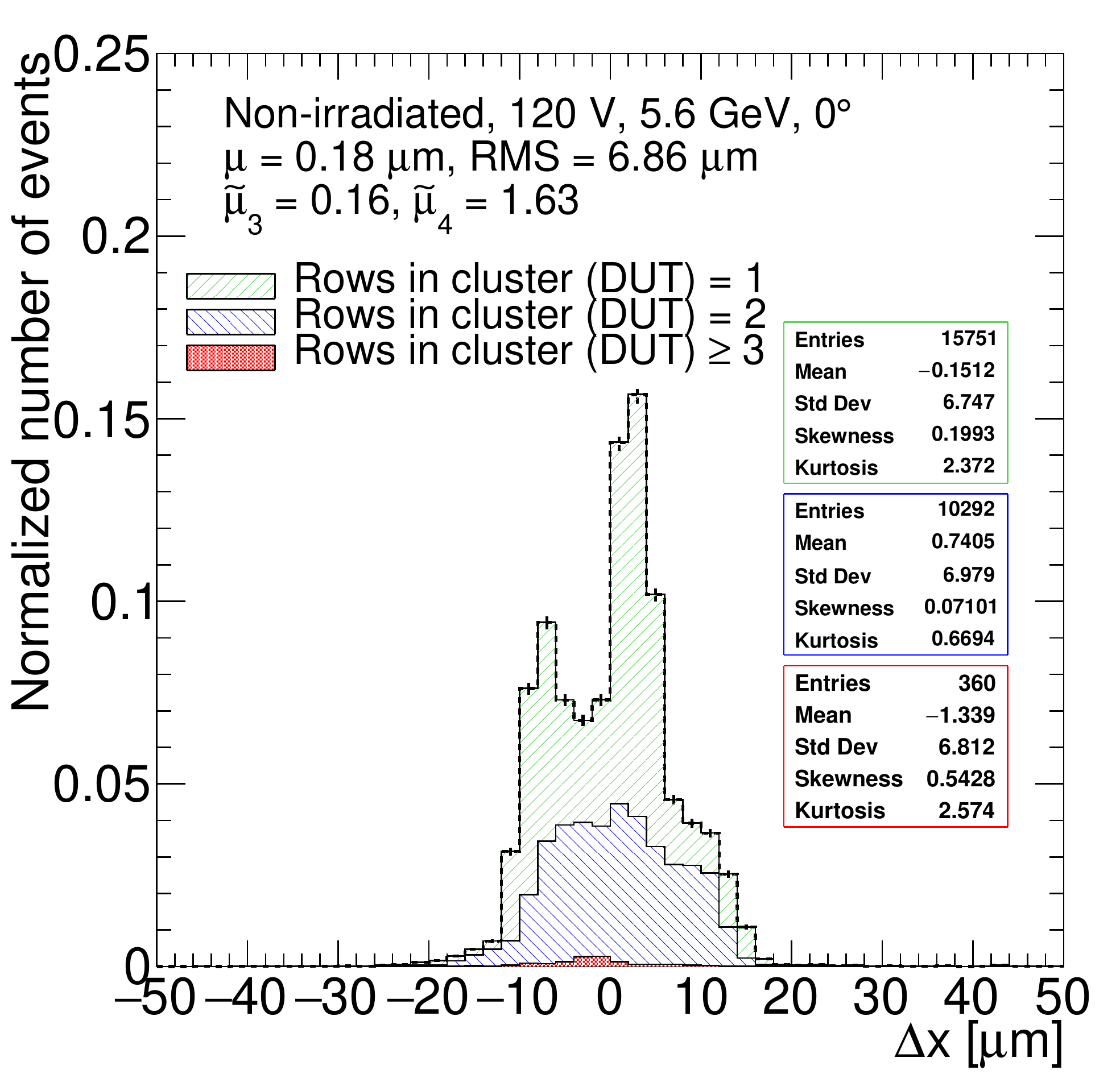}\\

\caption{Residuals, $\Delta x$, for a non-irradiated sensor at vertical beam incidence angle. Filled areas of different colors represent the stacked contribution of different size clusters. Mean ($\mu$), RMS, skewness ($\tilde{\mu}_3$) and kurtosis ($\tilde{\mu}_4$) combining all cluster sizes (legend) and for the different cluster sizes (boxes) are also reported.}
\label{fig:dx3nonirrVert}
\end{figure}

\begin{figure}[!t]
\centering

\subfloat[]{\label{fig:dx3nonirrbest}
\includegraphics[width=0.5\textwidth]{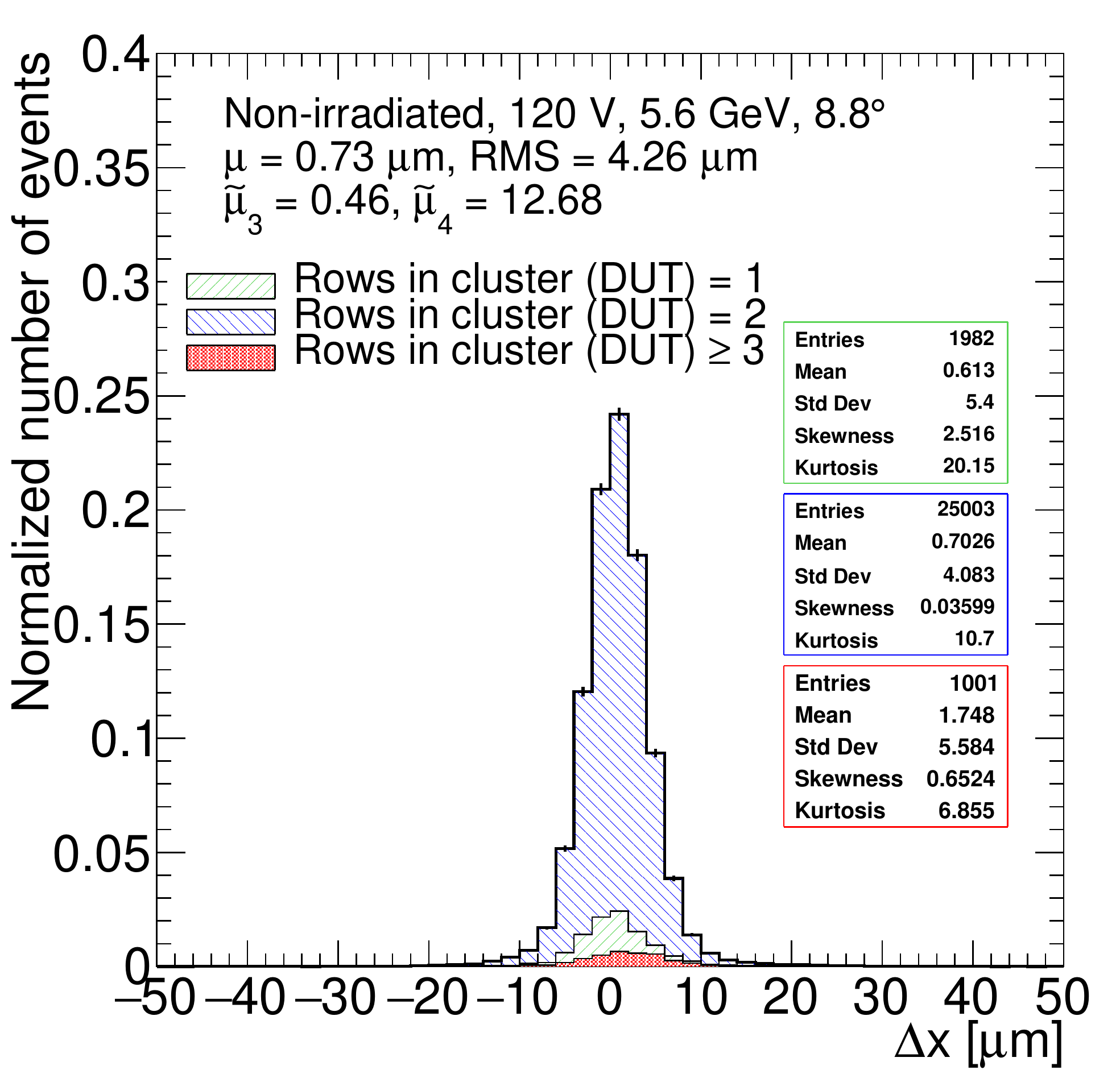}}\\
\subfloat[]{\label{fig:dx3nonirrShallow}
\includegraphics[width=0.5\textwidth]{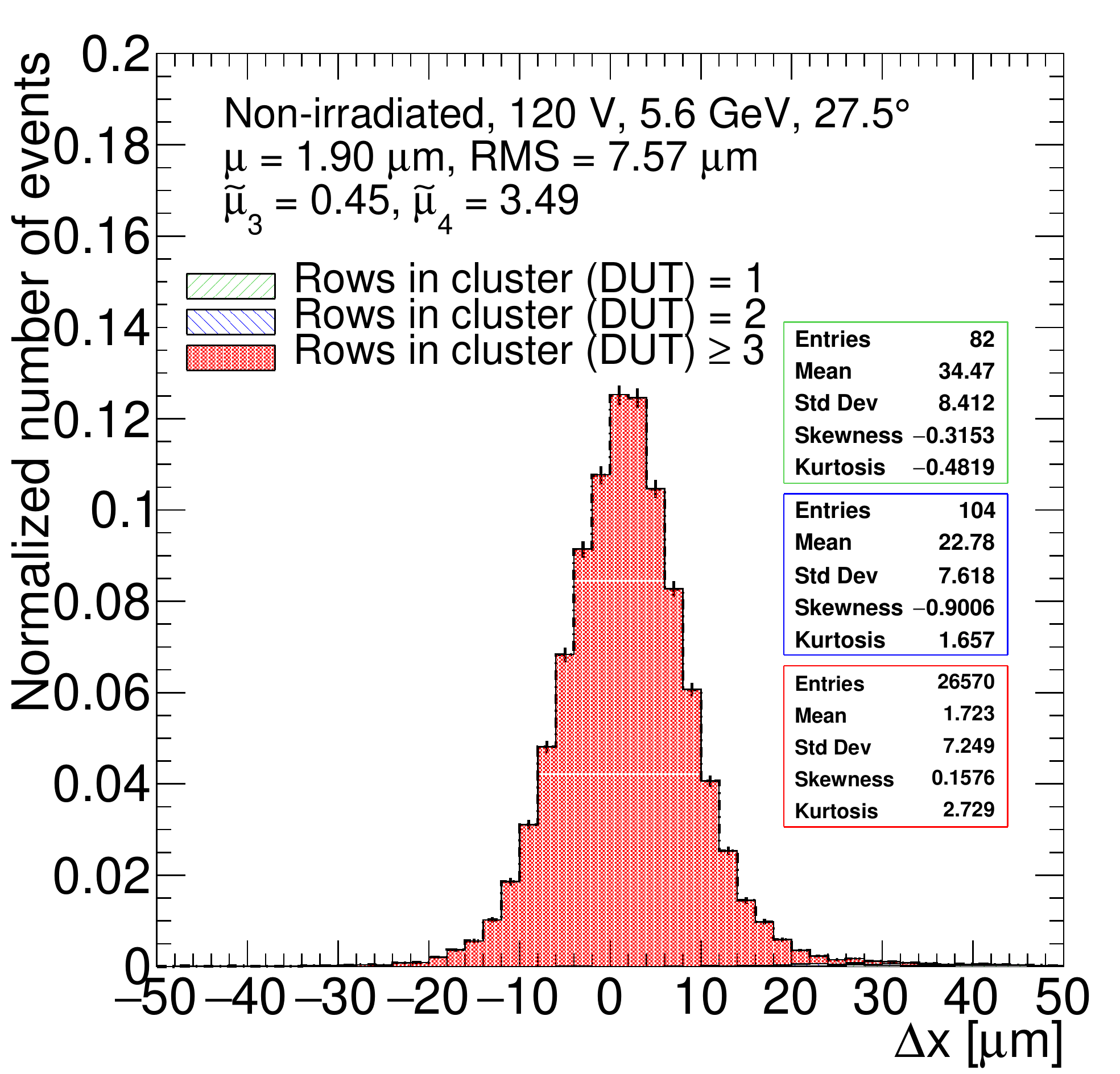}}

\caption{Residuals, $\Delta x$, for a non-irradiated sensor for the optimal (a) and a shallow beam incidence angle (b). Filled areas of different colors represent the stacked contribution of different size clusters. Mean ($\mu$), RMS, skewness ($\tilde{\mu}_3$) and kurtosis ($\tilde{\mu}_4$) combining all cluster sizes (legend) and for the different cluster sizes (boxes) are also reported.}
\label{fig:dx3nonirrcl}
\end{figure}

A problem related to the use of the CoG algorithm with threshold cuts can be observed in \f~\ref{fig:dx3nonirrVert}. The residual
\begin{equation}
\label{eq:deltax}
\Delta x = x_{\textrm{B}} - \frac{x_{\textrm{A}} + x_{\textrm{C}}}{2}
\end{equation}
distribution for events with cluster size 1 in the central plane has a 2-peak structure. The cluster size, here and in the following, is defined as the number of rows in a cluster. The peaks are separated by pitch$_x$/2 = \SI{12.5}{\micro\meter}. They are due to events with cluster size 1 in all three sensor planes which are reconstructed at the centres of the pixels independent of the particle position. If the angular divergence of the beam and the distance between the sensors are small, a fraction of events will have the same reconstructed position in all three sensors and peaks appear in the $\Delta x$ distribution. If the reconstructed position in one of the outer planes differs by pitch$_x$ and the position in the other outer plane remains the same, the prediction in the central plane changes by pitch$_x$/2, which explains the observed distance between the peaks. The actual position of the two peaks in the $\Delta x$ distribution and their frequency depends on the relative alignment of the pixels in the three sensor planes on the micrometer scale. Therefore, the determination of the resolution has a significant systematic uncertainty if the fraction of cluster size 1 events is large, which is the case for small angles of the particles to the normal of the sensors. It is assumed that for larger angles, in particular for the optimal angle, the effect of this correlation on the reconstructed positions is negligible, as the number of 1 pixel clusters is reduced. In \f~\ref{fig:dx3nonirrcl}, residual distributions at larger angles are presented, showing a single peak structure.

\subsection{Cluster quality requirements}
\label{sec:selections}
Only clusters fulfilling specific requirements are used in the track reconstruction.
First of all, to reduce spurious hits from noise, an offline threshold is applied to all pixels. A pixel is considered in the subsequent clustering step if its pulse-height after pedestal subtraction is higher than the offline threshold. The offline threshold has been selected as the one providing the best performance in terms of resolution at the optimal angle for charge sharing. A too low offline threshold results in including noisy pixels in the clusters. If it is too high, the detection efficiency will be degraded. An extensive threshold scan is presented in~\ref{sec:additional}. The effect of different thresholds on the resolution measurement is described in Sec.~\ref{sec:results}. The thresholds are expressed as a percentage of the Most Probable Value (MPV) of the cluster charge distribution. The MPV is obtained from the fit of a Landau function convolved with a Gaussian function to the cluster charge distribution. To improve the precision of the position measurements, different thresholds were applied depending on the type and level of irradiation to take into account the different amount of noise in the sensors: \thr \ for non-irradiated sensors, \thrp \ for proton-irradiated sensors at \phip \ and \thrn \ for neutron-irradiated sensors at \phin. 
The average seed pixel thresholds, introduced in Sec.~\ref{sec:datataking}, and offline thresholds are summarized in Table~\ref{table:adc}.

\begin{figure}[t]
\includegraphics[width=0.5\textwidth]{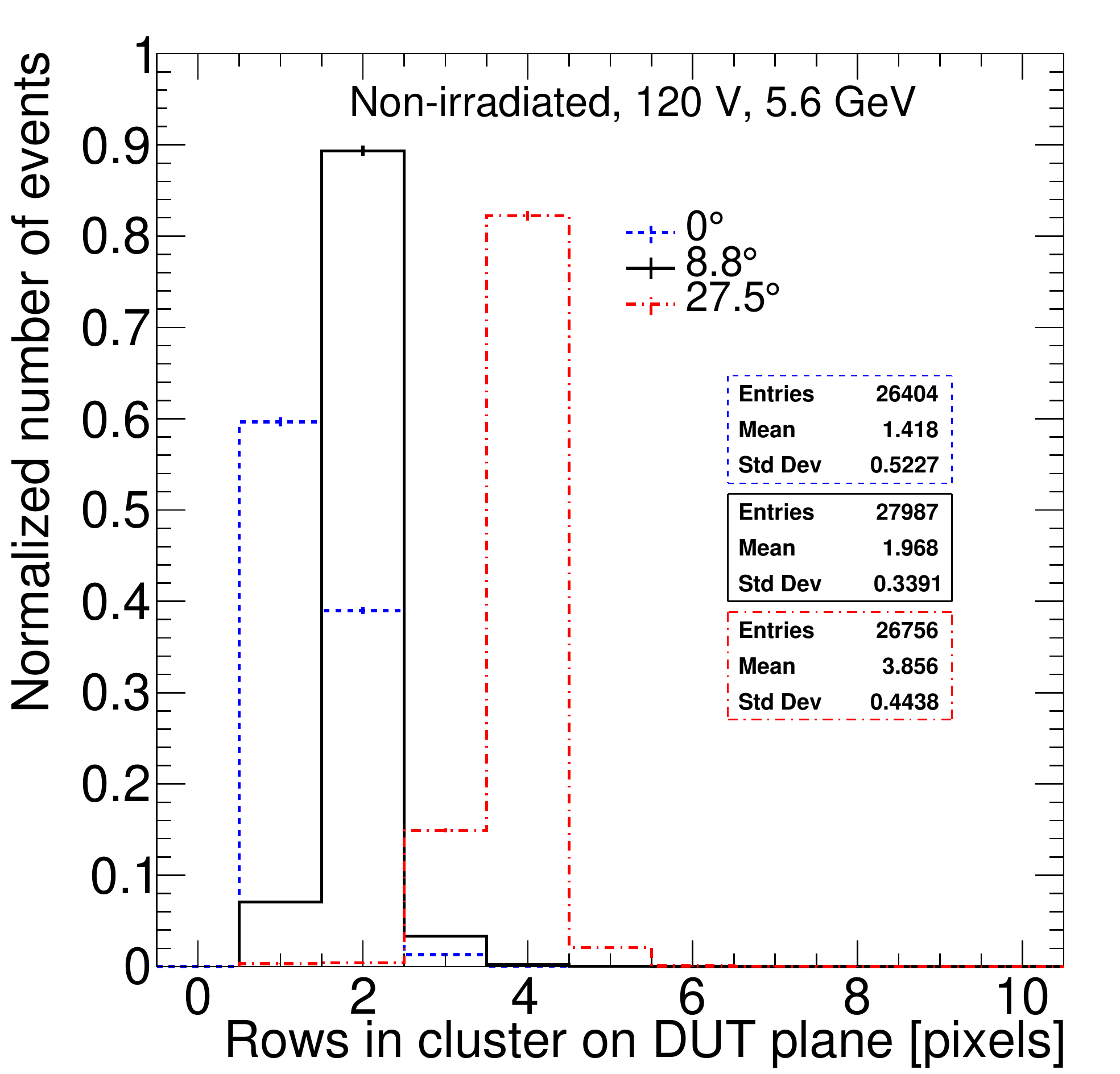}\\
\caption{Cluster size. Different line colors and styles refer to different beam incidence angles $\theta$. A threshold at \thr \ of the cluster charge MPV is applied.}
\label{fig:clsizenonirr}
\end{figure}

 \f~\ref{fig:clsizenonirr} shows the cluster size distributions for a non-irradiated sensor at different beam incidence angles $\theta$ with the \thr \ threshold applied.
 
\begin{table}[h]

\begin{tabular}{c|ccc}
Irradiation & Average seed   & Offline thr. & MPV\\
$[\phi_{\mathrm{eq}}$/\SI{e15}{\per\square\centi\meter}$]$ & pixel thr. [\%] &  [\%]  & [ADC counts]\\
\hline
0 & 7.5 & 6.6 & 183\\
 2.1, proton& 20.2 & 12.7 & 118 \\
 3.6, neutron & 11.6 &9.5 & 159 \\
\end{tabular}
\caption{The table lists the average seed pixel thresholds during data taking and the offline thresholds used in this analysis as a percentage of the cluster charge MPV, for the fluences under study. The MPV is obtained from the fit of a Landau function convolved with a Gaussian function to the cluster charge distribution at an angle of~\SI{8.8}{\degree} for the non-irradiated and the proton-irradiated sensor, and an angle of~\SI{12}{\degree} for the neutron-irradiated one. The non-irradiated module was operated at a bias voltage of 120~V and room temperature. The irradiated ones were measured at a bias voltage of 800~V and T $\approx$ \SI{-24}{\celsius}.}
\label{table:adc}

\end{table}

For a non-irradiated sensor, typically one cluster per event is recorded in each plane. In $\approx$5\% of the events with at least one cluster, also a second (noise) hit is recorded. The number of events with more than two hits is $<< 5$\%.

If the distance between any of the pixels in two different clusters is lower than \SI{600}{\micro\meter}, only the cluster with the higher charge is used in the track reconstruction to eliminate remaining noise contributions. Furthermore, since more clusters can be present outside the \SI{600}{\micro\meter} radius, only the cluster with the smallest $d_{xy} = \sqrt{(x_\textrm{B} -(x_\textrm{A}+x_\textrm{C})/2)^2+(y_\textrm{B}-(y_\textrm{A}+y_\textrm{C})/2)^2}$ is considered, as that is assumed to originate from a real particle track. 

Given the assumption in the alignment step of incoming particles parallel to the $z$-axis, the following requirements are also applied:
\begin{align*} 
|x_\textrm{A}-x_\textrm{C}|,~|y_\textrm{A}-y_\textrm{C}| &< d_{\textrm{AC}} \cdot 3~d_{\alpha} \cdot 5\text{[GeV]}/\textrm{p}_{\text{beam}} \nonumber \\
|y_\textrm{B}-0.5\cdot(y_\textrm{A}+y_\textrm{C})| &< 2\cdot \textrm{pitch}_y/\sqrt{12}, \stepcounter{equation}\tag{\theequation}\label{eq:parallel} \\
\end{align*}
where $d_{\textrm{AC}}$ is the \SI{40}{\milli\meter} distance between the A and C planes, the factor 3 is to allow a 3$\sigma$ tolerance on the beam divergence  $d_{\alpha}$ introduced above, 5[GeV]/$p_{\textrm{beam}}$ takes into account the dependence of the beam divergence from the beam momentum, normalized to \SI{5}{\giga\electronvolt}. This restricts the analysis to straight tracks and reduces the combinatorics in track seeding. 

\begin{figure}[!ht]
\centering

\includegraphics[width=0.5\textwidth]{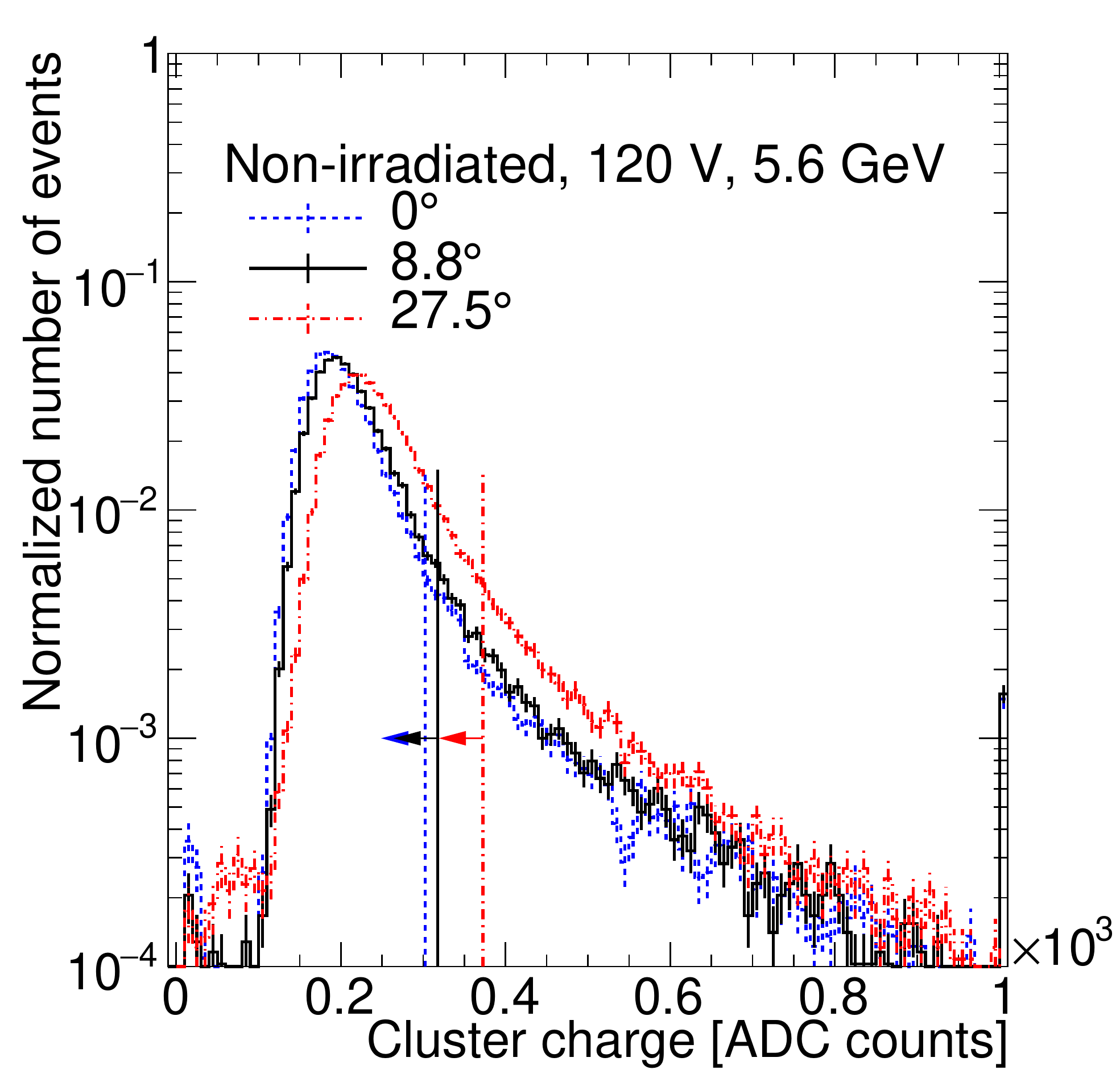}\\
\caption{Cluster charge of the DUT sensor. Different line colors and styles refer to different beam incidence angles $\theta$. The vertical lines and arrows indicate the parts of the distributions that are kept. Underflow and overflow bins are also shown. A threshold at \thr \  of the cluster charge MPV is applied.}
\label{fig:landaunonirr}
\end{figure}

To reduce the contribution of delta electrons on the resolution whilst maintaining high statistics, hits above the 90\% quantile of the cluster charge distribution are rejected on each of the three three-master planes. This is illustrated in \f~\ref{fig:landaunonirr} where cluster charge distributions for a non-irradiated sensor at different beam incidence angles $\theta$ are shown. The vertical lines and arrows indicate the parts of the distributions that are kept.  
The pixel charge distributions with and without this requirement in place are shown in~\ref{sec:additional}. 


Unless stated otherwise, the quality requirements illustrated in this section are always applied in the following.

\subsection{Spatial resolution definition}
\label{sec:spres}

Clusters fulfilling the quality requirements described in \\ Sec.~\ref{sec:selections} are used to reconstruct the tracks, defined as all possible combinations of straight lines linking a cluster in the A plane with a cluster in the C plane. 

 Each track is then interpolated to the central plane, in order to measure the residual difference $\Delta x$ between the measured impact point on the DUT sensor and the predicted impact point from the track reconstruction, defined in Eq.~\ref{eq:deltax}.

\begin{figure}[t]
\centering
\includegraphics[width=0.5\textwidth]{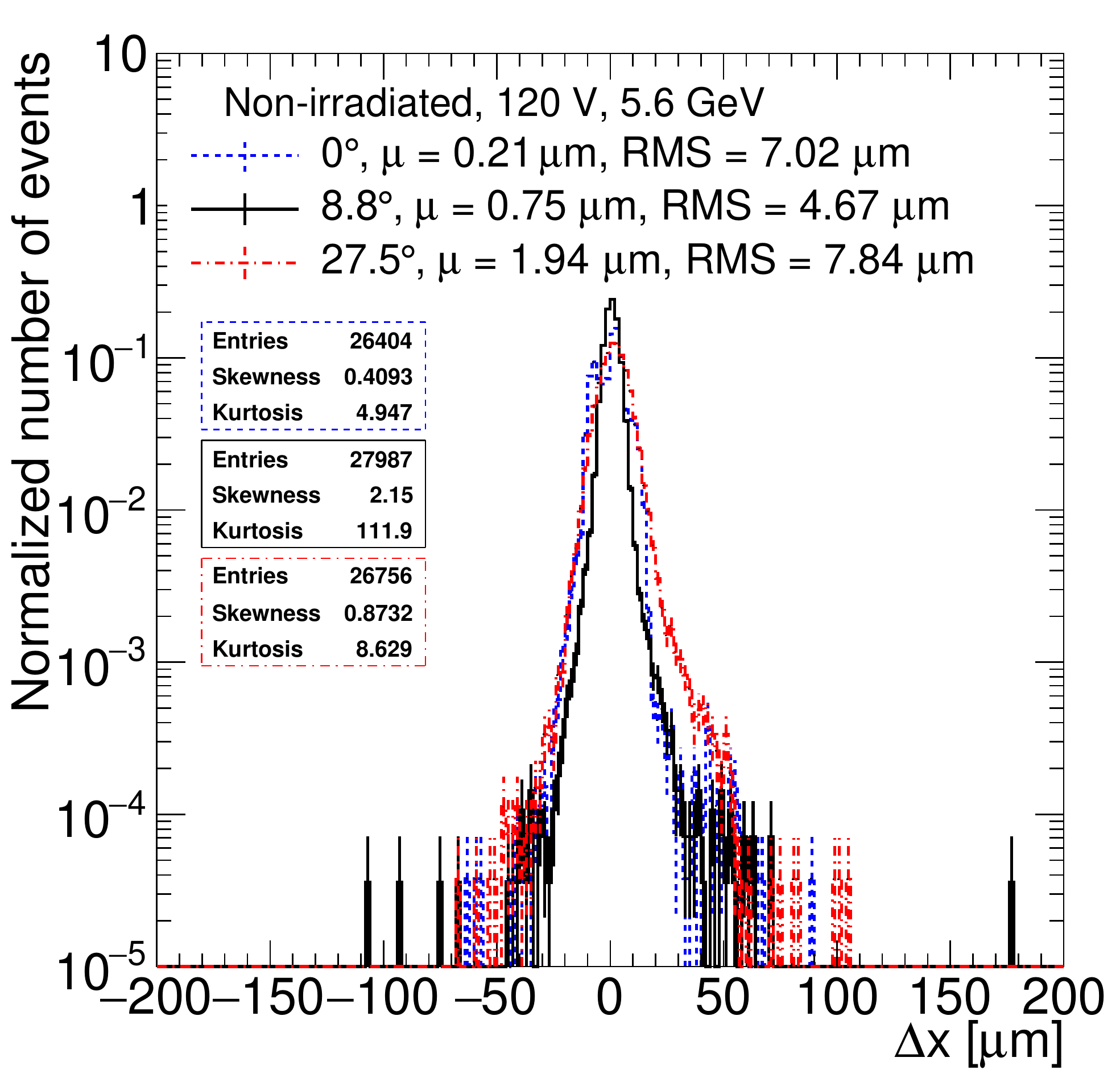}

\caption{Residual distributions, $\Delta x$. Different line colors and styles refer to different incidence angles $\theta$. A threshold at \thr \ of the cluster charge MPV is applied. A logarithmic scale is used to show the distribution tails.}
\label{fig:dx3lognonirr}
\end{figure}

\begin{figure}[t]
\centering
\includegraphics[width=0.5\textwidth]{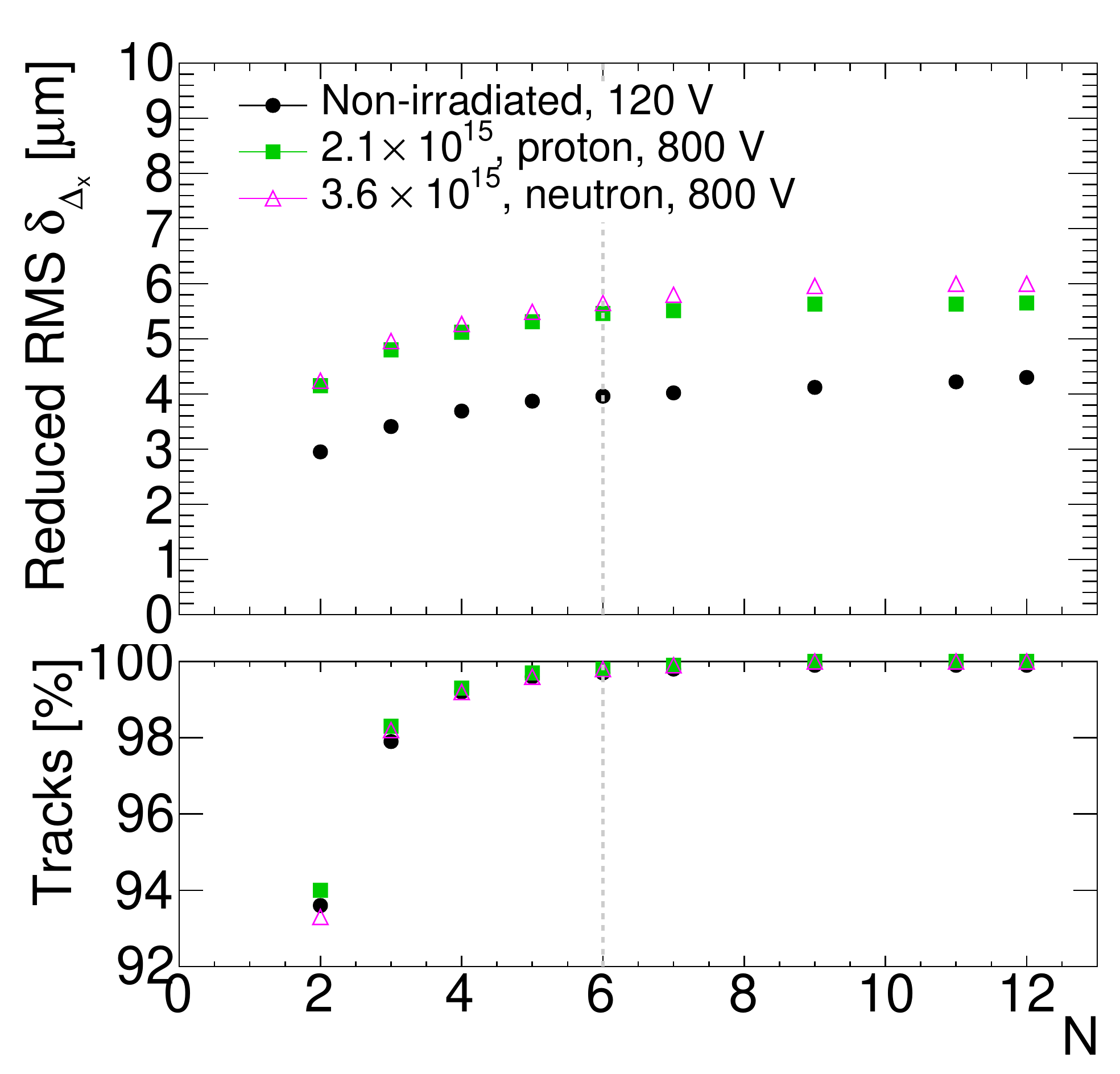}

\caption{Reduced RMS and number of tracks entering its calculation through the method described in Sec~\ref{sec:spres} as a function of $N$. Different markers refer to different fluences. The non-irradiated and proton-irradiated sensor are measured at $\theta =$~\SI{8.8}{\degree} and the neutron-irradiated one at $\theta =$~\SI{12}{\degree}. The vertical dashed line marks the value used throughout this paper.}
\label{fig:nrms}
\end{figure}

 Figure~\ref{fig:dx3lognonirr} shows example residual distributions after the above mentioned requirements are applied. 
In literature (see references of Table~\ref{table:resolution} in~\ref{appendix:reslit}), two main approaches are adopted to estimate the width of the residuals distribution: either a fit  with a Gaussian function (or a generalized version of it) is performed or the RMS (in a certain range) is evaluated. 
 In this paper, a {\em reduced RMS} as described below is used as an estimator of the spatial resolution due to non-Gaussian tails, for example because of delta electrons or the clustering algorithm. The use of the simple RMS can result in overestimating the width of the distribution for the impact of the tails. Thus, the reduced RMS  $\delta_{\Delta x}$ has been developed to reduce the influence of the tails in the evaluation of the distribution width. The reduced RMS is obtained recursively excluding the tracks with $|\Delta x|> N\cdot$RMS. The procedure is repeated until the RMS converges to a stable value $\delta_{\Delta x}$. For the measurements described in this paper $N = 6$ and $\approx$~99\% of the tracks satisfying the requirements in Sec.~\ref{sec:selections} have been used in the evaluation of the spatial resolution. The choice of $N$ = 6 maintains a stable number of tracks used in the resolution estimation for different fluences and beam incidence angles. Figure~\ref{fig:nrms} shows the reduced RMS and the number of tracks entering its calculation as a function of~$N$. For small $N$ values, a smaller range is considered, leading to better resolution but a lower number of tracks used. Increasing~$N$, both the estimated residual width and the number of tracks entering the estimation converge to stable values. 
 
In Table~\ref{table:reswidth} different methods of estimating the residuals distribution width are compared for a non-irradiated sample at the optimal angle. The same cluster quality requirements described in Sec.~\ref{sec:selections} are applied.
For the fit functions, the same definitions as in Ref.~\cite{thesis:simon} are employed. Two cases have been considered. In the first one, the fit has been performed over the full residuals distribution. Since the tails cannot be described by the fits, in the second case the fit has been restricted to $\pm~3\sigma$ of the width obtained in the first case.
Depending on the method, the width varies by up to 25\% (more than \SI{1}{\micro\meter}) with respect to the width definition adopted in this study, the reduced RMS $\delta_{\Delta x}$ ($\pm$ 6 RMS). 
\begin{table}[h]
\centering
\begin{tabular}{c|c}
Measurement method & Width $[\upmu$m$]$\\
\hline
Gaussian & 3.40 $\pm$ 0.02\\
Gaussian ($\pm~3\sigma$) &  3.14 $\pm$ 0.02 \\
Generalized Error Function & 2.98 $\pm$ 0.03 \\
Generalized Error Function ($\pm~3\sigma$) & 3.13 $\pm$ 0.05\\
Student's t & 3.02 $\pm$ 0.02\\
Student's t ($\pm~3\sigma$) &  3.12 $\pm$ 0.03 \\
RMS & 4.86 $\pm$ 0.02\\
RMS ($\pm$ pitch) & 3.97 $\pm$ 0.02\\
RMS ($\pm$ pitch/2) & 3.60 $\pm$ 0.02\\
$\delta_{\Delta x}$ ($\pm$ 3 RMS) & 3.41  $\pm$ 0.02\\
$\delta_{\Delta x}$ ($\pm$ 6 RMS) & 3.96 $\pm$ 0.02\\
\end{tabular}
\caption{Spatial resolution for a non-irradiated sensor for a beam incidence angle of \SI{8.8}{\degree}. Different residual width definitions are compared. The range in which the width has been estimated is specified in brackets. If nothing is specified the full range has been used.} 
\label{table:reswidth}

\end{table}

\begin{figure}[t]
\centering

\includegraphics[width=0.5\textwidth]{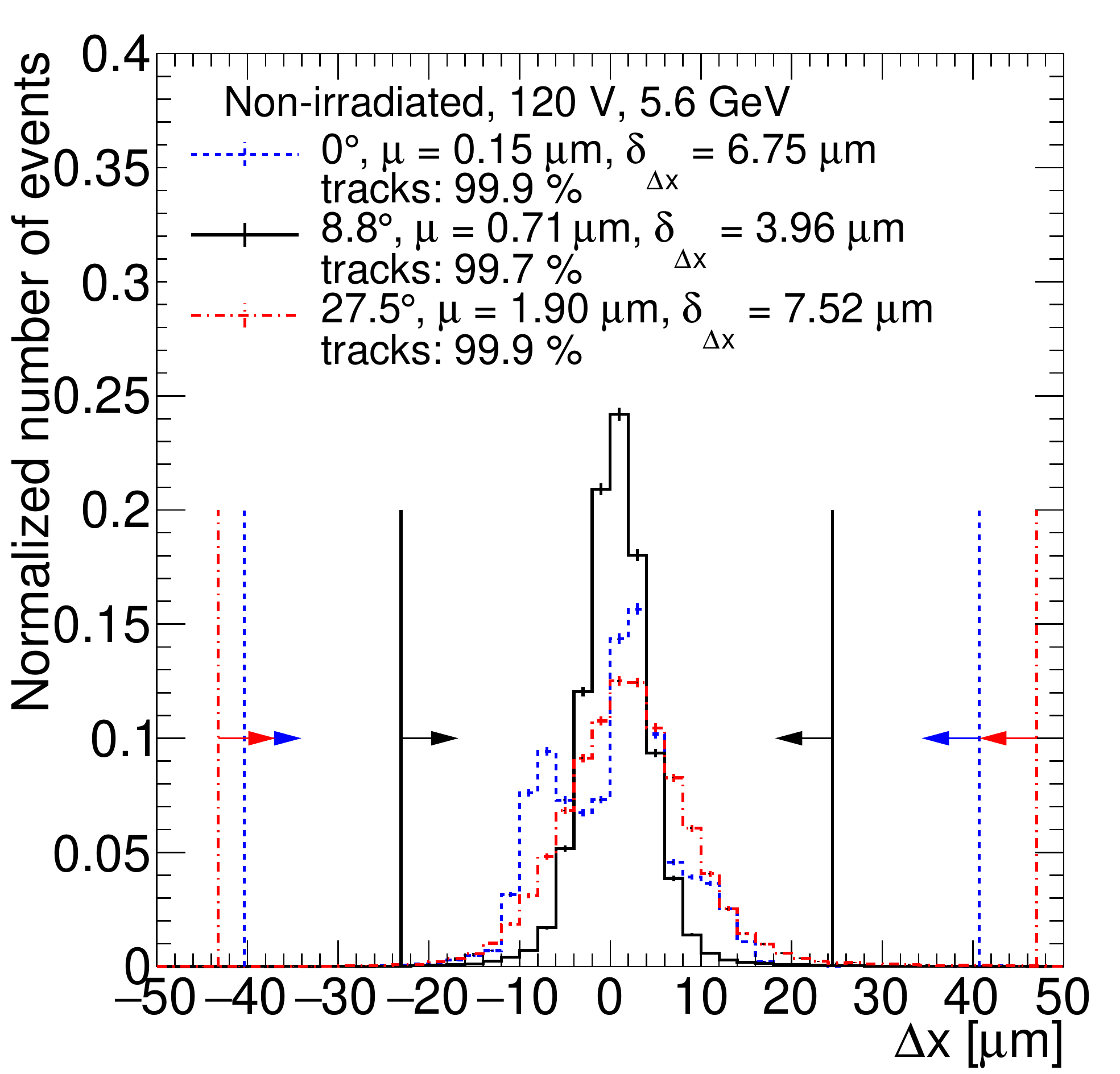}

\caption{Residual $\Delta x$ distribution. Different line colors and styles refer to different incidence angles $\theta$. A threshold at \thr \ of the cluster charge MPV is applied. Vertical lines indicate the range in which the RMS converges to the stable values reported in the legend and used to estimate the three-master resolution. The legend also contains the mean $\mu$ of the residual distribution and the fraction of tracks in the range indicated by the vertical lines.}
\label{fig:dx3nonirrfinal}
\end{figure}

 Figure~\ref{fig:dx3nonirrfinal} shows again the residual distribution, this time with vertical lines indicating the range in which the RMS converges to the stable values reported in the legend. The double peak structure at $\theta =$ \SI{0}{\degree} is due to the impact of single pixel clusters, as shown in \f~\ref{fig:dx3nonirrVert} and explained in Sec.~\ref{sec:recalign}.

The reduced RMS $\delta_{\Delta x}$ of the residual distribution from measurements taken under the same conditions yields the three-master resolution, with the contribution of all the three planes.

 With the three-master device, the measurement of the spatial resolution $\sigma_x$ in the plane of the central inclined sensor can be obtained. 
 Under the assumptions that the positions determined in the three planes of the three-master have the same resolution $\sigma^0_x$ and that they are uncorrelated, which however is not the case for small angles, $\sigma^0_x$ can be obtained from $\delta_{\Delta x}$:

\begin{equation}
\label{eq:non-irr}
 \sigma^0_x = \sqrt{\frac{2}{3}}\delta_{\Delta x}.
\end{equation} 
The assumption of same resolution in each plane is valid for three non-irradiated sensors.
The assumption of uncorrelated resolutions is valid for tracks formed by hits reconstructed as center-of-gravity of a cluster having a size $\geq$ 2. This is, for small angles, not the majority of the cases, as shown in \f~\ref{fig:clsizenonirr}. For a proper treatment of clusters with size 1 a correction of the introduced bias should be applied first, as presented for instance in Ref.~\cite{paper:robert}.
In addition, only Gauss functions have the property that the convolution of two Gauss functions is again a Gauss function. However, the resolution function of a single sensor can be very different from a Gauss function, as discussed above, and multiple scattering causes significant non-Gaussian tails. Equation~\ref{eq:non-irr} is therefore not always correct and its applicability should be carefully considered and verified against simulation. 
 
 When the central plane is irradiated a similar calculation can be used, considering the single hit uncertainty of the central plane as different from the one of the two reference planes. The spatial resolution $\sigma^{\phi}_{x} $ is obtained from the following formula, using for each setup configuration the resolution of the non-irradiated planes from Eq.~\ref{eq:non-irr}, 
\begin{equation}
\label{eq:irr}
\sigma^{\phi}_{x} = \sqrt{\delta_{\Delta x}^2 - \dfrac{1}{2} {\sigma^{0}_{x}}^2}.
\end{equation} 

It has to be noted that Equations~\ref{eq:non-irr}--\ref{eq:irr} do not take into account the significant multiple scattering contribution, present at beam momenta at which the measurements are performed. MS contributions are addressed in the next section (e.g. Eq.~\ref{eq:fit}).

\section{Results}
\label{sec:results}

\begin{figure}[ht]
\centering

\includegraphics[width=0.5\textwidth]{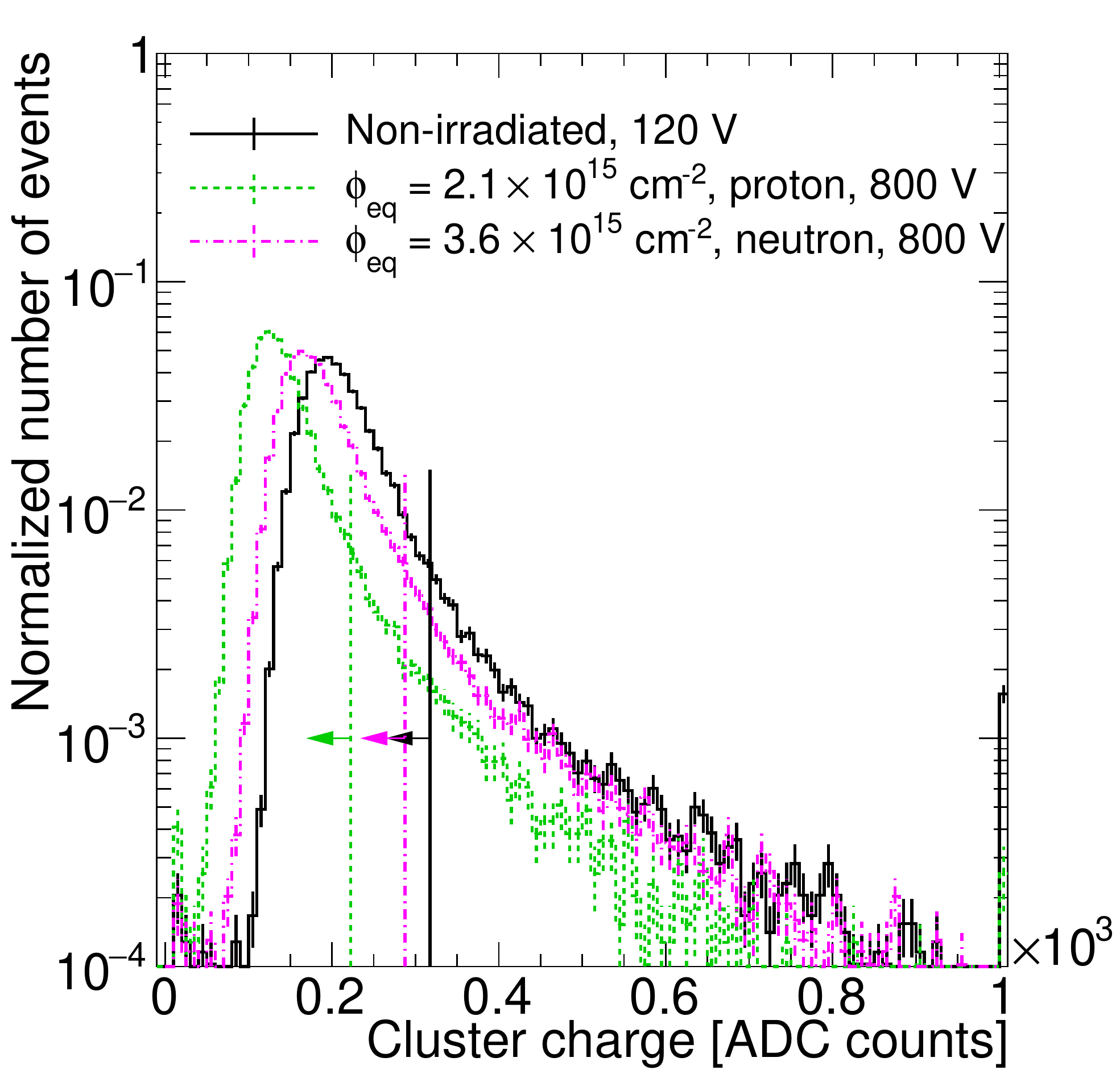}
\caption{Cluster charge of the DUT sensor. Different line colors and styles refer to different irradiations. A threshold at \thr \ of the cluster charge MPV is applied on the non-irradiated sensor, \thrp \ on the proton-irradiated sensor at \phip \ and \thrn \ on the neutron-irradiated sensor at \phin. The non-irradiated and proton-irradiated sensor are measured at $\theta =$~\SI{8.8}{\degree} and the neutron-irradiated one at $\theta =$~\SI{12}{\degree}. The vertical lines and arrows indicate the parts of the distributions that are kept.  Underflow and overflow bins are also shown.}
\label{fig:landaucomparison}
\end{figure}

\begin{figure}[ht]
\centering

\includegraphics[width=0.5\textwidth]{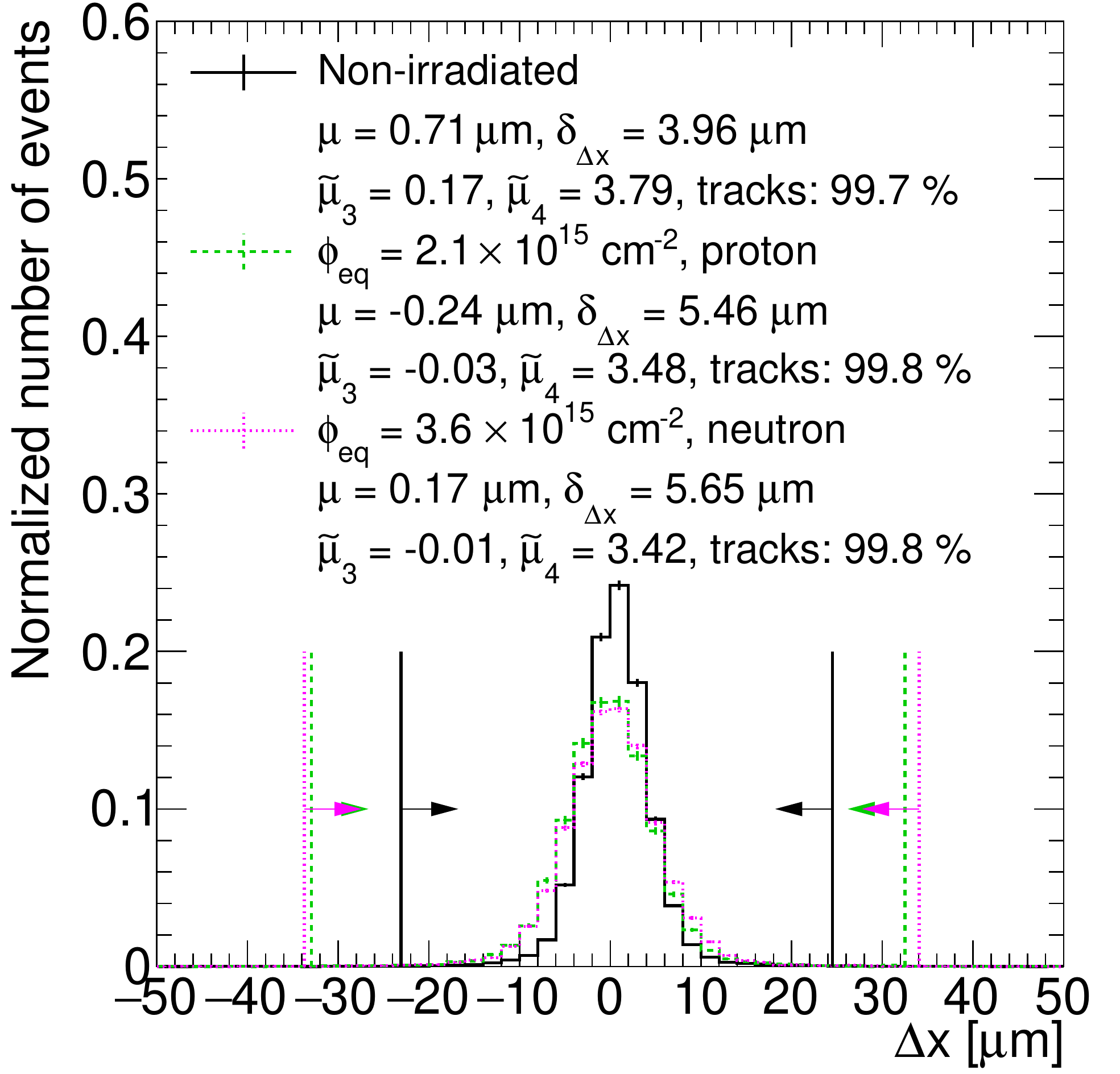}

\caption{Residual $\Delta x$ distribution. The non-irradiated and proton-irradiated sensor are measured at $\theta =$~\SI{8.8}{\degree} and the neutron-irradiated one at $\theta =$~\SI{12}{\degree}. Different line colors and styles refer to different irradiations. A threshold at \thr \ of the cluster charge MPV is applied for the non-irradiated sensor, \thrp \ for the proton-irradiated sensor at \phip \ and \thrn \ for the neutron-irradiated sensor at \phin. Vertical lines indicate the range in which the RMS converges to the stable value reported in the legend and used to estimate the three-master resolution. Mean ($\mu$), reduced RMS $\delta_{\Delta x}$, skewness ($\tilde{\mu}_3$) and kurtosis ($\tilde{\mu}_4$) are also reported, as well as the fraction of tracks entering the reduced RMS computation.}
\label{fig:dx3log}
\end{figure}

In this section the spatial resolution for different beam and setup parameters is presented.
As a preliminary step in the comparison of the performance of non-irradiated and irradiated sensors, control distributions of the cluster charge and residuals are shown in \f~\ref{fig:landaucomparison} and \f~\ref{fig:dx3log}. 

The non-irradiated sensor was operated at a bias voltage of \SI{120}{\volt}. The design is the one shown in \f~\ref{fig:pstoprd53Apads}. Measurements with the other designs from Sec.~\ref{sec:pixels} are in agreement within the sub-micron uncertainties and therefore not shown. All irradiated sensors have the design shown in \f~\ref{fig:pstopdefault} and were operated at a bias voltage of \SI{800}{\volt}, unless stated otherwise.

\begin{figure}[ht]
\centering
\includegraphics[width=0.5\textwidth]{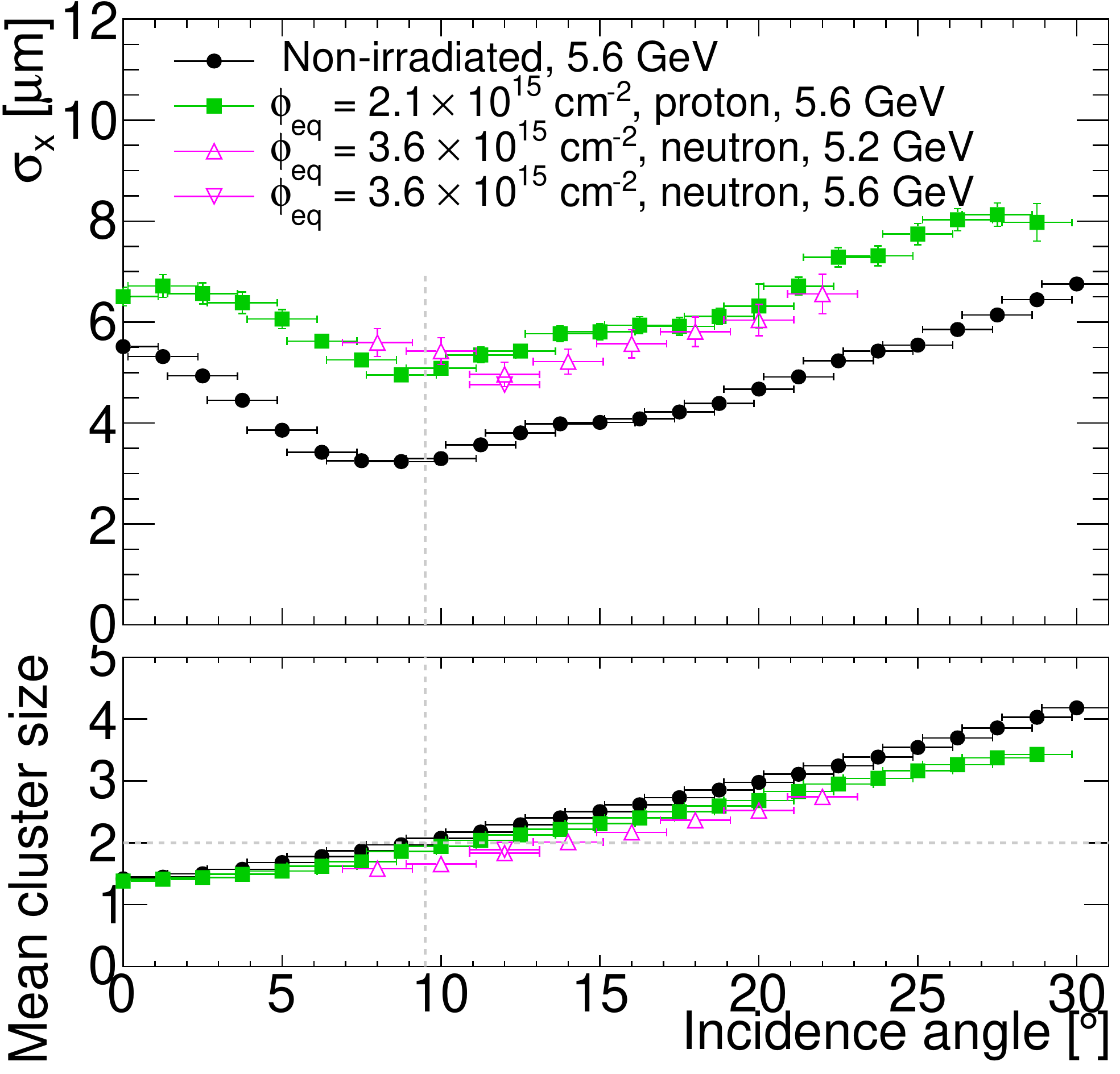}
\caption{Spatial resolution and average cluster size as a function of the beam incidence angle with respect to the normal to the sensors. Different markers refer to different irradiations. The electron beam momentum at which the measurements were taken is stated in the legend. A threshold at \thr \ of the cluster charge MPV is applied to the non-irradiated sensor, operated at 120V, \thrp \ on the proton-irradiated sensor at \phip \ and \thrn \ on the neutron-irradiated sensor at \phin. The irradiated sensors were operated at 800 V. The vertical (horizontal) dashed line marks the expected optimal angle of \SI{9.5}{\degree} given by $\arctan{(\textrm{pitch}_x/t)}$ (the cluster size of two where the optimal resolution is expected).}
\label{fig:anglescan}
\end{figure}

In \f~\ref{fig:anglescan} the spatial resolution along the \yv \ pitch direction and the average cluster size are shown as a function of the beam incidence angle with respect to the normal to the sensors ($\theta$ in \f~\ref{fig:setup}). 
The beam momentum was \SI{5.6}{\giga\electronvolt} for the non-irradiated and the proton-irradiated sensors. For the neutron-irradiated sensor module, measurements for beam momenta of \SI{5.2}{\giga\electronvolt} and \SI{5.6}{\giga\electronvolt} are reported. Vertical error bars represent the statistical Poisson uncertainties, horizontal error bars show the uncertainty on the angle determination: $\pm$ \SI{1}{\degree} due to the mechanical precision of the rotating device and its alignment in the mounting process, and a statistical uncertainty of $\pm$ \SI{0.5}{\degree}.

The incidence angle at which a minimum in the spatial resolution ({\em best resolution}) is achieved is indicated as the optimal angle in the following. At this angle the average cluster size is about 2.
Within uncertainty, the optimal angle is consistent with \SI{9.5}{\degree}, the angle maximizing the number of two pixels clusters given by $\arctan{(\textrm{pitch}_x/t)}$, for both the non-irradiated and proton-irradiated samples. 
The best resolution for the neutron-irradiated sample corresponds to an angle of \SI{12}{\degree}, with an average cluster size of about 2. Differences between neutron and proton irradiated samples can be due to different shapes of the electric fields in the sensors, providing a less uniform charge collection in neutron-irradiated sensors~\cite{paper:irrdiff,paper:klanner}, and more severe ionising energy loss to the readout chip for proton-irradiated modules. Nevertheless, the measured spatial resolution at the respective optimal angle is the same.

\begin{figure}
\centering
\includegraphics[width=0.5\textwidth]{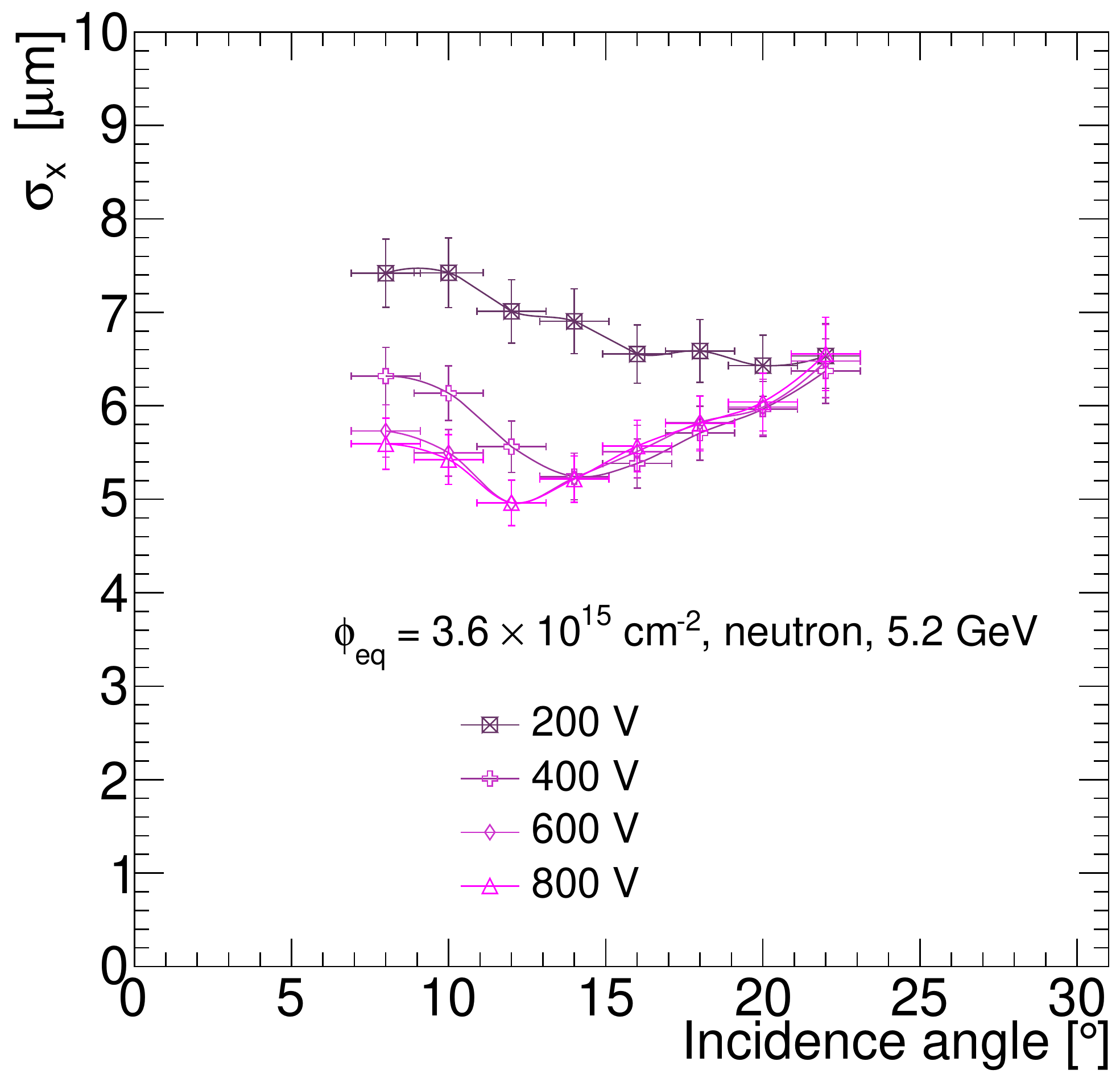}

\caption{Spatial resolution for a neutron-irradiated sensor module as a function of the beam incidence angle with respect to the sensor normal. Different markers refer to different bias voltages.}
\label{fig:biasscan}

\end{figure}

To further investigate the dependence of the resolution on the beam angle for the neutron-irradiated sensor, the measurement was repeated at lower bias voltages, as presented in \f~\ref{fig:biasscan}. The curves have the only purpose to guide the eyes. The angle for optimal resolution increases and the spatial resolution degrades with decreasing bias voltage. Above a bias voltage of \SI{600}{\volt} both the resolution and the angle at which the minimum is situated converge to stable values.

\begin{figure}[!h]
\centering
\includegraphics[width=0.5\textwidth]{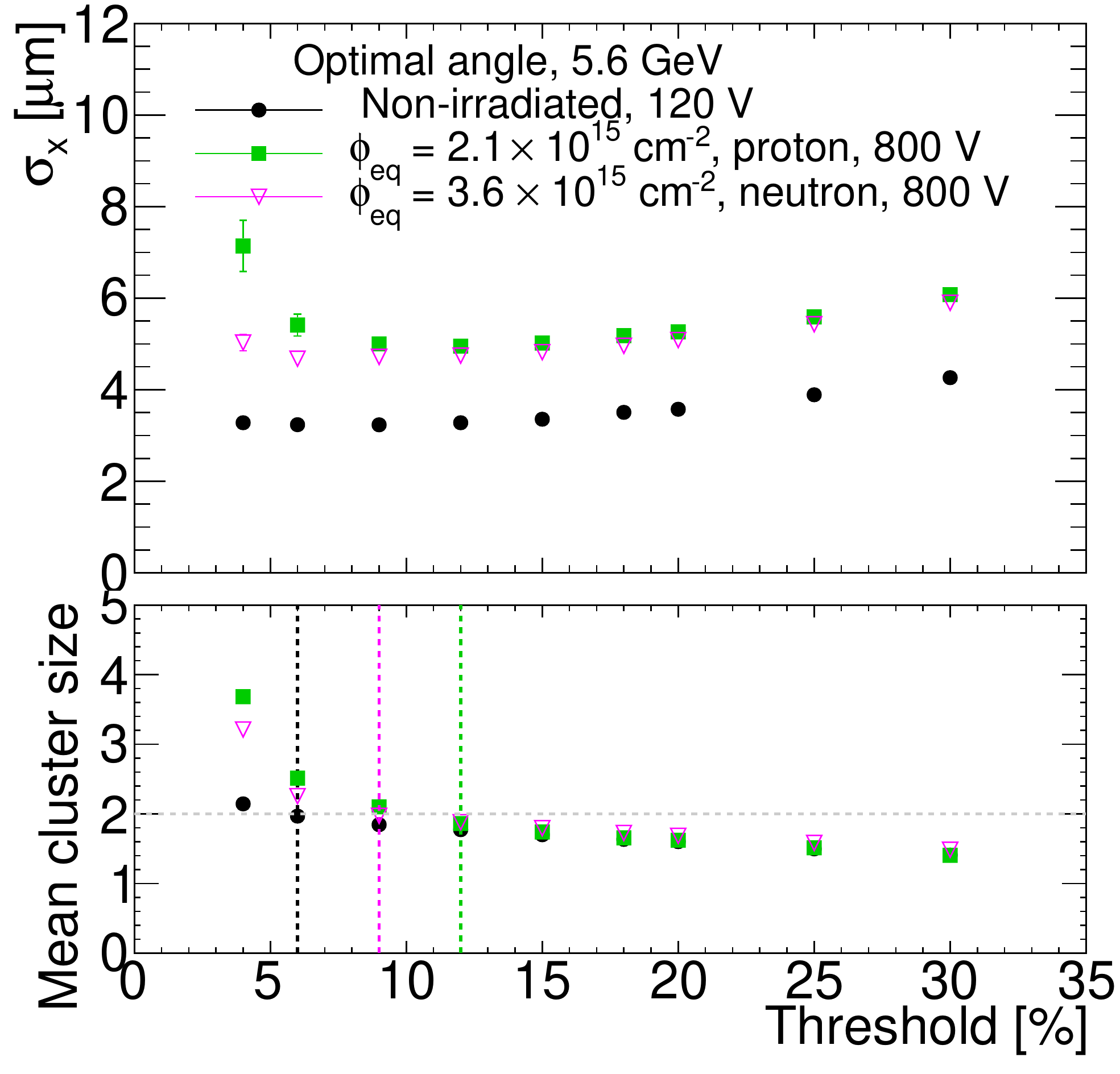}

\caption{Resolution and cluster size as a function of threshold. Different markers refer to different irradiations. Vertical lines show the thresholds used in Figs.~\ref{fig:clsizenonirr}--\ref{fig:biasscan} and Figs.~\ref{fig:chargecut} and~\ref{fig:momentumscan} for the different fluences studied. For an easier comparison of the performance of sensors with different irradiations, threshold percentages have been rounded.} 
\label{fig:threshold}
\end{figure}

The effect on the resolution of different requirements on the cluster charge has also been studied. The offline threshold requires a minimum charge to be collected by a single pixel to be considered in the clustering algorithm. In \f~\ref{fig:threshold} the impact of the applied threshold, expressed as a percentage of the cluster charge distribution MPV, on the resolution and the cluster size on sensors with different irradiations is shown. The non-irradiated sensor module is the least affected by different thresholds, but at higher thresholds a shift of the optimal angle towards higher values has been observed. The proton-irradiated sensor is the one showing a larger degradation of the resolution and increase in cluster size at low thresholds. The effect of applying the same threshold on all sensors despite the different irradiation level has been checked. If a \thr \ threshold, preferred for the non-irradiated sensor, had been applied to the irradiated sensors, the measurement would be affected by a high noise level: at such threshold the average cluster size of the proton-irradiated sensor is 25\% higher than expected from charge sharing. Raising the threshold of the non-irradiated sensors to \thrp, preferred for the proton-irradiated sensor, would not affect the result at small beam incidence angles but would deteriorate the resolution at shallow angles by $\approx$ 5\%.


Table~\ref{table:res} compares the spatial resolution excluding (``90\% cut'', default configuration in this study) and not excluding (``No cut'') the clusters above the 90\% quantile of the cluster charge distribution on the central plane of the three-master in the reduced RMS calculation. When the ``90\% cut'' requirement is not applied, the resolution is degraded by 10-20\%, especially at shallow incidence angle.
The resolution at \SI{0}{\degree} for the non-irradiated sample, with the ``90\% cut" applied, is the one quoted in Figs.~\ref{fig:literature} and~\ref{fig:literatureratio}.

\begin{figure}[!ht]
\centering

\subfloat[]{\label{fig:reschcut}
\includegraphics[width=0.5\textwidth]{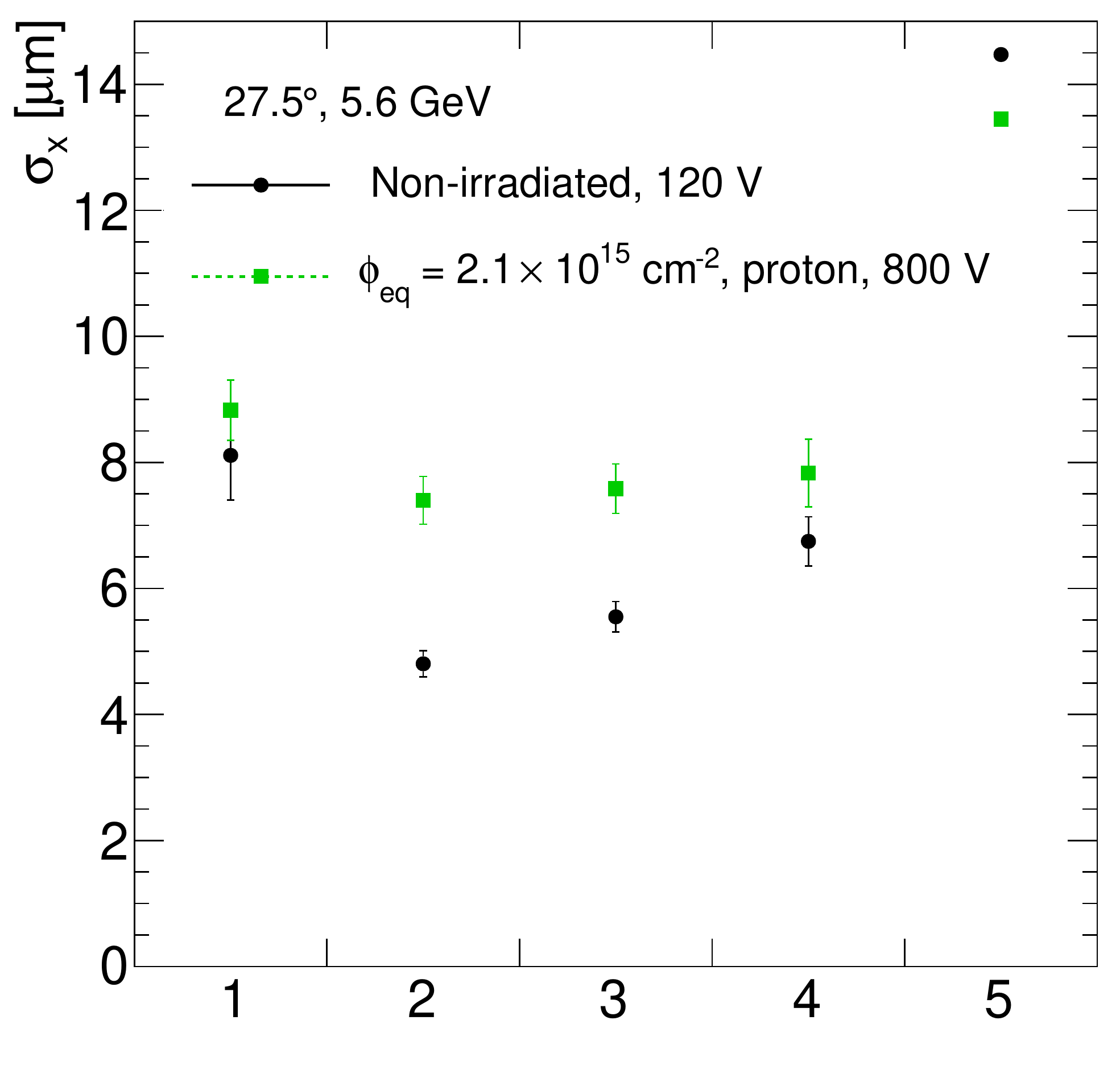}}\\
\subfloat[]{\label{fig:landauchcut}
\includegraphics[width=0.5\textwidth]{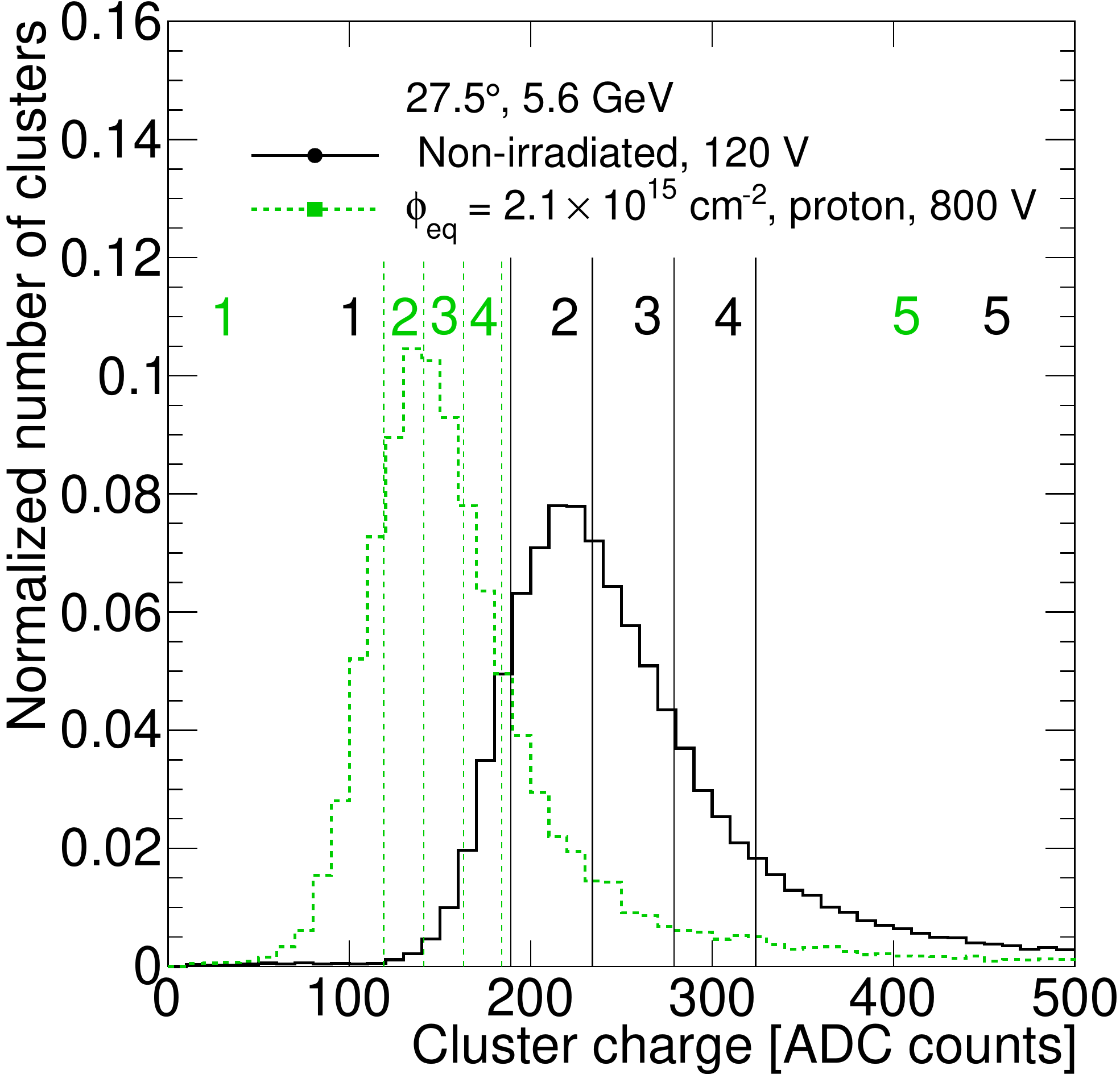}}

\caption{Top (a): spatial resolution for the central three-master sensor in different cluster charge intervals, indicated in (b). Bottom (b): cluster charge distribution shown for reference, with vertical lines highlighting the considered charge intervals. Black markers and lines refer to a non-irradiated sensor and green markers and lines to a proton-irradiated sensor at \phip.  The lower cluster charge threshold is fixed as discussed in \f~\ref{fig:threshold}. The intervals are defined as: 1 = [0, MPV-$\sigma$], 2 = [MPV-$\sigma$, MPV+$\sigma$], 3 = [MPV+$\sigma$, MPV+3$\sigma$], 4 = [MPV+3$\sigma$, MPV+5$\sigma$], 5 = ADC counts $>$MPV+5$\sigma$.
}
\label{fig:chargecut}
\end{figure}

The spatial resolution has also been measured in different ranges of the central three-master plane's cluster charge distribution. 
The impact on the resolution once again depends on the beam incidence angle. At shallow angles, the resolution is more sensitive to requirements on the cluster charge as the incoming electrons are traversing more material, therefore increasing both the MPV and the width of the cluster charge distribution. The observed shift of the MPV is consistent with the expected energy deposition in the increased material traversed by the incoming particles at shallow incidence angles. 
In \f~\ref{fig:reschcut} the spatial resolution as a function of each considered charge interval is shown for a shallow beam incidence angle. The cluster charge distribution of the central  three-master plane is also shown as reference (Fig.~\ref{fig:landauchcut}). 
For both the non-irradiated and the proton irradiated sensors, the resolution has a minimum in the interval [MPV-$\sigma$, MPV+$\sigma$] (interval 2), where $\sigma$ is the width of the cluster charge distribution obtained from the fit of a Landau function convolved with a Gaussian function. The cluster charge distribution for irradiated sensors is narrower than for non-irradiated ones, reducing the effect on the resolution of requiring a specific cluster charge interval.  


\begin{table}[h]

\begin{tabular}{c|ccc|ccc}
 \multicolumn{1}{c|}{Irradiation} &\multicolumn{6}{c}{Spatial resolution $[\upmu$m$]$}\\
$[\phi_{\mathrm{eq}}$/\SI{e15}{\per\square\centi\meter}$]$ & \multicolumn{3}{c|}{90\% cut} &  \multicolumn{3}{c}{No cut} \\
\multicolumn{1}{c|}{} & \SI{0}{\degree} &\SI{8.8}{\degree} & \SI{27.5}{\degree} & \SI{0}{\degree} &\SI{8.8}{\degree} & \SI{27.5}{\degree} \\
\hline
0 & 5.5 & 3.2 & 6.1 & 6.3 & 4.1 & 7.6\\
 2.1, proton& 6.5 & 5.0 & 8.1 & 7.2 & 5.8 & 9.4 \\
\hline
\hline
 & \SI{8}{\degree} &\SI{12}{\degree} & \SI{22}{\degree} & \SI{8}{\degree} &\SI{12}{\degree} & \SI{22}{\degree} \\
\hline
3.6, neutron & 5.6 & 5.0 & 6.6 & 6.5 & 5.9  & 7.2 \\
\end{tabular}
\caption{Spatial resolution for a non-irradiated sensor, a proton-irradiated (\phip) sensor, and a neutron-irradiated (\phin) sensor excluding (``90\% cut'') and not excluding (``No cut") the clusters above the 90\% quantile of the cluster charge distribution on the central plane of the three-master. The resolution at \SI{0}{\degree} for the non-irradiated sample, with the ``90\% cut" applied, is the one quoted in Figs.~\ref{fig:literature} and~\ref{fig:literatureratio}.} 
\label{table:res}

\end{table}

In the following analysis the ``90\% cut'' is applied on each three-master plane. 

\begin{figure}[!htb]
\centering
\includegraphics[width=0.5\textwidth]{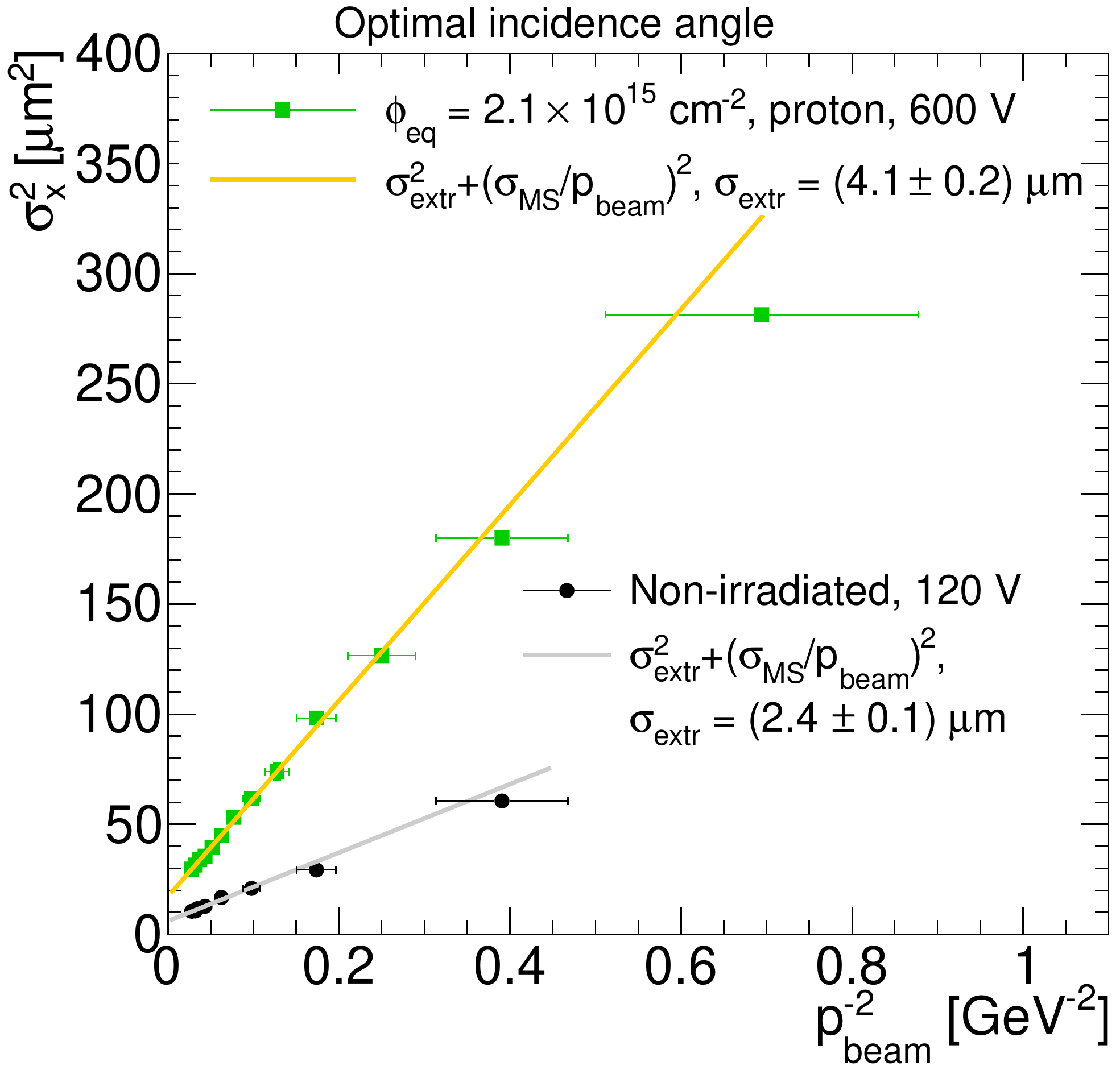}

\caption{Spatial resolution squared as a function of the inverse of the beam momentum squared $p^{-2}_{\text{beam}}$ for a non-irradiated sensor and a proton-irradiated sensor. A threshold at \thr \ of the cluster charge MPV is applied on the non-irradiated sensor and \thrp \ on the proton-irradiated sensor at \phip. The  function $\sigma^2_x~=~\sigma_{\text{extr}}^2 + (\sigma_{\text{MS}}/p_{\text{beam}})^2$ has been fitted to the data (solid lines). The uncertainties on $p_{\text{beam}}$ represent the absolute momentum spread $\sigma_p$ introduced in Sec.~\ref{sec:method}.} 
\label{fig:momentumscan}

\end{figure}

At beam momenta used for these measurements, the multiple Coulomb scattering contribution has a significant impact on the resolution. At infinite momentum the effect of MS is zero. With a momentum scan between 1--\SI{6}{\giga\electronvolt} the results are extrapolated to infinite momentum. 
The spatial resolution squared as a function of the beam momentum squared is shown in \f~\ref{fig:momentumscan}, for a non-irradiated sample and a proton-irradiated sample at \phip, operated at 120 V and 600 V, respectively. The uncertainties on $p_{\text{beam}}$ represent the absolute momentum spread $\sigma_p$ introduced in Sec.~\ref{sec:method}.

 The  function 
\begin{equation}
\label{eq:fit}
\sigma^{2}_x(p_{\text{beam}}) = \sigma_{\text{extr}}^2 + (\sigma_{\text{MS}}/p_{\text{beam}})^2
\end{equation}
has been fitted to the data (solid lines), where $\sigma_{\text{extr}}$ is the extrapolated resolution at infinite beam momentum  and $\sigma_{\text{MS}}$ the contribution to the resolution due to multiple scattering. $\sigma^{2}_x(p_{\text{beam}})$ corresponds to $\sigma^0_{x}$ ( $\sigma^{\phi}_{x}$) defined in Eq.~\ref{eq:non-irr} (Eq.~\ref{eq:irr}) in the case of a non-irradiated (irradiated) central sensor. 
Equation~\ref{eq:irr} is valid if $\sigma^0_{x}$ and $\sigma^{\phi}_{x}$ are obtained under the same conditions, thus to extract $\sigma^{\phi}_{x}(p_{\text{beam}})$, $\sigma^0_{x}$ should be measured at the same $p_{\text{beam}}$. Since $\sigma^0_{x}$ has been sampled in larger $p_{\text{beam}}$ intervals than $\sigma^{\phi}_{x}$, a different  procedure has been used to extract $\sigma^{\phi}_{x}$: $\sigma^{0}_{x}$ is not directly obtained from measurements but from the fit function in Eq.~\ref{eq:fit} at a given $p_{\text{beam}}$. 


A spatial resolution $\sigma_{\text{extr}}^0$ of $\ensuremath{2.4 \pm \SI{0.1}{\micro\meter}}$ was found for the non-irradiated sensor at the optimal angle when extrapolating to infinite beam momentum. For the proton-irradiated sample the extrapolated spatial resolution $\sigma_{\text{extr}}^{\phi}$ is $\ensuremath{4.1 \pm \SI{0.2}{\micro\meter}}$.  

The neutron-irradiated sample showed a similar performance as the proton-irradiated sample in terms of resolution values (Fig.~\ref{fig:anglescan}). Thus, the measurement at different beam momenta were not repeated.

\section{Conclusions}
In this study spatial resolution measurements for planar n$^+$p sensors with an active thickness of \yceci \ and a pixel size of \venticinque \ produced by Hamamatsu Photonics are presented for non-irradiated, proton-irradiated and neutron-irradiated sensors bump-bonded to ROC4Sens analog readout chips. 

Measurements of sensors irradiated to two fluences were performed in the DESY test beam. Four different sensor designs are compared using non-irradiated samples and no significant difference in the spatial resolution is found.
Beam incidence angles between \SI{0}{\degree} and \SI{30}{\degree} were investigated. 

The Center-of-Gravity method with threshold cuts is used to calculate the position of the beam particle, which causes uncertainties of the determination of the resolution at small angles. 

The spatial resolution determined varies between 3 and \SI{7}{\micro\meter} for the non-irradiated sensor and between 5 and \SI{8}{\micro\meter} for the irradiated sensors.
The angle expected for the optimal resolution of a non-irradiated detector is $\textrm{\SI{9.5}{\degree} }= \arctan{(\text{pitch}_x/t)}$, which agrees with the values observed for the non-irradiated sensor and the proton-irradiated sensor at \phip. For the neutron-irradiated sensor with \phin \ the optimal resolution is observed at \SI{12}{\degree}. Proton and neutron irradiated sensors achieve similar spatial resolutions.

Measurements for the neutron-irradiated sensor at \phin \ were repeated at lower bias voltages, showing a deterioration of the position resolution and a shift of the optimal angle to higher values. Above a bias voltage of \SI{600}{\volt} both the resolution and the optimal angle converge to stable values. 

The effects on the spatial resolution of different thresholds and requirements on the cluster charge have also been presented.

A spatial resolution of $\ensuremath{2.4 \pm \SI{0.1}{\micro\meter}}$ was found for a non-irradiated sensor at the optimal angle when extrapolating to infinite beam momentum. For the proton-irradiated sample the extrapolated spatial resolution is $\ensuremath{4.1 \pm \SI{0.1}{\micro\meter}}$.
Excellent resolution is therefore maintained up to the tested fluence of \phip \ and \phin, which correspond to the lifetime fluence of layer 3 and more than 70\% of the lifetime fluence of layer 2 of the CMS Phase-2 Inner Tracker, respectively.

The results presented in this paper show that the tested sensors are suitable candidates for high precision measurements at the HL-LHC. 
Results in Ref.~\cite{paper:finn} show that the sensors also maintain a high hit efficiency after irradiation. 
The sensors employed in this paper and in Ref.~\cite{paper:finn} are bump-bonded to an analog readout chip with 12 bit resolution to allow for detailed studies of the sensor properties. The chip that will be used in the Phase-2 IT adopts a Time Over Threshold approach and has a 4 bit resolution. Sensors which are similar to the ones investigated in this paper are also being tested with such readout~\cite{paper:georg} to demonstrate their suitability for the Phase-2 upgrade of the CMS experiment at the HL-LHC.

\section*{Acknowledgements}
 \label{sect:Acknowledgement}
The measurements leading to these results have been performed at the Test Beam Facility at DESY Hamburg (Germany), a member of the Helmholtz Association (HGF). 
 
This work was supported by the German Federal Ministry of Education and Research (BMBF) in the framework of the ``FIS-Projekt - Fortf\"{u}hrung des CMS-Experiments zum Einsatz am HL-LHC: Verbesserung des Spurdetektors fur das Phase-II- Upgrade des CMS-Experiments'' and supported by the H2020 project AIDA-2020, GA no.\ 654168.


The authors thank C. Muhl (DESY) for the design of the three-master.

I. Zoi and A. Hinzmann gratefully acknowledge funding in the Emmy-Noether program (HI 1952/1-1) of the German Research Foundation DFG.



\newpage
\quad    \newline
\newpage

\appendix
\begingroup
\let\clearpage\relax 
\onecolumn 

\section{Resolution measurements in literature}
\label{appendix:reslit}
\begin{sideways} 

\begin{minipage}{1.2\textwidth}


\centering
\resizebox{\textwidth}{!}{

\begin{tabular}{ccccccccccccccccc}
\hline
Legend & Ref. & Pixel cell [$\upmu $m$^2$] & Res. [$\upmu $m] & Def.  &Cluster size &  Sensor type & Thickness [$\upmu $m] & Bias Voltage [V] &Threshold [ke] & S/N& Hit Algorithm & Unc. [$\upmu $m] &Beam & TB Setup & ROC & T [\textdegree{}C] \\
\hline

\textcolor{mediumblue}{$\blacksquare$} &\cite{thesis:simon} & 100$\times${\bf 150} & 50.2 & Generalized error function & all & n$^+$-in-n, planar &285 & 200 & 2.0 & - &CoG (\ddag) & 0.08 (stat), 0.28 (syst) (*)& DESY, e, 5.2 GeV & EUDET, 3.4 $\upmu $m~\cite{man:eudet} &PSI46digV2.1-r & 17 \\


\textcolor{mediumblue}{$\blacksquare$} &\cite{thesis:simon}  & {\bf 100}$\times$150 & 32.7 & Generalized error function & all & n$^+$-in-n, planar &285 & 200 & 2.0 & - &CoG  (\ddag) &0.09 (stat), 0.21 (syst) (*)& DESY, e, 5.2 GeV& EUDET, 3.4 $\upmu $m~\cite{man:eudet} &PSI46digV2.1-r & 17 \\

\Plus & \cite{thesis:atlas} & 25$\times${\bf 100} & $\sim$ 28.9 & RMS & 1 & n-in-p, planar & 150 &  50 & 1.5 &- & Pixel center & - & 
CERN,  $\pi$, k and p, 120 GeV & EUDET, 6.8$\upmu $m~\cite{man:eudet}   & RD53A (LIN FE) & -  \\

\includegraphics[width=0.02\textwidth]{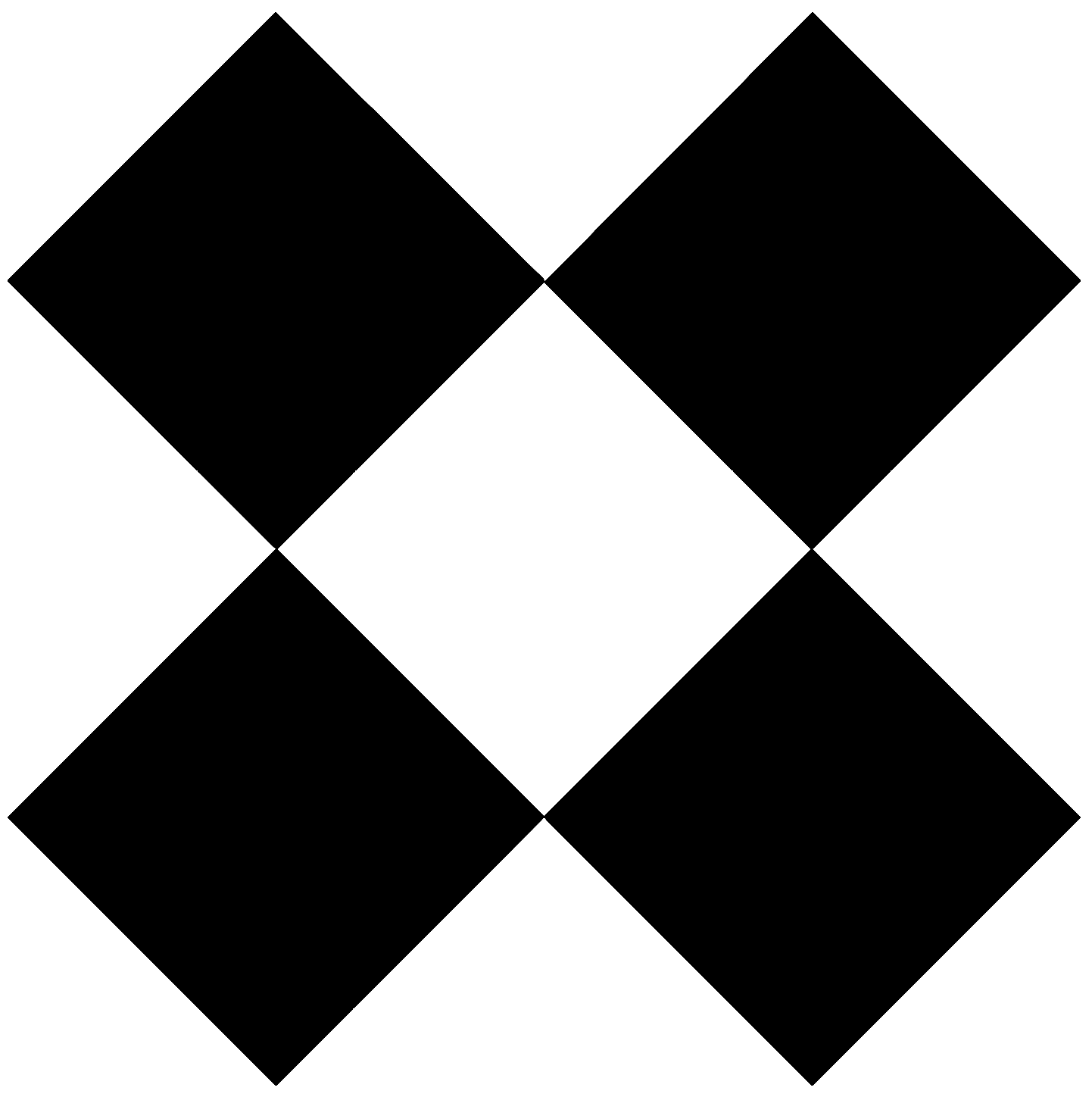} &\cite{paper:caterina}& {\bf 100}$\times$150 & 26.5& Gauss & all & n-in-n, planar & 285 &150 & 2.4 & - & CoG &- &FNAL, p, 120 GeV & Pixel telescope~\cite{man:fnalpixel}, $\sim$ 8 $\upmu $m &  PSI46digV2.1-r& RT \\

\XSolidBold &\cite{master:irene}& {\bf 100}$\times$150 & 22.9 & RMS in $\pm$ 1/2 pitch& all & n-in-n, planar & 285 &150 & $\sim$ 1.5 & - & CoG &- &FNAL, p, 120 GeV & Pixel telescope~\cite{man:fnalpixel}, $\sim$ 6 $\upmu $m &  PSI46digV2.1-r& RT \\


\includegraphics[width=0.02\textwidth]{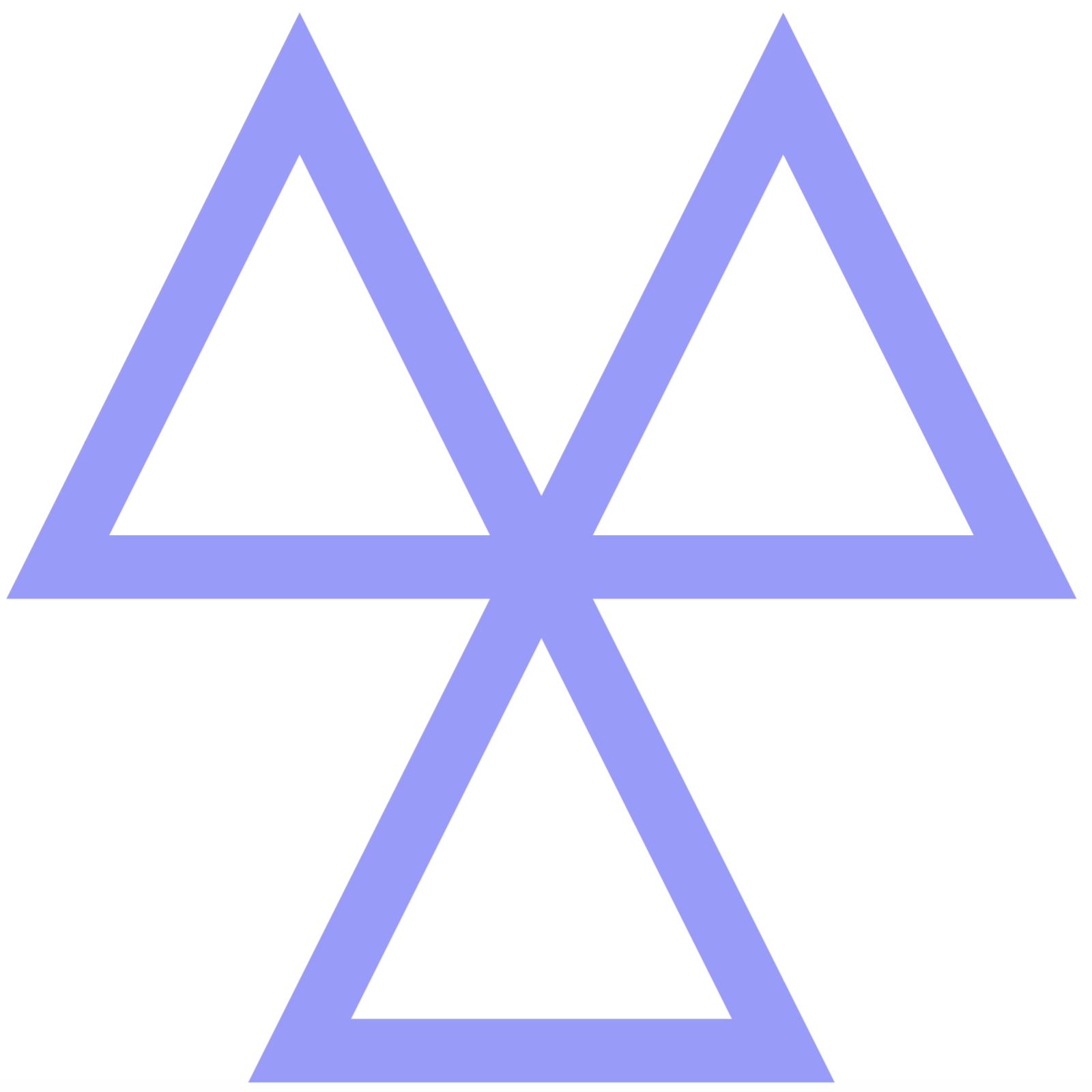} & \cite{master:boronat} & {\bf 75}$\times$50 & 20 & Reduced RMS, N=2& 1 & DEPFET & 50 &- & - ($\diamond$) & $\approx$ 40&CoG +$\eta$&-&(\textsection)&EUDET/AIDA~\cite{man:eudetold}& - &- \\

\includegraphics[width=0.02\textwidth]{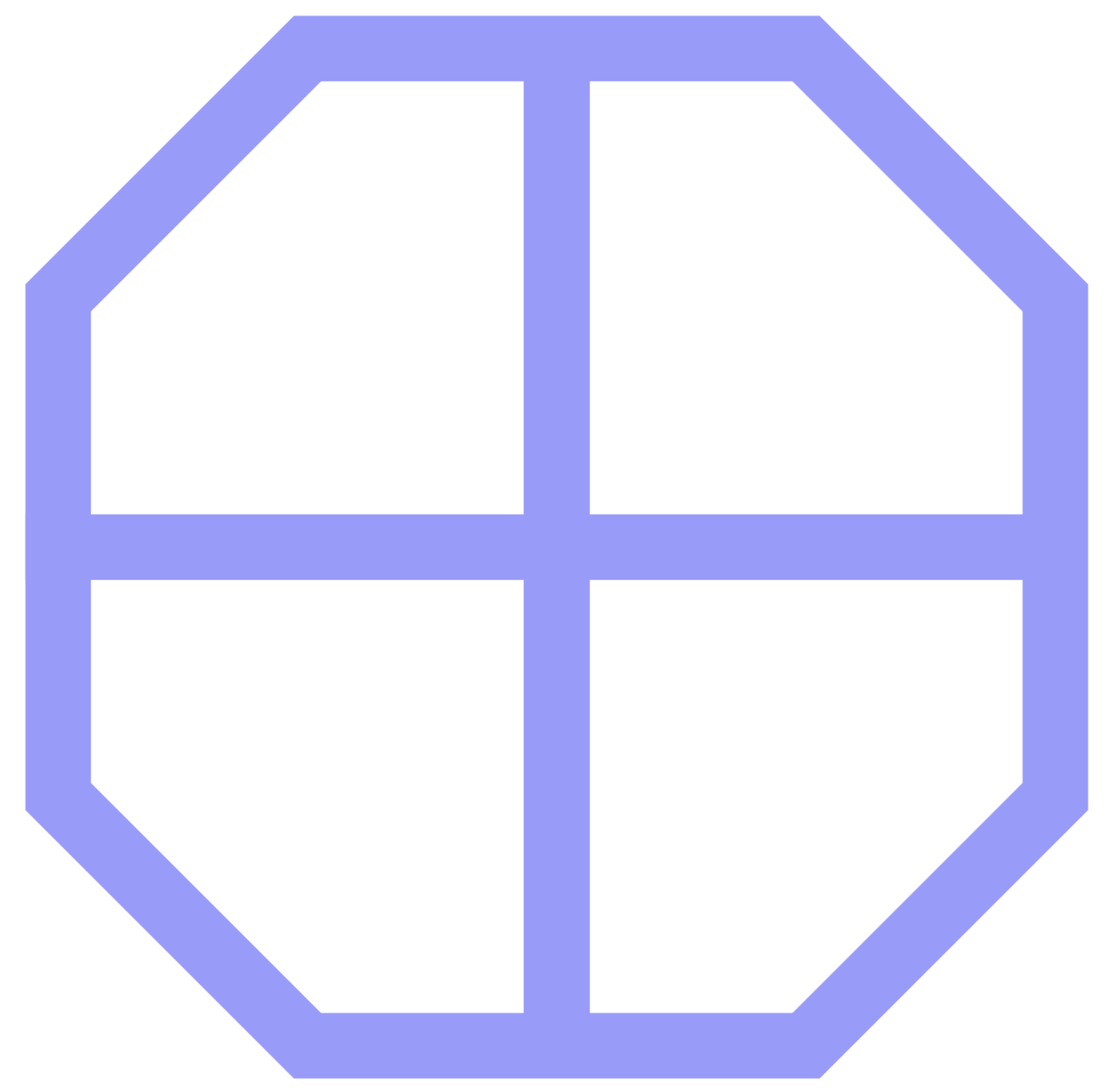} & \cite{master:boronat} & {\bf 75}$\times$50 & 10.8 & Reduced RMS, N=2 & 2 &DEPFET & 50 &-&- ($\diamond$)& $\approx$ 40&CoG +$\eta$&-&(\textsection)&EUDET/AIDA~\cite{man:eudetold}& - &- \\

\includegraphics[width=0.02\textwidth]{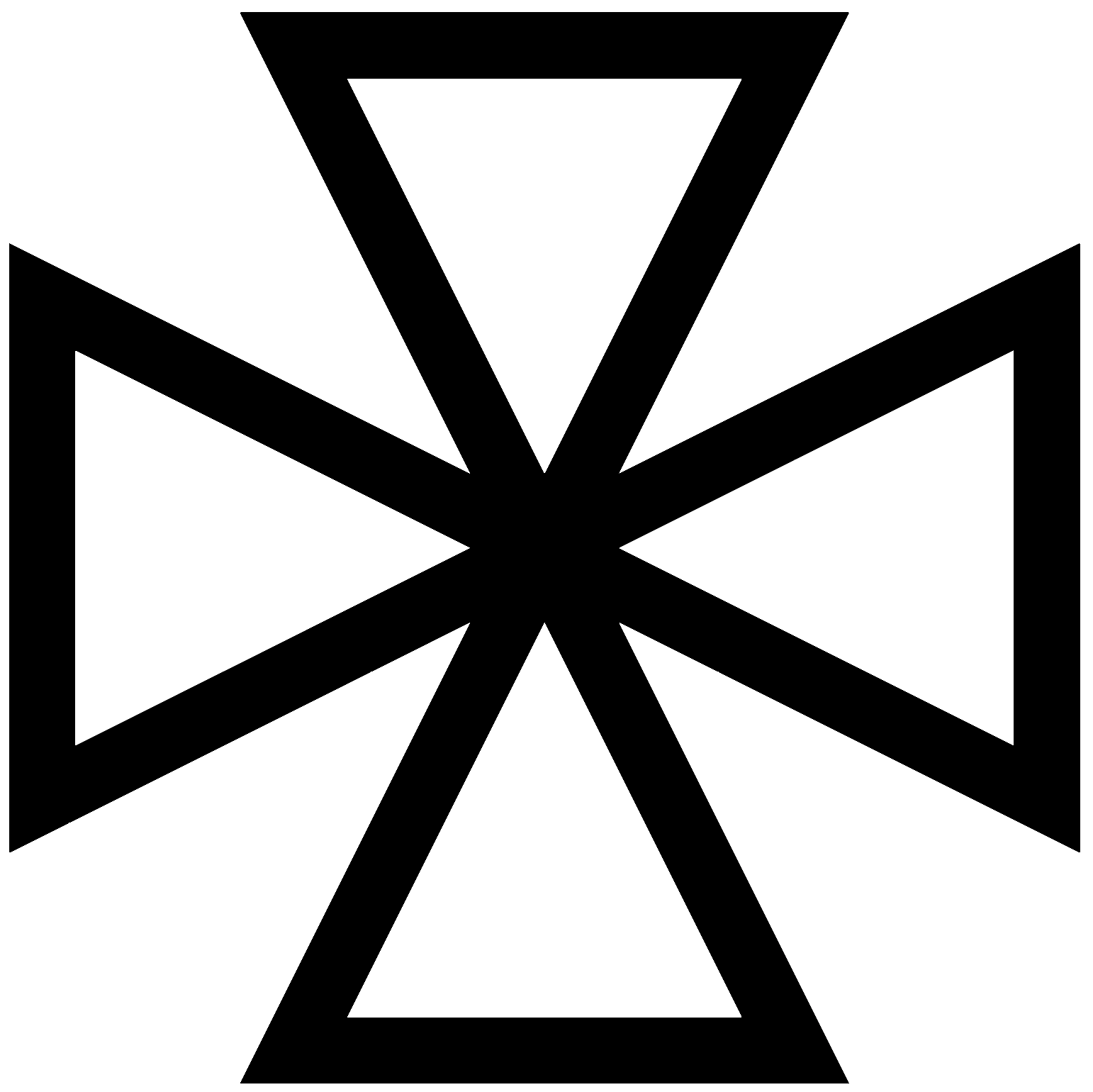}&\cite{paper:lhcbwhiterose} & {\bf 55}$\times${\bf 55} & 16 & Gauss& all& n, 3D& 285& 10&  1.52 & - & CoG+$\eta$&0.2  & CERN ($\pi$ 120 GeV ) & ($\ominus$) & Timepix3 ASIC (ToT)&  RT- \\

$\bigcirc$& \cite{paper:lhcbvelo} & {\bf 55}$\times${\bf 55} & 14 & Gauss & - & n-on-p, planar & 200 & 200 & - & - & - & - & SPS, CERN & ($\ominus$)  & Timepix3 ASIC (ToT)&  - \\

\includegraphics[width=0.018\textwidth]{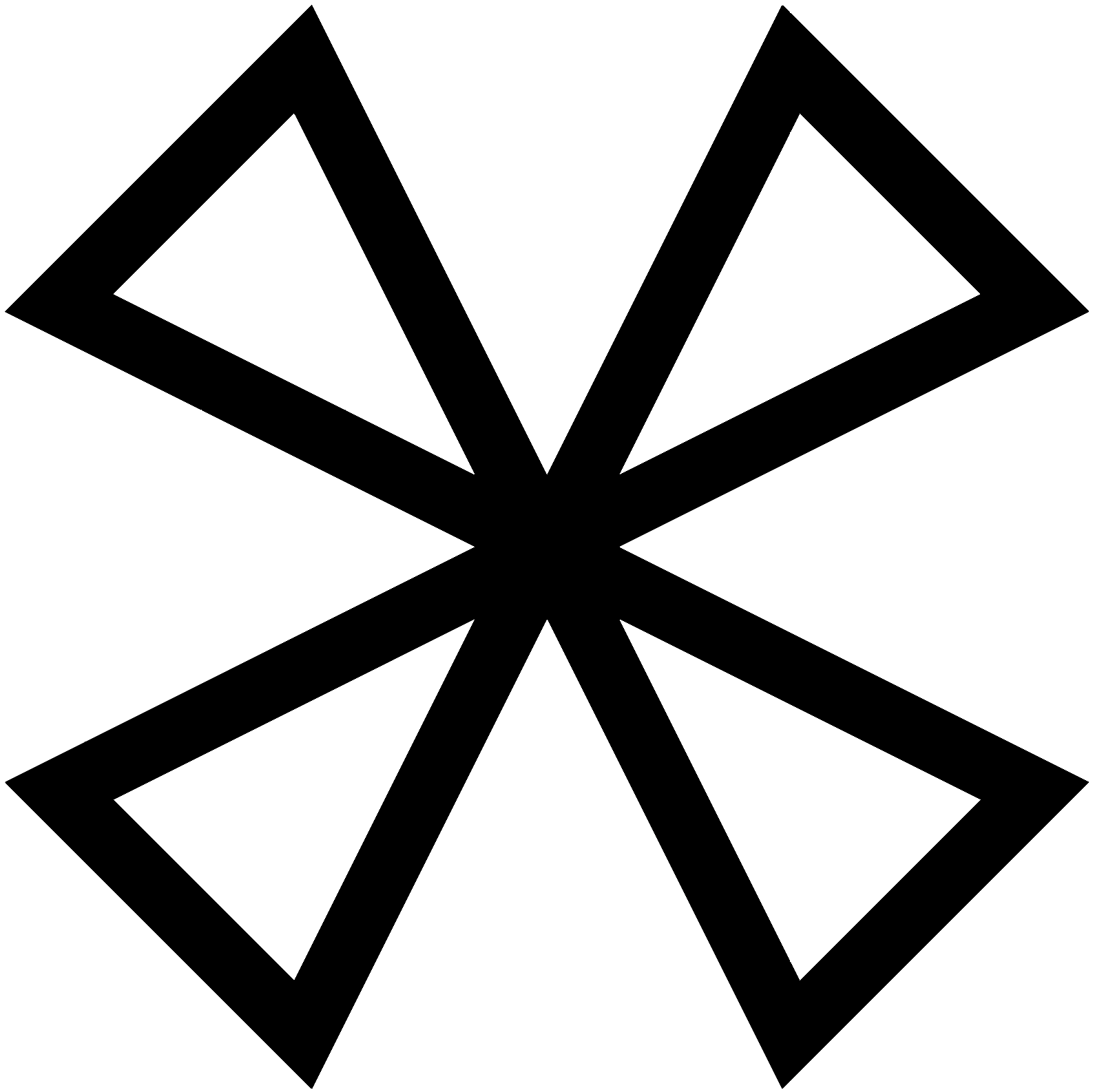}& \cite{paper:lhcbwhiterose}& {\bf 55}$\times${\bf 55} & 10 & Gauss &all & p$^+$-on-n, planar & 300 & 100 & 1.52 & - & CoG+$\eta$&  0.1 & CERN ($\pi$ 120 GeV ) & ($\ominus$) & Timepix3 ASIC (ToT)&  - \\

\includegraphics[width=0.02\textwidth]{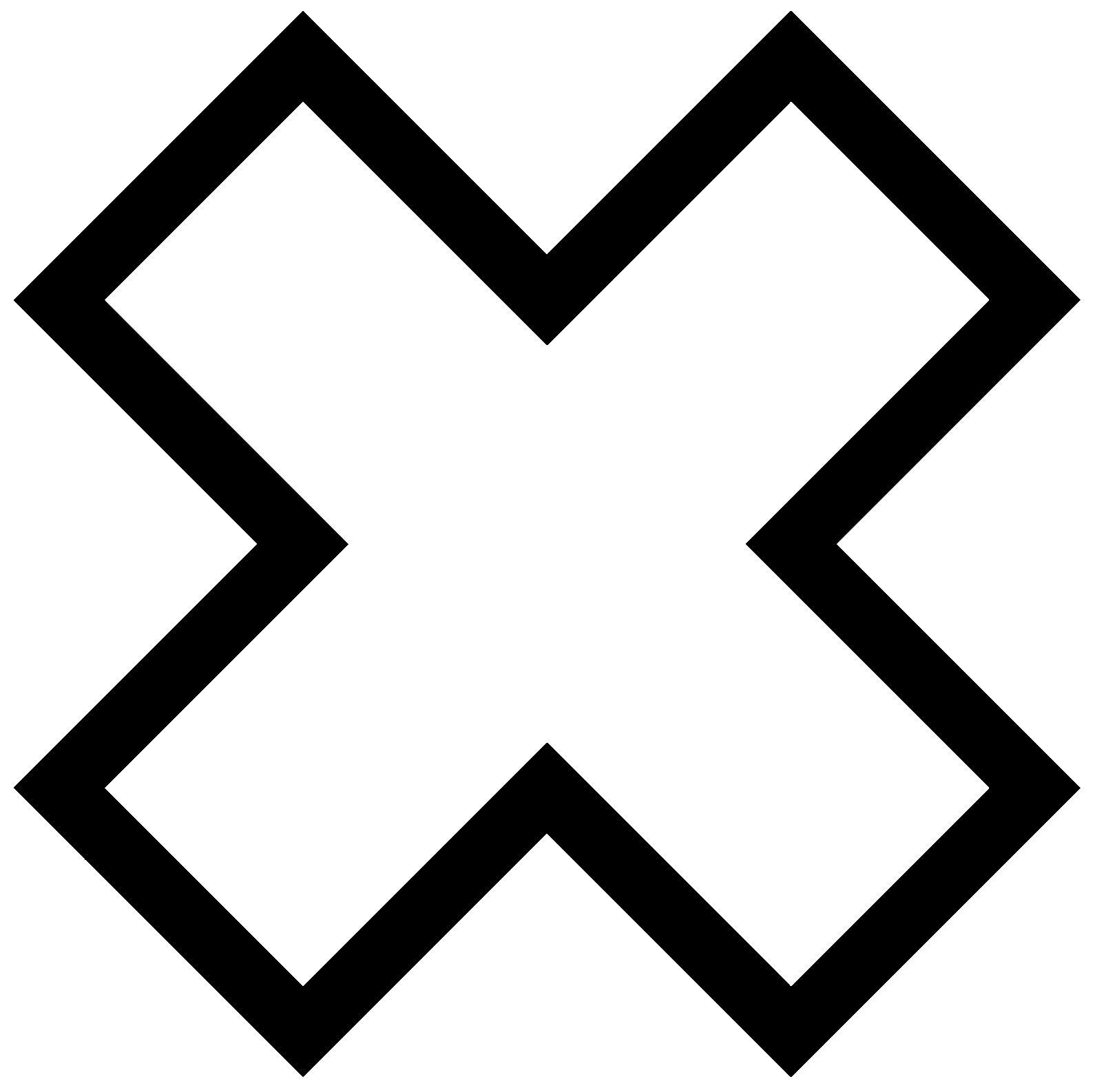}& \cite{paper:lhcbwhiterose}& {\bf 55}$\times${\bf 55} & 4.4 & Gauss &all & p$^+$-on-n, planar & 300 & 10 & 1.52 & - & CoG+$\eta$&  0.2&  CERN ($\pi$ 120 GeV ) & ($\ominus$) & Timepix3 ASIC (ToT)&  - \\

\textcolor{mediumblue}{$\blacktriangle$} &\cite{paper:georg}& {\bf 50}$\times${\bf 50} & 15 & Generalized error function & all & n-in-p, planar & 150 & 120 & 0.7 & - & CoG & - & DESY, e, 5.2 GeV & EUDET, 3.4 $\upmu $m~\cite{man:eudet}  & RD53A & RT \\

$\blacklozenge$ & \cite{paper:atlaspixel} & {\bf 50}$\times$400 & 14 ($\boxast$) & RMS & all & n-in-n, planar (SSGb) & 200 & 150 & $\sim$ 3.0 &- & CoG & 0.3 (stat) & CERN, $\pi$, 180 GeV & microstrip 3-5 $\upmu $m & FE-A/FE-B & -9 \\

\includegraphics[width=0.02\textwidth]{figures/markers/cross9.png}& \cite{master:boronat} & 75 or 50$\times${\bf 50} & 12.8 & Reduced RMS, N=2 & 1 &DEPFET& 50&-&- ($\diamond$)& $\approx$ 40&CoG +$\eta$&-&(\textsection)&EUDET/AIDA~\cite{man:eudetold}& - &- \\

\includegraphics[width=0.02\textwidth]{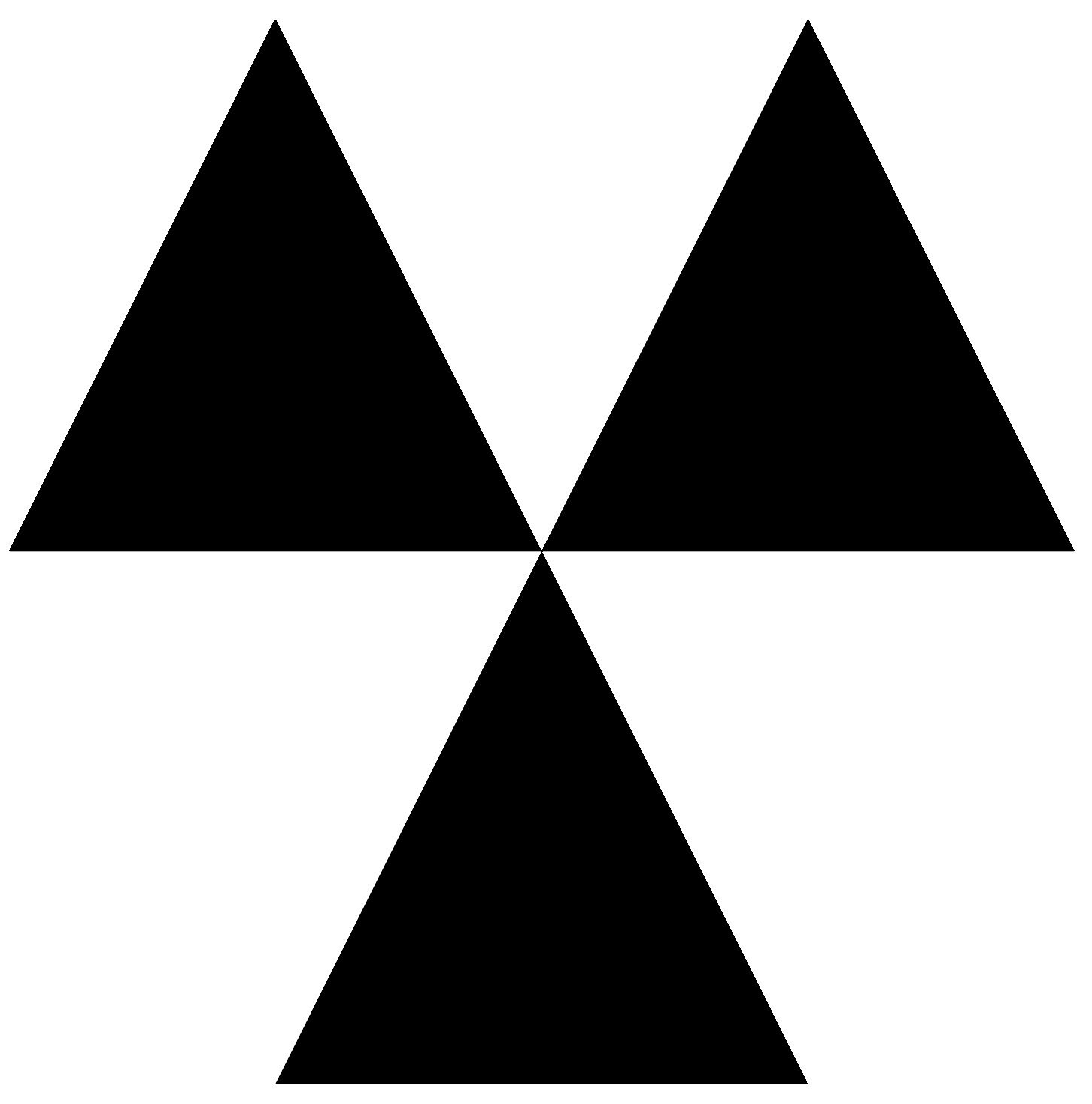}& \cite{paper:atlaspixel} & {\bf 50}$\times$400 &  12.7 ($\boxast$) & Gauss & all & n-in-n, planar (SSGb)  & 200 & 150 & $\sim$ 3.0 & - &CoG & 0.3 (stat) & CERN, $\pi$, 180 GeV & microstrip 3-5 $\upmu $m & FE-A/FE-B & -9 \\

\includegraphics[width=0.02\textwidth]{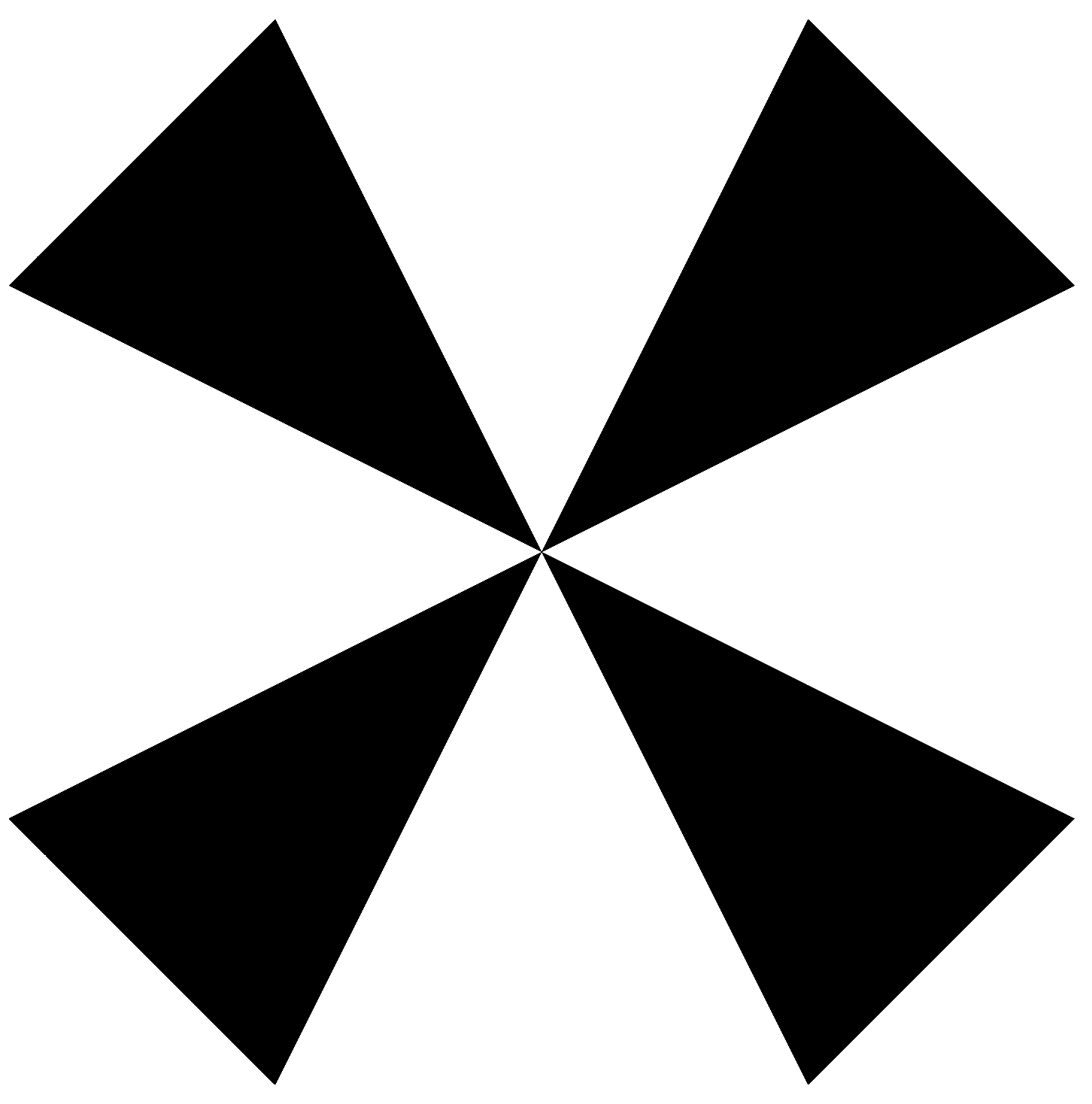}& \cite{paper:atlaspixel} & {\bf 50}$\times$400 & 12.1 ($\boxast$) & Gauss & all & n-in-n, planar (ST2) & 280 & 150 & $\sim$ 3.0 & - &CoG & 0.3 (stat) & CERN, $\pi$, 180 GeV & microstrip 3-5 $\upmu $m & FE-A/FE-B & -9 \\

\includegraphics[width=0.02\textwidth]{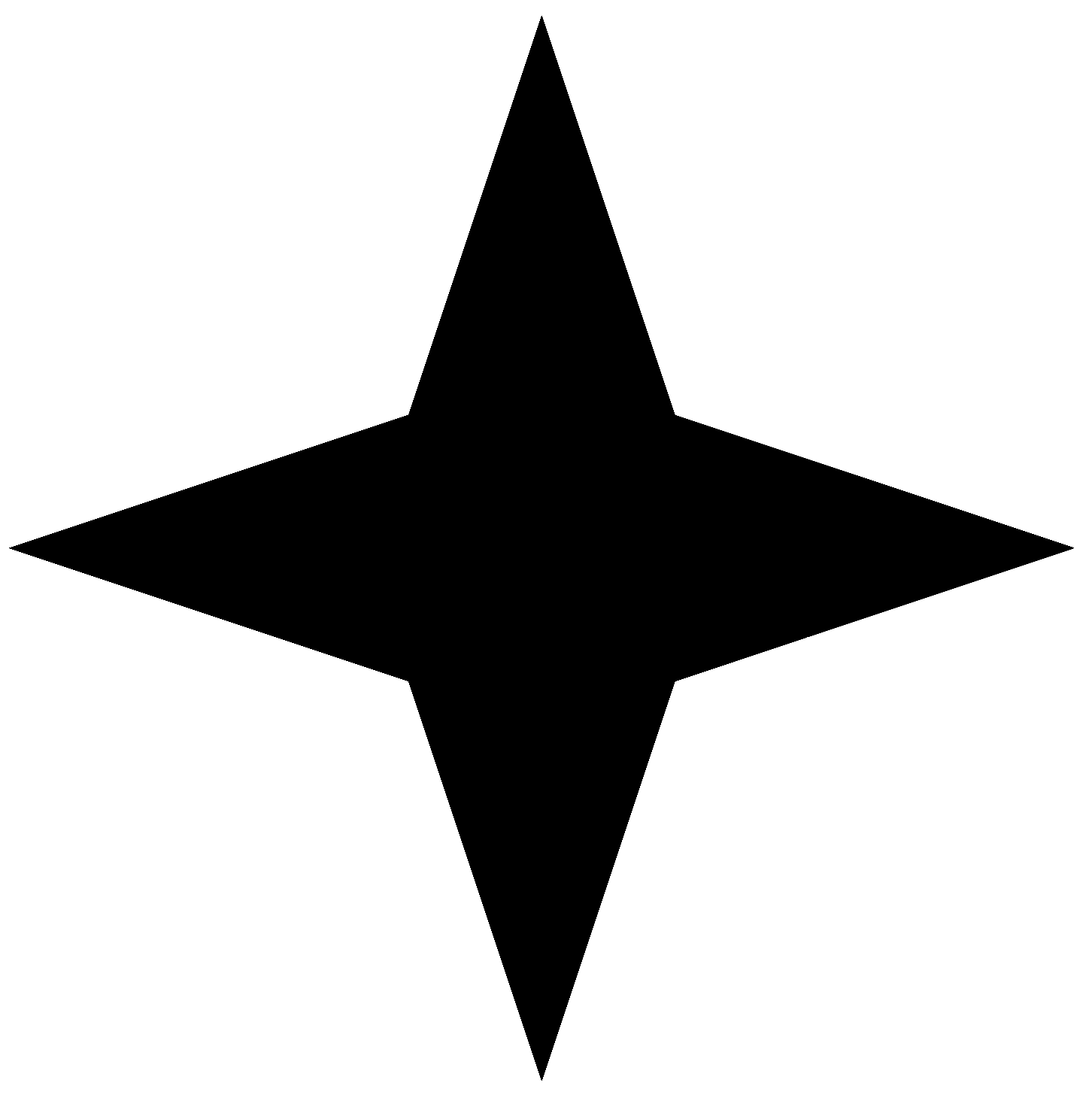} & \cite{paper:atlaspixel} & {\bf 50}$\times$400 & 10.7 ($\boxast$) & Gauss & all & n-in-n, planar (ST1) & 280 & 150 & $\sim$ 3.0 & - &CoG & 0.3 (stat) & CERN, $\pi$, 180 GeV & microstrip 3-5 $\upmu $m & FE-A/FE-B & -9 \\

\includegraphics[width=0.02\textwidth]{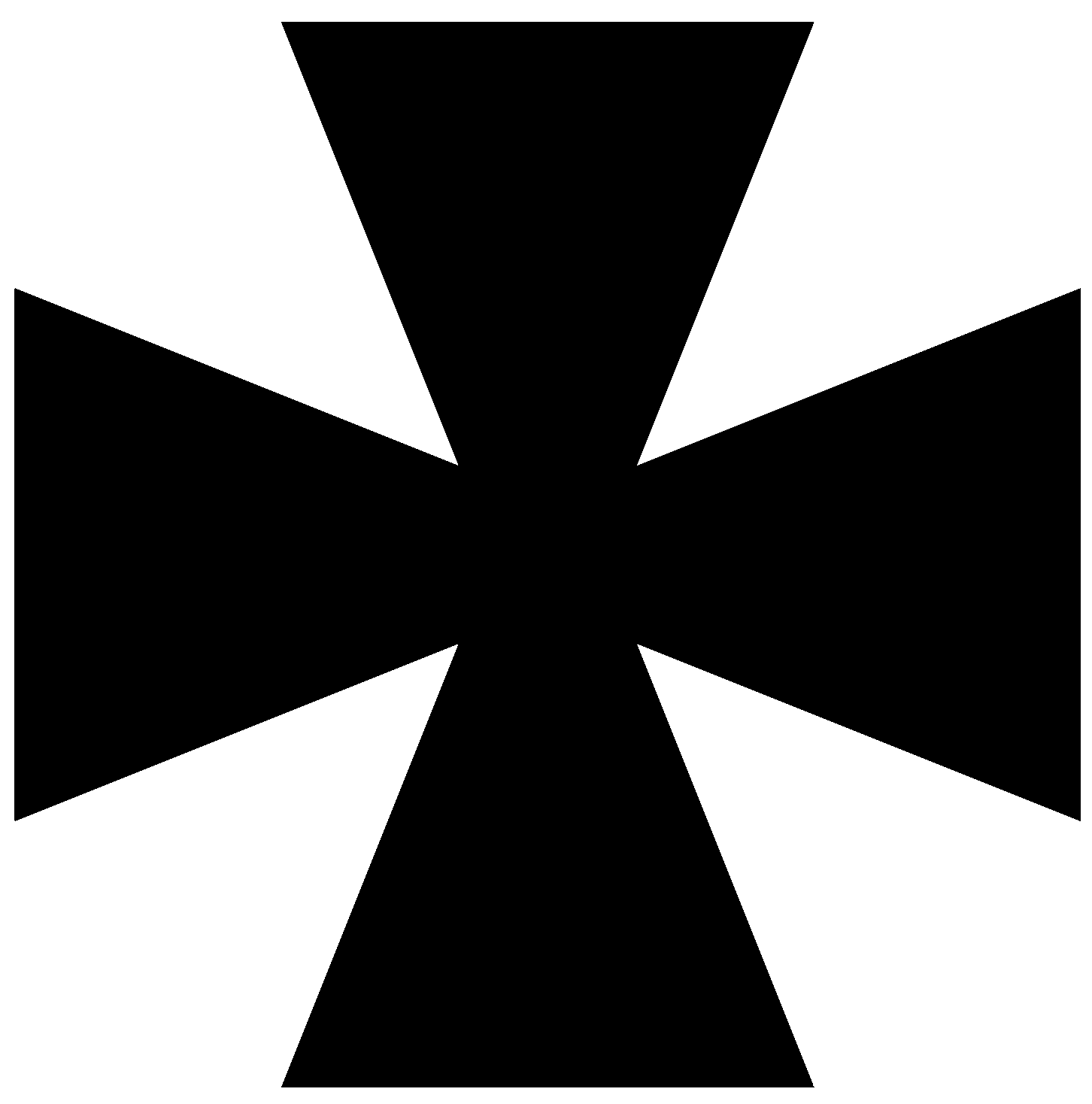}& \cite{paper:atlaspixel} & {\bf 50}$\times$400 & 10.5 ($\boxast$) & Gauss & all & n-in-n, planar (SSG) & 280 & 150 & $\sim$ 3.0 & - &CoG & 0.3 (stat) & CERN, $\pi$, 180 GeV & microstrip 3-5 $\upmu $m & FE-A/FE-B & -9 \\


\Plus & \cite{thesis:atlas} & {\bf 50}$\times${\bf 50} & $\sim$ 14.5 & RMS & 1 & n-in-p, planar & 100 &  50 & 1.0 & - & Pixel center & - & 
CERN,  $\pi$, k and p, 120 GeV & EUDET, 6.7$\upmu $m~\cite{man:eudet}   & RD53A (LIN FE) & - \\

\textcolor{rootred}{\CircleSolid} & This paper & {\bf 50}$\times${\bf 50} & 12.18 &Reduced RMS, N=6 & all & n-in-p, planar & 150 & 120 & $\sim$ 0.4 &100 & CoG& 0.04 (stat)  & DESY, e, 5.2 GeV &EUDET~\cite{man:eudet} & ROC4Sens & RT\\


\XSolidBold &\cite{master:irene}& {\bf 50}$\times$300 & 9.8 & Gauss & all & n-in-n, planar & 285 &150 & $\sim$ 1.5 & - & CoG &- &FNAL, p, 120 GeV & Pixel telescope~\cite{man:fnalpixel}, $\sim$ 6 $\upmu $m &  PSI46digV2.1-r& RT \\

\includegraphics[width=0.02\textwidth]{figures/markers/cross11.png} &\cite{paper:caterina}& {\bf 50}$\times$300 & 8.9& Gauss & all & n-in-n, planar & 285 &150 & 2.4 & - & CoG &- &FNAL, p, 120 GeV & Pixel telescope~\cite{man:fnalpixel}, $\sim$ 8 $\upmu $m &  PSI46digV2.1-r& RT \\

\includegraphics[width=0.02\textwidth]{figures/markers/cross10.png}& \cite{master:boronat} & 75 or 50$\times${\bf 50} & 5.7 & Reduced RMS, N=2 & 2 &DEPFET &50& - &- ($\diamond$)& $\approx$ 40&CoG +$\eta$&-&(\textsection)&EUDET/AIDA~\cite{man:eudetold}& - &- \\

\textcolor{lightblue}{\Square} &\cite{paper:clicseveral} & 130$\times${\bf 40} & 13 & - & - & ($\oslash$) & - & - & - & - & - &- & - & - &ATLASpix\_Simple & -\\ 

\Square & \cite{paper:depfet} & {\bf 32}$\times$24 & 1.6 & RMS&  2$\times$2 & DEPFET & 450 &200 & - & 120 &  CoG +$\eta$ & 0.1 & CERN, $\pi$, 120 GeV   & 6 sensors telescope & - & RT \\

\includegraphics[width=0.02\textwidth]{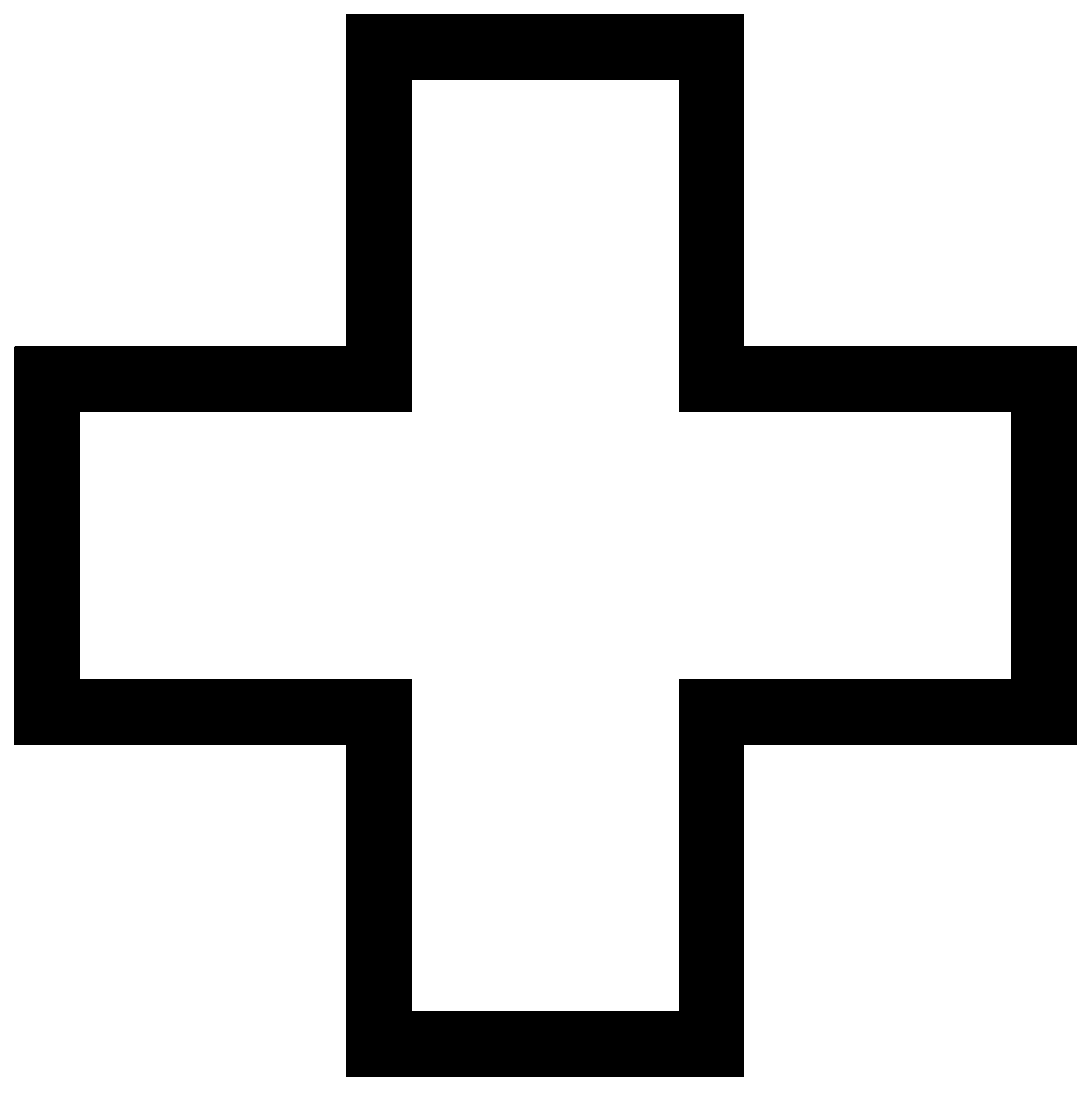}& \cite{paper:soi} ($\divideontimes$) & {\bf 30}$\times${\bf 30} & 4.3 & Gauss & all & SOI sensors & 500 & 130 & (\textbullet)  &$\approx$ 100  & CoG &  0.1 &  CERN, $\pi$, 120 GeV &  ($\ominus$)& - & RT \\

$\boxtimes$ &\cite{report:eudet} &  {\bf 30}$\times${\bf 30} & 2.7 & Gauss & all & MimoTel & 14 ($\odot$)  & - & - &- & - & - & CERN, $\pi$, 120 GeV & EUDET & - &RT\\

\includegraphics[width=0.02\textwidth]{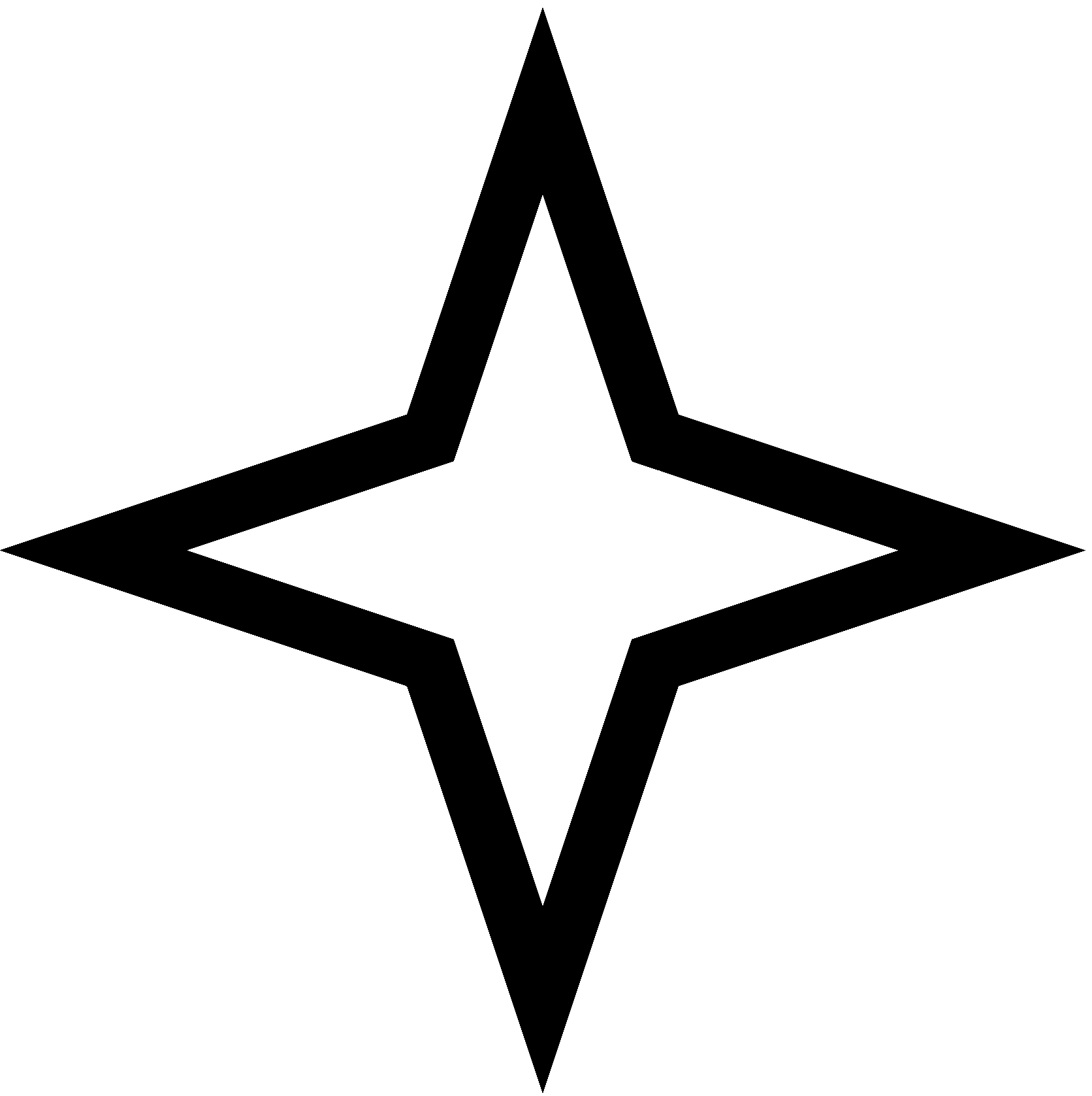} &\cite{paper:clic2} & {\bf 28}$\times${\bf 28} & $\sim$ 6 & RMS In range $\pm$ 20 $\upmu $m & all &  High-Resistivity CMOS  & - & 6 & 0.05 ($\circledast$) & - & CoG + $\eta$ & - & CERN $\pi$, k and p, 120 GeV &($\ominus$) & Investigator chip  & -  \\

\textcolor{lightblue}{\large $\triangle$} & \cite{paper:clicseveral} &{\bf 25}$\times${\bf 25} & 8  & - & - & High-Voltage CMOS C3PD & 50 - & - & - & - & - & - & - & - & - & -  \\

\XSolidBold &\cite{master:irene}& {\bf 25}$\times$600 & 7.7 & Gauss & all & n-in-n, planar & 285 &150 & $\sim$ 2.0 & -  & CoG &- &FNAL, p, 120 GeV & Pixel telescope~\cite{man:fnalpixel}, $\sim$ 6 $\upmu $m &  PSI46digV2.1-r& RT \\

\Plus & \cite{thesis:atlas} & {\bf 25}$\times$100 & $\sim$ 7.2 & RMS & 1 & n-in-p, planar & 150 &  50 & 1.5 &- & Pixel center & - & 
CERN,  $\pi$, k and p, 120 GeV & EUDET, 6.8$\upmu $m~\cite{man:eudet}   & RD53A (LIN FE) & -  \\

\textcolor{mediumblue}{$\blacktriangledown$} &\cite{paper:marco}\ & {\bf 25}$\times$100 & $\sim$ 6.2 & Student-t  &  all & p DRIE column, 3D & 130 &30 &0.9 &-& CoG &-& DESY, e, 5.2 GeV & EUDET~\cite{man:eudet}, 3.8-6.2 $\upmu $m & RD53A & RT \\

\includegraphics[width=0.02\textwidth]{figures/markers/cross11.png}&\cite{paper:caterina}& {\bf 25}$\times$600 & 5.8& Gauss & all & n-in-n, planar & 285 &150 & 2.4 & - & CoG &- &FNAL, p, 120 GeV & Pixel telescope~\cite{man:fnalpixel}, $\sim$ 8 $\upmu $m &  PSI46digV2.1-r& RT \\

\textcolor{mediumblue}{$\blacktriangle$} &\cite{paper:georg} & {\bf 25}$\times$100 & 5.8 & Generalized error function & all & n-in-p, planar & 150 & 120 & 1.0 & - & CoG & - & DESY, e, 5.2 GeV & EUDET, 3.4 $\upmu $m~\cite{man:eudet}  & RD53A & RT \\

\textcolor{rootred}{$\bigstar$} & This paper & {\bf 25}$\times$100 & 5.61 & Reduced RMS, N=6 & all & n-in-p, planar & 150 & 120 & $\sim$ 0.4 & 77 $\pm$ 6&CoG& 0.03 (stat) & DESY, e, 5.6 GeV & Dreimaster & ROC4Sens & RT \\

\textcolor{lightblue}{$\triangledown$} &\cite{paper:simonvertex} & {\bf 25}$\times${\bf 25}  & $\sim$ 5 &- &- & CLICpix2 prototype &  130 & - & - & - & - &- & -&- & ($\div$) &- \\

\includegraphics[width=0.02\textwidth]{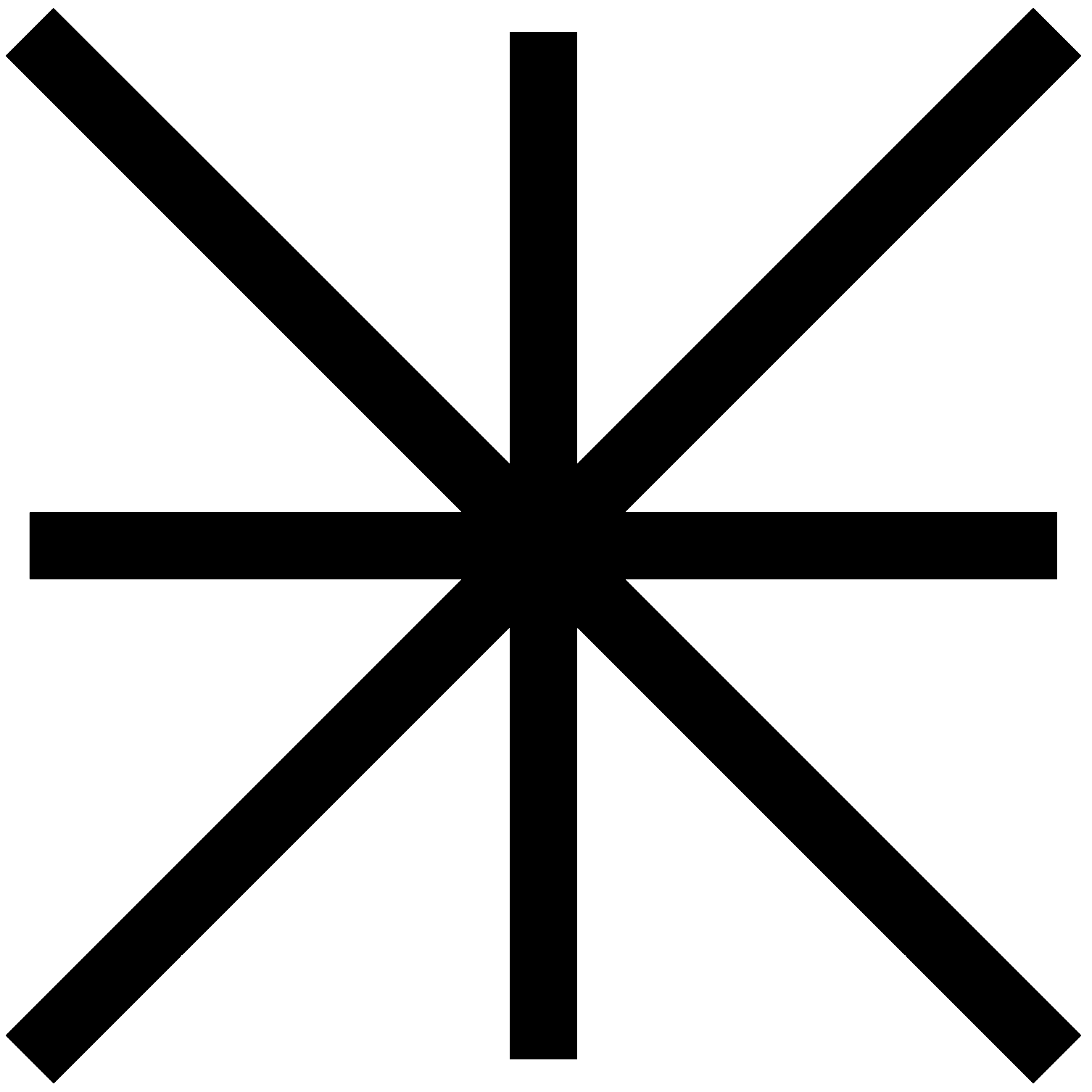} & \cite{paper:depfetlaser} &{\bf 24}$\times${\bf 24} & 4.2 &-&-& DEPFET & 450 &200 &  - & 110 & CoG (\dag) & 0.16 & CERN, $\pi$, 180 GeV & 5 sensors & - & - \\

\Square & \cite{paper:depfet} & 32$\times${\bf 24} & 1.2 &RMS& 2$\times$2 & DEPFET & 450 &200 & - & 120 & CoG +$\eta$ & 0.1 & CERN, $\pi$, 120 GeV  & 6 sensors telescope & - & RT \\

\textcolor{lightblue}{$\lozenge$} &\cite{paper:clic21}& {\bf 21}$\times${\bf 21} & $\sim$ 3  & Gauss & - & ($\triangleleft$)  & 15 & 60-100 &- & 27 &  - &- & CERN SPS & Eudet telescope 2.3~\cite{man:eudet} & - & -\\

\FiveStarOpen &\cite{paper:sofist} &{\bf 20}$\times${\bf 20} & 1.35 & -& 5$\times$5 & SOI 0.2 $\upmu $m & 500 & -& - &300 & - & - & - &- &-& -\\

\Square & \cite{paper:depfet}& {\bf 20}$\times${\bf 20} & 1 &RMS& 3$\times$3 & DEPFET & 450 &200&-  & 200& CoG +$\eta$ & 0.1 & CERN, $\pi$, 120 GeV &6 sensors telescope & - &  RT \\

\textcolor{mediumblue}{$\boxtimes$}&\cite{man:eudet}&  {\bf 18.4}$\times${\bf 18.4} & $\sim$ 2-3 & Gauss & all & ($\boxdot$)  & 50 & - & (\#) & - & - & 0.01 (stat), 0.08 (syst) & DESY, e, 6 GeV &EUDET DATURA~\cite{man:eudet} & - & 18\\

\rotatebox{45}{$\boxtimes$}& \cite{paper:soi2} & {\bf 13.75}$\times${\bf 13.75} & 1.12 & Gauss & >1 & SOI 0.2 $\upmu $m & 114$\pm$6 & 60 & (\$) & ($\star$) & CoG+$\eta$& 0.03 (stat+syst) & CERN, $\pi$, 200 GeV & 3 SOImager-2 sensors&  SOImager-2 & RT\\

$\boxtimes$ &\cite{report:eudet} &  {\bf 10}$\times${\bf 10} & 0.85 & Gauss & all & Mimosa18 & 14 ($\odot$)  & -& - &- & - & - &CERN, $\pi$, 120 GeV & EUDET & -&RT\\

$\bigcirc$ & \cite{paper:fpixsoi} & {\bf 8}$\times${\bf 8} & 0.6 & Gauss& - &FPIX SOI 0.2  $\upmu $m & 400 &70 & - &310 &CoG & 0.01 & FNAL, proton, 120 GeV & 3 FPIX telescope& -& -\\
\hline

\end{tabular}
} 
\captionof{table}{{\scriptsize Selection of spatial resolution measurements from the available literature, performed using non-irradiated pixel detectors of different designs and pitch sizes. The measurements are ordered by decreasing pixel size. The particle beams were perpendicular to the sensors. When in a paper more measurements are available, e.g.  different cluster sizes or algorithms used, the most simple and general one is reported here. Bold in the pitch column highlights the direction of the measurement. The track fitting algorithm has been omitted as the procedures could not be shortly summarized in the table and we refer to the primary papers for this information. \newline
(\ddag) Measurement using cluster skewness as parametrization are also described and improve the position resolution significantly. \newline
(*) For measurement at 27.1\textdegree{}. These measurements were not optimized for data taking at vertical incidence, but at specific incidence angles matching the conditions of the target experiment. \newline
(\textsection) CERN ($\pi$, 120 GeV) / DESY (e, 4 GeV).\newline
($\ominus$) Timepix3 Telescope, \SI{2.3}{\micro\meter}~\cite{man:timepix3}. \newline
($\boxast$) Telescope extrapolation uncertainty not subtracted. \newline
($\divideontimes$) Later publication~\cite{paper:soilater} with generalization of the standard $\eta$-correction adapted for arbitrary cluster sizes obtains \SI{1.5}{\micro\meter} for the FZ-n wafer and about \SI{3.0}{\micro\meter} for the Double SOI Czochralski type p wafer. \newline
($\oslash$) Monolithic High-Voltage CMOS.\newline
(\dag) Measurements with $\eta$ corrections and laser calibration are also present. Spatial resolution has been further improved, in a later publication a 2D correction and energy split correction are added~\cite{paper:depfetbelle}.\newline
(\textbullet) Two Seed Method (TSM). \newline
($\circledast$) Analysis threshold, triggering threshold is 0.15 ke.\newline
($\div$) 5-bit time-over-threshold (ToT). \newline
($\triangleleft$) n in p, High Voltage CMOS.\newline
($\boxdot$) AMS 350 nm CMOS technology.\newline
(\#) Charge in a single pixel $\geq$ 6$\times$noise. \newline 
(\$) Double Threshold method. \newline
($\star$) seed 5 + neighbour 3. \newline
($\diamond$) The seed pixel threshold is indicated as a multiple of the pixel noise, the threshold on the neighbouring pixels is 12 times the noise.\newline
($\odot$) Active thickness. The total thickness is \SI{700}{\micro\meter}.}}
\label{table:resolution}

\end{minipage}
\end{sideways}
%


\endgroup
\newpage
\begingroup
\let\clearpage\relax 
\twocolumn


\section{Modules used in this study}
\label{sec:3mcombo}

Table~\ref{table:3Mcombo} lists the sensors used for all the measurements presented in this study. The detailed description of the designs is presented in Ref.~\citep{paper:designJoern}. The external planes A and C are always non-irradiated. The fluence of the B plane is also reported in the table.

\begin{sidewaystable}
\centering
\begin{tabular}{cc|ccc|cc|c}
 \multicolumn{2}{c}{plane A} & \multicolumn{2}{c}{plane B (DUT)}& Irradiation  & \multicolumn{2}{c}{plane C} & Measurement  \\
 Nr. & Design& Nr. & Design&$[\phi_{\mathrm{eq}}$/\SI{e15}{\per\square\centi\meter}$]$ &Nr. & Design&\\
\hline
148 &  P-stop RD53A routing (FDB)  & 150 &P-stop basic (FDB)&0 & 163 &  P-spray RD53A routing (FDB)&  ``small" angle scan\\
109 & P-stop basic (FTH) & 148& P-stop RD53A routing (FDB)&0 & 110& P-stop common punch-through (FTH)& full angle scan \\
148& P-stop RD53A routing (FDB) &146& P-spray basic (FDB) & 0&163&  P-spray RD53A routing (FDB)& ``small" angle scan\\
148& P-stop RD53A routing (FDB)&163&  P-spray RD53A routing (FDB)&0&  150 &P-stop basic (FDB) & momentum scan  \\
\hline
 148 &  P-stop RD53A routing (FDB) &120 & P-stop basic (FTH) &2.1, proton &163&  P-spray RD53A routing (FDB)& full angle scan \\
148& P-stop RD53A routing (FDB)  & 130&P-stop basic (FDB) &2.1, proton &109 & P-stop basic (FTH)& momentum scan \\
148& P-stop RD53A routing (FDB) &194&P-stop basic (FDB)&3.6, neutron &150 &P-stop basic (FDB)& full angle scan  \\
\end{tabular} 
\caption{The table lists the three-master combinations used in this paper. Each sensor is uniquely identified by the listed number. ``Small" angle scan refers to an angle scan performed around the optimal angle for resolution, to identify it, and compare the resolution. The combinations used for the full angle scan have been used also for the control plots presented in this paper. FDB (Si-Si direct bonded) and FTH (Physical thinned) refer to two different sensor substrates. The detailed description of the designs and the substrates is given in Ref.~\citep{paper:designJoern}.}
\label{table:3Mcombo}
\end{sidewaystable} 


\endgroup

\section{Requirements on the collected charge}
\label{sec:additional}
\begin{figure}[!h]
\centering
\includegraphics[width=0.5\textwidth]{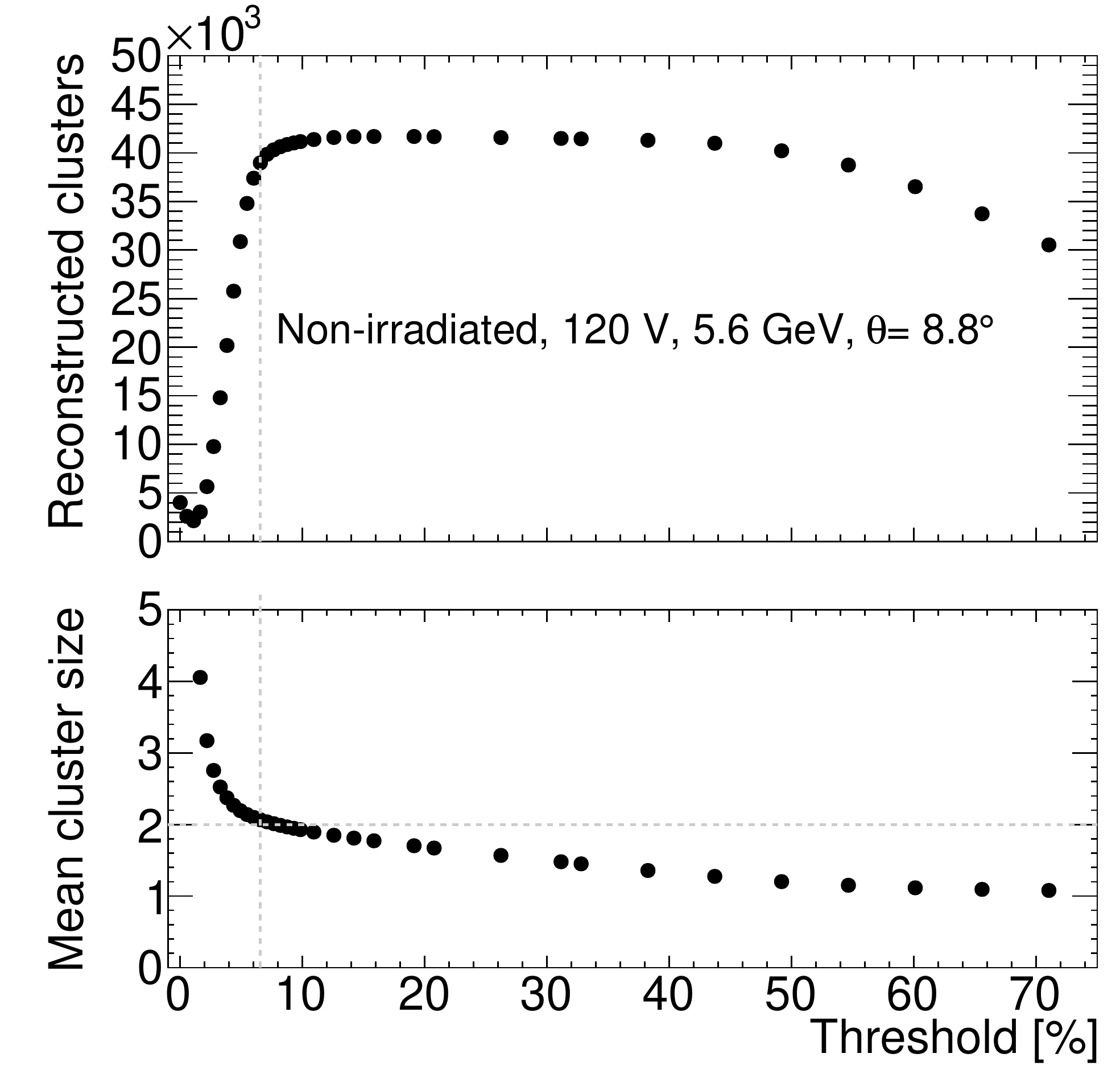}

\caption{Number of reconstructed clusters and average cluster size as a function of the offline threshold. The horizontal dashed line marks the expected cluster size from charge sharing at the beam incidence angle at which the measurement was performed. The vertical dashed line shows the offline threshold used throughout this study.}
\label{fig:thr}
\end{figure}

\f~\ref{fig:thr} shows the number of reconstructed clusters and the average cluster size as a function of the offline threshold. An offline threshold too low results in including noisy pixels in the clusters. If it is too high, the detection efficiency will be degraded. The thresholds are expressed as a percentage of the MPV of the cluster charge distribution.  

\begin{figure}[!t]
\centering

\includegraphics[width=0.5\textwidth]{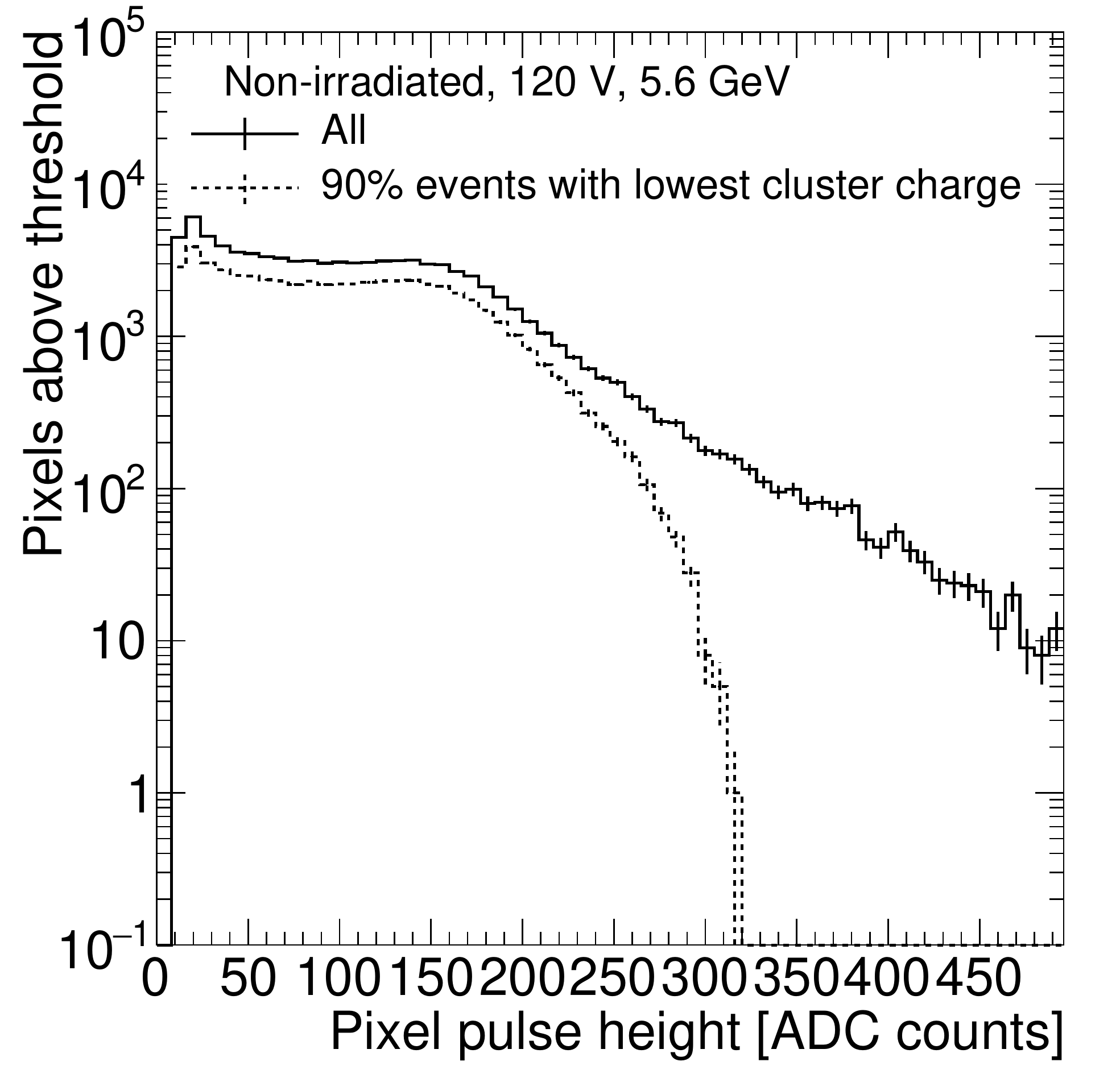}

\caption{Pixel charge of all clusters above threshold on the central three-master plane. A threshold at \thr \ of the cluster charge MPV is applied. The solid line shows all clusters above threshold. The dashed line indicates the pixel charge after hits above the 90\% quantile of the cluster charge distribution are rejected on each of the three three-master planes.}
\label{fig:pxcharge}
\end{figure}

In \f~\ref{fig:pxcharge}, the solid line shows the pixel charge of all clusters above thresholds on the central three-master plane. The dashed line indicates the pixel charge after hits above the 90\% quantile of the cluster charge distribution are rejected on each of the three three-master planes.

\end{document}